\documentclass[journal]{IEEEtran}
\usepackage{amsmath,graphicx}
\usepackage{graphicx} 
\usepackage{bm}
\usepackage{amsmath}
\usepackage{amssymb}
\usepackage{overpic}
\usepackage{caption}
\usepackage{rotating}
\usepackage{algorithm} 
\usepackage{algorithmic} 
\usepackage{array}
\usepackage{booktabs}
\usepackage{multirow} 
\usepackage{caption}
\usepackage{graphicx}
\usepackage{subfigure}
\usepackage{float}
\usepackage{tikz}
\usetikzlibrary{spy} 
\usepackage{color}
\usepackage{wrapfig}
\usepackage[absolute,overlay]{textpos}
\usepackage{hyperref}

\usepackage{tikz}
\usetikzlibrary{positioning}
\tikzset{>=stealth}

\newcommand{\x}{\mathbf x}
\newcommand{\y}{\mathbf y}
\newcommand{\A}{\mathbf A}
\newcommand{\omg}{\mathbf \Omega}
\newcommand{\tabincell}[2]{
	\begin{tabular}{@{}#1@{}}#2\end{tabular}
} 
\renewcommand{\P}{\mathbf P}
\newcommand{\z}{\mathbf z}

\newcommand{\Z}{\mathbf Z}
\newcommand{\Y}{\mathbf Y}
\newcommand{\W}{\mathbf W}

\newcommand{\I}{\mathbf I}

\newcommand{\M}{\mathbf M}

\newcommand{\R}{\mathsf R}

\newcommand{\eps}{\bm \epsilon}

\newcommand{\diag}{\mathsf{diag}}

\begin{document}
%
\title{DECT-MULTRA: Dual-Energy CT Image Decomposition With Learned Mixed Material Models and  Efficient Clustering}
%
%

\author{Zhipeng~Li,~\IEEEmembership{Student Member,~IEEE,} Saiprasad Ravishankar,~\IEEEmembership{Member,~IEEE,} Yong~Long$^*$,~\IEEEmembership{Member,~IEEE,} and~Jeffrey~A.~Fessler,~\IEEEmembership{Fellow,~IEEE}
\thanks{This work was supported in part by NSFC (61501292) and NIH grant U01 EB018753. (Corresponding author: Yong Long.) }


\thanks{Z. Li and Y. Long are with the University of Michigan - Shanghai Jiao Tong
	University Joint Institute, Shanghai Jiao Tong University, Shanghai 200240,
	China (email: zhipengli@sjtu.edu.cn, yong.long@sjtu.edu.cn).}

\thanks{S. Ravishankar is with the Department of Computational Mathematics,
	Science and Engineering, and the Department of Biomedical Engineering,
	Michigan State University, East Lansing, MI, 48824 USA (email: ravisha3@msu.edu).}
\thanks{ J. A. Fessler is with the Department of Electrical Engineering and Computer Science, University of Michigan, Ann Arbor, MI, 48109 USA (email: fessler@umich.edu).}}

%
%


\maketitle
\begin{abstract}
 Dual energy computed tomography (DECT) imaging plays an important role in advanced imaging applications due to its material decomposition capability. Image-domain decomposition operates directly  on CT images using linear matrix inversion, but the decomposed material images can be severely degraded by noise and artifacts. This paper proposes a new method dubbed DECT-MULTRA for image-domain DECT material decomposition that combines conventional penalized weighted-least squares (PWLS) estimation with regularization based on a mixed union of learned transforms (MULTRA)
 model. Our proposed approach pre-learns a union of common-material sparsifying transforms from patches extracted from all the basis materials, and a union of cross-material sparsifying transforms from multi-material patches. The common-material transforms capture the common properties among different material images, while the cross-material transforms capture the cross-dependencies.  The proposed PWLS formulation is optimized efficiently by alternating between an image update step and a sparse coding and clustering step, with both of these steps having closed-form solutions. The effectiveness of our method is validated with both XCAT phantom and clinical head data. The results demonstrate that our proposed method provides superior material image quality and decomposition accuracy compared to other competing methods.
\end{abstract}

\IEEEpeerreviewmaketitle

\section{Introduction}
X-ray computed tomography (CT) is a popular imaging technique used in many clinical applications. Compared to conventional X-ray CT, dual energy CT (DECT) provides two sets of attenuation measurements by exploiting two different energy spectra. DECT enables enhanced tissue characterization due to its ability to produce images of different constituent materials such as soft-tissue and bone in scanned objects, known as material decomposition \cite{Mendonca2014A,Mccollough2015Dual}. This decomposition of a mixture into multiple basis materials depends on the principle that the attenuation coefficient is material and energy dependent. DECT is of great interest in many clinical and industrial applications such as iodine quantification \cite{Li2013Iodine,Chandarana2011Iodine}, kidney stone characterization \cite{Primak2007Noninvasive}, virtual monoenergetic imaging \cite{Lifeng2011Virtual}, and security inspection \cite{Liu2010Feasibility}. 
DECT measurements are also used to obtain element decompositions (of Hydrogen, Carbon, Nitrogen, Oxygen, etc.) for radiation therapy applications that require atomic compositions and densities for treatment planning \cite{Lalonde2016,shen182,shen189}.

\vspace{-0.03in}
\subsection{Background}

Methods for DECT material decomposition can be characterized into direct decomposition \cite{long:14:mmd}, projection-domain \cite{noh:09:ssr}, and image-domain \cite{xue:2017:statistical} methods. 
Direct decomposition greatly reduces the noise and cross-talk artifacts in the reconstructed basis material images and improves decomposition accuracy, but it is computationally expensive because of the repeated forward and back-projections required between basis material images and DECT sinograms.
Projection-domain decomposition converts the low- and high-energy measurements into sinograms of basis materials, from which the material images are then reconstructed. 
Although these methods have the theoretical advantage of avoiding  beam-hardening artifacts, they require accurate system calibrations that use nonlinear models \cite{MM:11:accu}.   
Projection-domain and direct decomposition methods require sinograms that are not available to users of current commercial DECT scanners. Image-domain methods directly decompose the readily available reconstructed high- and low-energy \emph{attenuation images} into basis material images, and are more efficient than projection-domain and direct decomposition techniques in terms of computational cost, but their efficacy may be limited due to sensitivity to noise and artifacts. Image-domain DECT methods may be substantially improved by exploiting learned or adapted prior information from existing big databases of CT images. Many related methods have been proposed in this regard, such as spectral prior image constrained compressed sensing (PICCS) \cite{yu:16:spi}. 


Xu \textit{et al.} \cite{xu:12:ldx} developed dictionary learning (DL) methods for conventional low-dose CT (LDCT) image reconstruction by combining the PWLS approach with regularization involving a pre-learned redundant dictionary. 
Later, dictionary learning was applied to DECT for denoising \cite{mechlem:16:dbi} and reconstruction \cite{Li2012Dual,zhao:12:ddl}.
 Recently, a tensor dictionary learning (TDL) \cite{zhang:15:tbd} method was proposed for spectral CT reconstruction by accommodating sparsity in both spatial and spectral dimensions. Although the TDL-based scheme showed promise in preserving fine tissue features, it underperformed in terms of preserving edge information and removing artifacts and noise. To overcome the limitations of the TDL approach, an $\ell_0$ \mbox{``norm"} of the image gradient was incorporated ($\ell_0$TDL) \cite{WU2018538} to recover edge information by penalizing the number of non-zeros in the gradient domain rather than the image gradient magnitudes. However, the sparse coding step in the dictionary model is computationally expensive, and the dictionary learning problem is typically NP-Hard in general. Recently, a generalized analysis dictionary model called sparsifying transform (ST) model was investigated in \cite{Ravishankar2015,ravishankar:13:lst}, where sparse coefficients are efficiently obtained in the transform domain by thresholding-type operations. Learned sparsifying transforms have recently shown promise for LDCT image reconstruction compared to nonadaptive methods \cite{Zheng2016Low,Ye:17:adaptive}. 
Zheng \textit{et al.} \cite{zheng2018} generalized the single square sparsifying transform learning-based reconstruction approach to a union of sparsifying transforms scheme and showed its promise for LDCT reconstruction. 
That approach pre-learned a collection of sparsifying transforms such that image patches are adaptively assigned (clustered) to their best-matching sparsifying transforms during the reconstruction process. 
Apart from dictionary and sparsifying transform learning-based methods, deep learning techniques have received attention in the field of DECT material decomposition recently. Deep learning algorithms usually learn deep filtering models to achieve high image quality for specific datasets. 
Liao \textit{et al.} \cite{YuCNN} proposed a deep learning-based framework to obtain basis material images via cascaded deep convolutional neural networks (CD-ConvNet) that approximately capture a non-linear mapping from the measured energy-specific CT images to the desired decomposed basis material images. Zhang \textit{et al.} \cite{Niu:18:butterfly} developed a model-based butterfly network to perform image-domain material decomposition for DECT. This network has a double-entry double-output crossover architecture that exploits the relationship between the CT data model and the neural network. These methods are all fully supervised learning methods requiring long time and large datasets for training and also tend to be less generalizable, which may limit their use. These approaches also do not consider material models and properties. In this work, we explicitly exploit the common properties and cross-dependencies between different basis materials to improve the performance of conventional DECT decomposition.

\vspace{-0.05in}
\subsection{Contributions}
Considering the common properties (e.g., each material image could be modeled as piece-wise smooth) and cross-dependencies (e.g., the material images share similar boundary structures) among different basis material images, here we propose a new image-domain DECT material decomposition method dubbed DECT-MULTRA that combines conventional PWLS estimation with regularization based on a mixed union of learned sparsifying transforms (MULTRA) model. In this MULTRA framework, we first efficiently pre-learn unions of sparsifying transforms from image patches extracted from a dataset of material density images. One group of transforms (dubbed common-material transforms) is learned to sparsify features common across different basis material density images, and another group (dubbed cross-material transforms) is learned to sparsify the dependencies between such material density maps. 
The DECT-MULTRA formulation effectively incorporates the pre-learned material models in a clustering-based framework.
We propose an exact and efficient alternating minimization algorithm for the image-domain DECT-MULTRA decomposition problem that alternates between an image update step and a sparse coding and clustering step, each of which has closed-form solutions. 

We compare our proposed DECT-MULTRA method with recent image-domain techniques such as DECT-ST \cite{li2018image} as well as other related methods. Numerical experiments with XCAT phantom and patient data demonstrate that the proposed method significantly improves the quality of decomposed material images compared to other techniques. Moreover, material models learned with the XCAT phantom generalize well to patient data.

\vspace{-0.05in}
\subsection{Organization}
The rest of this paper is organized as follows. \mbox{Section \ref{Sec:formulation}} describes the formulations for image-domain DECT decomposition with regularization based on learned sparsifying transforms. Section \ref{Sec:algorithm} derives the algorithms for learning the mixed union of sparsifying transforms model and  for material image decomposition. Section \ref{Sec:experiment} presents detailed experimental results on XCAT phantom and patient data along with comparisons. Section \ref{Sec:conclusion} concludes the paper and mentions areas of future work.

\section{Problem Formulation for Material Image Decomposition}
\label{Sec:formulation}
This section discusses the proposed formulation for image-domain decomposition incorporating the MULTRA model and its variations.
\vspace{-0.05in}
\subsection{DECT-MULTRA Formulation}
For image-domain DECT, we start with two scanner reconstructed images at each energy and form a stacked two-channel image vector $\y=(\y_H^T, \y^T_L)^T \in \mathbb R^{2N_p}$, where $\y_H$ and $\y_L$ are the attenuation maps (images) at high and low energy, respectively, and $N_p$ is the number of pixels in each map. Vector $\x=( \x_1^T, \x_2^T)^T \in \mathbb R^{2  N_p }$ denotes the stacked material density images (unknown), where $\x_l=(x_{l1},\dots,x_{ln},\dots,x_{lN_p})^T \in \mathbb R^{N_p}$ represents the $l$th material for $l = 1, 2$.\footnote{This work focuses on decomposing a mixture of two materials.} The stacked attenuation maps are related to the stacked densities as $\y \approx \A \x$, where $\A \in \mathbb{R}^{2N_p \times 2 N_p}$ is a mass attenuation coefficient matrix that is a Kronecker product of $\A_0$ and the identity matrix $\I_{N_p}$ \cite{xue:2017:statistical}, i.e., $\A = \A_0 \otimes \I_{N_p}$, where $\A_0$ is a $2\times 2$ material decomposition matrix defined as follows:
\vspace{-0.09in}
\begin{equation}
\A_0=  \Bigg(
\begin{array}{cc}
\varphi_{1H} & \varphi_{2H} \\
\varphi_{1L} & \varphi_{2L} \\
\vspace{-0.22in}
\end{array}
\Bigg),
\vspace{-0.1in}
\end{equation}
where $\varphi_{lH}$ and $\varphi_{lL}$ denote the mass attenuation coefficient of the $l$th material at high and low energy, respectively. In this paper, we obtain these four values (for two materials) as $\varphi_{lH} = \mu_{lH}/ \rho_l$ and $\varphi_{lL} = \mu_{lL}/ \rho_l$, where $\mu_{lH}$ and $\mu_{lL}$ denote the linear attenuation coefficient of the $l$th material at high and low effective energy, respectively, and $\rho_l$ denotes the density of the $l$th material. For the density $\rho_l$, we use the theoretical value 1 g/cm$^3$ for water and 1.92 g/cm$^3$ for bone. To obtain the value of $\mu_{lH}$ and $\mu_{lL}$,  we manually select two uniform areas in $\y_H$ and $\y_L$ that contain the basis materials and then compute the average pixel values in these areas \cite{niu2014iterative}.

Directly solving for $\x$ in $\y\approx \A\x$, called direct matrix inversion decomposition, would produce significant noise in the result. Our proposed approach models the underlying basis material densities using a common-material and a cross-material image sparsity model. The models apply to image patches, where we say that a patch $\mathbf{q}$ is sparsifiable by a transform or operator $\omg$ if $\omg \mathbf{q} \approx \z$, where $\z$ has many zeros and the error in the sparse approximation is small \cite{ravishankar:13:lst}. In the common-material model, we extract patches from all basis materials' images independently and assume they are sparse under a common union (or collection) of sparsifying transforms \cite{Wen2015Structured}.\footnote{Such unions of transforms provide enhanced sparsification of images and are a richer model than a single transform.} Every patch extracted from some material image is assumed to be best sparsified (or best matched) by a particular transform in the collection. The common-material transforms capture features that sparsify the common properties among various basis materials. In the cross-material model, we extract patches from the same spatial location of different basis materials and stack them to form larger multi-material (3D) patches\footnote{We focus on 2D images here, and 3D means 2D with one more channel direction consisting of the different materials.}, that we assume are sparsified by a collection of cross-material sparsifying transforms. These transforms sparsify the cross-dependencies among the material images, and may be particularly suited for patches that straddle the boundaries between multiple materials, whereas spatial regions where only one material is present may be less suited to the cross-material model (or more suited to the common-material model). 

Based on the above model, we propose to obtain the image-domain DECT decomposed images by solving the following optimization problem:

\begin{equation}
\label{Eq:cost_func}
\min_{ \x \in \mathbb R^{2 N_p}}  \frac{1}{2} \| \y- \A \x\|^2_\W  + \R(\x), \tag{P0}
\end{equation}
where we define the regularizer $\R(\x)$ as
\begin{equation}
\label{Eq:cost_func_reg}
\begin{split}
\hspace{-0.1in}
{\hspace{-0.03in} \min \limits_{ \{ \z_{j}, C_{k_r}^r \} }\hspace{-0.03in} \sum_{r=1}^2   \sum_{k_r=1}^{K_r} \hspace{-0.03in} \sum_{j\in C_{k_r}^r } \hspace{-0.03in} \beta_r \Big\{ \left\| \omg_{r,k_r} \P_{j}  \x - \z_{j} \right\|_2^2+  \gamma_r^2 \left\|  \z_{j}\right\|_0 \Big\}, \hspace{-0.03in}} 
\end{split}
\end{equation}
and $r=1,2,$ represent the common-material and cross-material models, respectively. The operator $\P_{j}\in \mathbb R^{2m\times 2N_p}$ is the patch extraction operator that extracts the $j$th patch of materials as a vector $\P_{j} \x$. 
The patch is constructed by stacking together the 2D patches extracted from the same spatial location of different basis materials. Each such multi-material patch is grouped with the best matching (sparsfiying) transform in either the common or cross material models. Parameter $K_r$ denotes the number of clusters in the $r$th model and  $C_{k_r}^r$ denotes the indices of all the patches matched to the $k_r$th transform (class) in the $r$th model. 
$\{\omg_{1,k_1}\}_{k_1=1}^{K_1}$ denotes a pre-learned union of common-material transforms and $\{\omg_{2,k_2} \}_{k_2=1}^{K_2}$ denotes a union of cross-material transform matrices, where each individual transform is assumed \mbox{unitary. \footnote{The unitary assumption simplifies the proposed algorithm in Section \ref{Sec:algorithm}.} }
For $r=1$, each transform $\mathbf{\Omega}_{1,k_1} \in \mathbb R^{2m\times 2m}$ is a block diagonal matrix that sparsifies individual material images' patches independently without mixing them. All the smaller constituent block matrices are of size $m\times m$, which are learned from vectorized individual material patches and then used to form the larger matrix $\mathbf{\Omega}_{1,k_1}$.  On the other hand, for $r=2$, each sparsifying transform $\mathbf{\Omega}_{2,k_2} \in \mathbb R^{2m\times 2m}$ is a general matrix learned from stacked material patches, and is used to sparsify the entire 3D patches.
Each patch $\P_j \x$ is mapped to the best matching transform domain, where it is approximated by the sparse vector $\z_j \in \mathbb{R}^{2m}$. The $\ell_0$ \mbox{``norm"} enforces sparsity by penalizing the number of non-zeros in $\z_j$, with $\gamma_r$ (a different sparsity penalty weight for each $r$) controlling the sparsity level. The parameters $\beta_r>0$ control the balance between noise and image resolution in the decomposition.


We model the acquired attenuation maps $\y$ with additive Gaussian distributed noise $\mathbf{\eps}\in \mathbb R^{2N_p}$ as $\y = \A \x +\eps$. Assuming that the noise is uncorrelated between the high and low energy attenuation images \cite{Zhang2014Model}, the statistical weight matrix $\W \in \mathbb R^{2N_p\times 2N_p}$ in (\ref{Eq:cost_func}) is a block-diagonal matrix. We also approximate the noise in each attenuation image as being independent and identically distributed (i.i.d.) over pixels \cite{niu2014iterative}. Thus, $\W$ is expressed as $\W = \W_j \otimes \I_{N_p}$, where $\W_j = \diag{(\sigma_H^2, \sigma_L^2)}^{-1}$, and $\sigma_H^2$ and $\sigma_L^2$ denote the noise variances for pixels in the high and low energy attenuation maps, respectively. 

\vspace{-0.05in}
\subsection{Variations of (\ref{Eq:cost_func})}
\label{Sec:II-B}
While (\ref{Eq:cost_func}) uses a mixed (common and cross material) model, a simpler alternative formulation would involve only the cross-material component (dubbed DECT-CULTRA), with $\R(\x)$ defined as follows:
\begin{equation}
\label{eq:cultra_Reg}\hspace{-0.1in}
\R(\x)\hspace{-0.03in} = \hspace{-0.08in}\min \limits_{ \{ \z_{j}, C_{k_2} \} }\hspace{-0.03in} \beta_2 \hspace{-0.03in} \sum_{k_2=1}^{K_2} \hspace{-0.01in} \sum_{j\in C_{k_2}}\hspace{-0.06in} \left\{ \left\| \omg_{2,k_2}  \P_{j}  \x - \z_{j} \right\|_2^2+\gamma_2^2 \left\|  \z_{j}\right\|_0 \hspace{-0.03in} \right\}\vspace{-0.03in}, \hspace{-0.03in}
\end{equation}
where the multi-material patches are all sparsified by cross-material transforms, i.e., $\beta_1=0$ in (\ref{Eq:cost_func_reg}).

Another simpler regularizer was proposed in our recent conference work \cite{li2018image}, dubbed DECT-ST, where
\begin{equation}
\label{Eq:DECT_ST} \hspace{-0.07in}
\R(\x)\triangleq \min \limits_{ \{ \z_{lj} \} } \sum_{l=1}^2  \sum_{j=1}^{N_p} \beta_l \left\{ \left\| \omg_l  \P_{lj}  \x - \z_{lj} \right\|_2^2+\gamma_l^2 \left\|  \z_{lj}\right\|_0 \right\}.
\end{equation}
This regularizer employs one transform for the 2D patches of each material, with $\P_{lj} \x$ denoting the $j$th patch of the $l$th material, $\z_{lj}$ denoting the sparse vector for the $j$th patch of the $l$th material, and $\omg_1$ and $\omg_2$ denoting the transforms for patches of the two materials, respectively. DECT-ST does not exploit cross-dependencies between material images. Rather it involves a common-material model with two transforms (corresponding to $K_1=2$ and $K_2=0$ in (\ref{Eq:cost_func_reg})) and a specific clustering of the 2D patches (i.e., patches from each material are grouped together). 

\begin{figure}[htb]
	\vspace{-0.1in} 
	\centering
	\begin{tikzpicture}
	[spy using outlines={rectangle,red,magnification=2,width=15mm, height =10mm, connect spies}]				
	\node {\includegraphics[width=0.22\textwidth]{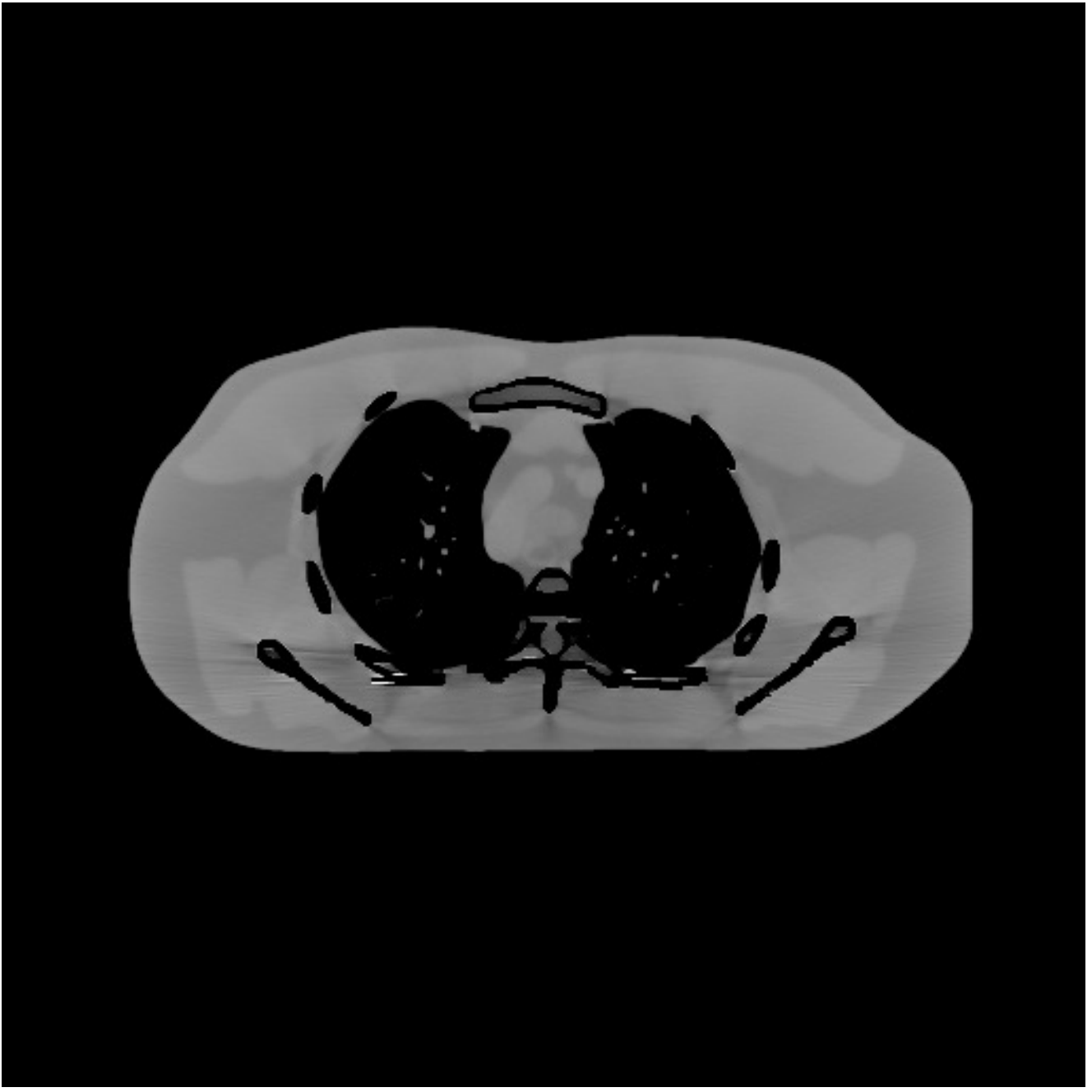}	};				
	\spy on (-0.69,-0.44) in node [right] at (-1.95,-1.45);	
	\spy on (1.1,0.15) in node [right] at (0.46,1.45);	
	\end{tikzpicture} 	
	\begin{tikzpicture}
	[spy using outlines={rectangle,red,magnification=2,width=15mm, height =10mm, connect spies}]				
	\node {\includegraphics[width=0.22\textwidth]{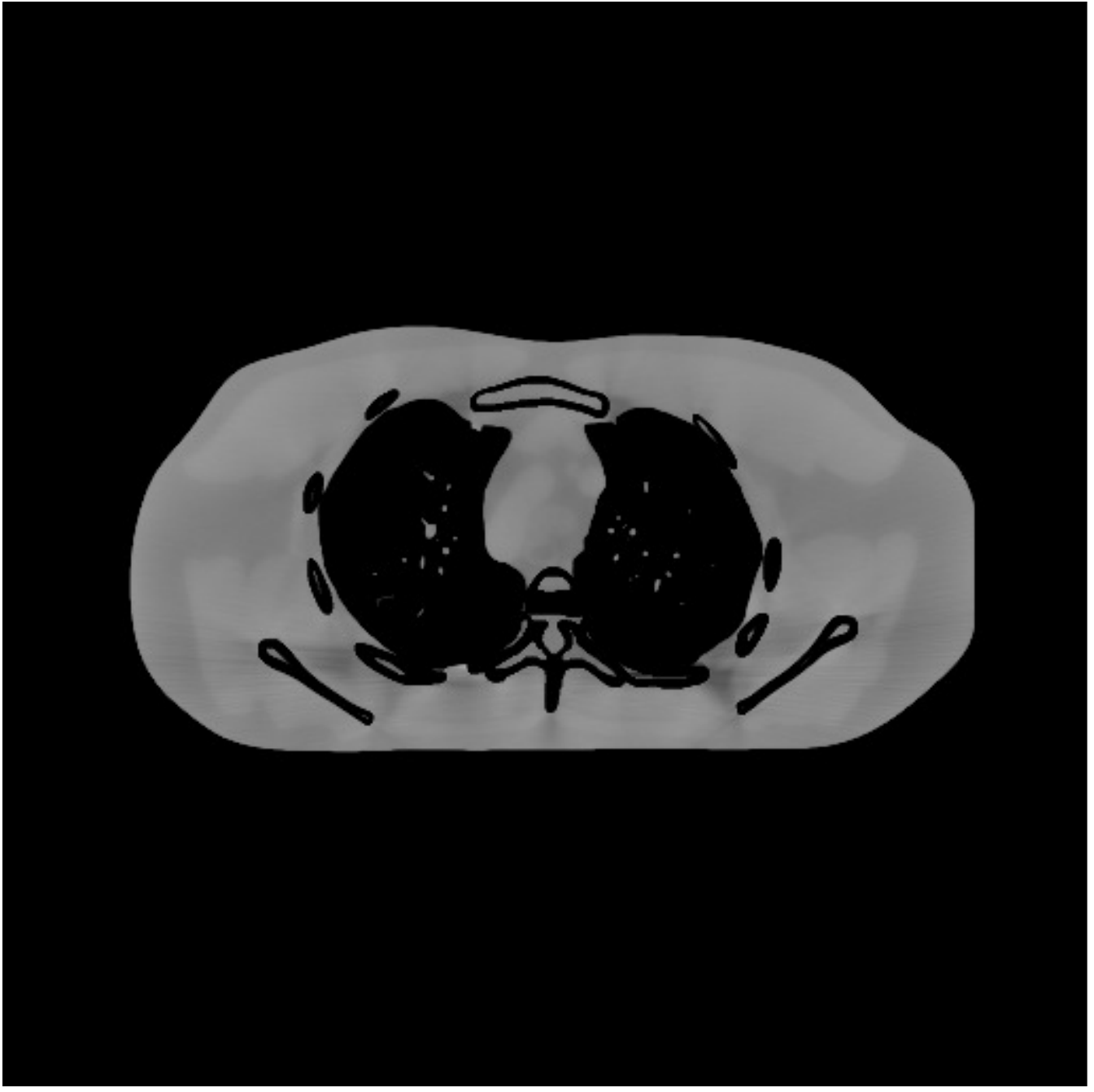}	};				
	\spy on (-0.69,-0.44) in node [right] at (-1.95,-1.45);	
	\spy on (1.1,0.15) in node [right] at (0.46,1.45);	
	\end{tikzpicture} \\
	\begin{tikzpicture}
	[spy using outlines={rectangle,red,magnification=2,width=15mm, height =10mm, connect spies}]				
	\node {\includegraphics[width=0.22\textwidth]{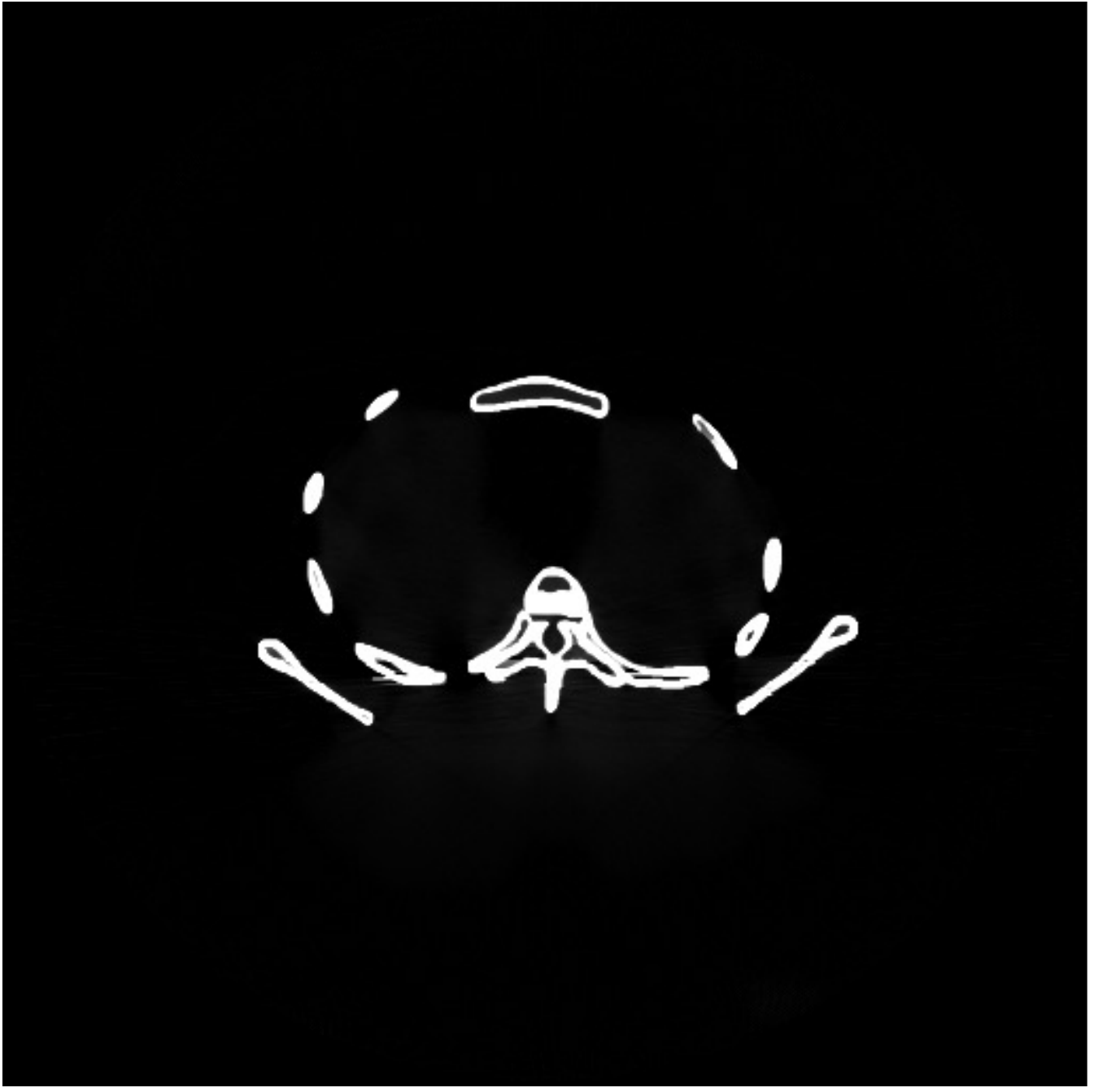}	};				
	\spy on (-0.69,-0.44) in node [right] at (-1.95,-1.45);	
	\end{tikzpicture} 
	\begin{tikzpicture}
	[spy using outlines={rectangle,red,magnification=2,width=15mm, height =10mm, connect spies}]				
	\node {\includegraphics[width=0.22\textwidth]{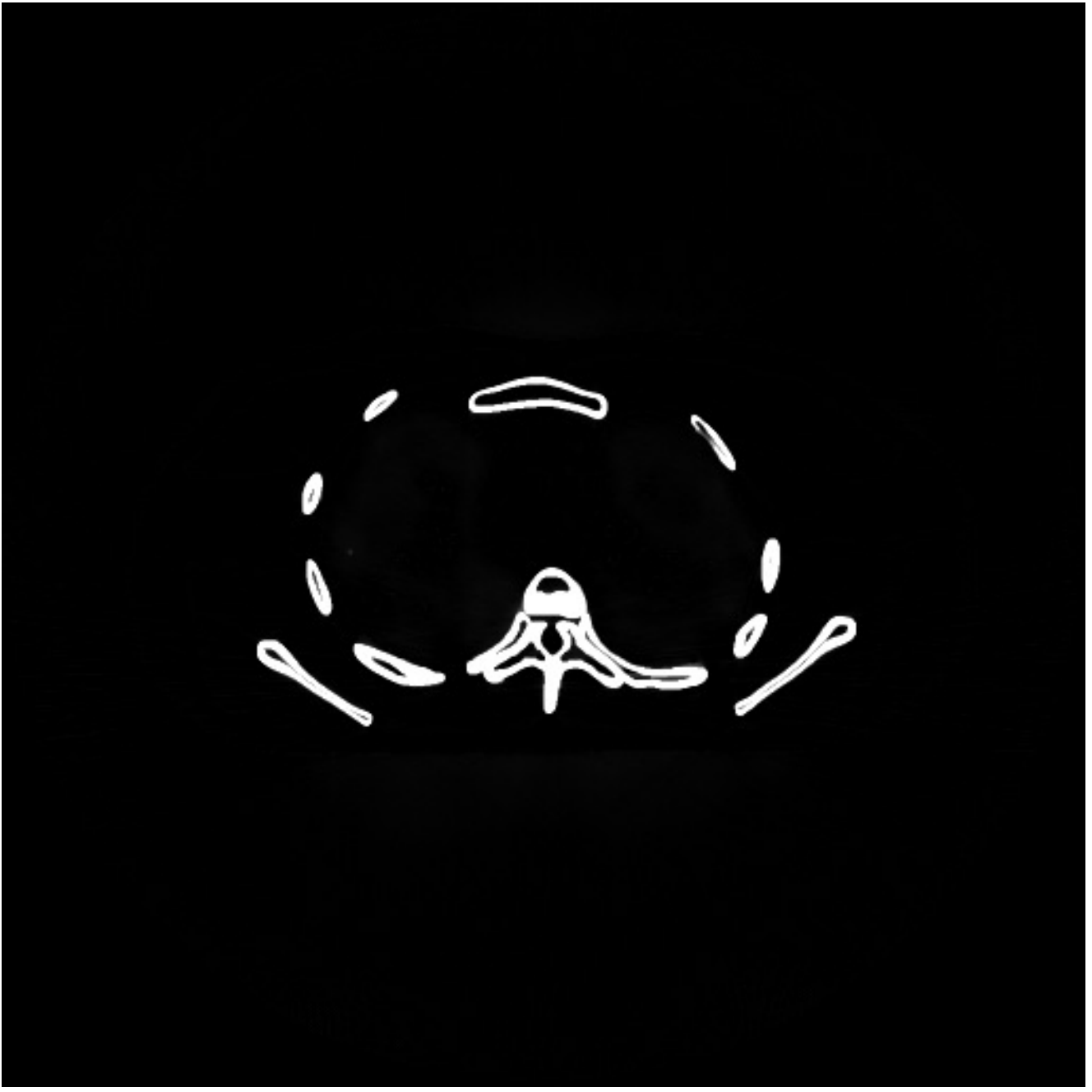}	};				
	\spy on (-0.69,-0.44) in node [right] at (-1.95,-1.45);	
	\end{tikzpicture} 	
	\caption{First to second column: material images decomposed by DECT-ST and DECT-CULTRA, respectively.  Water and bone images are in the top and bottom rows with display windows [0.7  1.3]\,g/cm$^3$ and [0 0.8]\,g/cm$^3$, respectively.} 
	\label{Fig:comp_STClutra}
	\vspace{-0.08in}
\end{figure}

Compared to the regularizers in (\ref{eq:cultra_Reg}) and (\ref{Eq:DECT_ST}), the MULTRA regularizer (\ref{Eq:cost_func_reg}) simultaneously promotes both the common and cross material models with an adaptive patch-dependent clustering. To motivate the benefits of the richer MULTRA model better,   Fig.~\ref{Fig:comp_STClutra} shows example decomposition results obtained by DECT-ST and DECT-CULTRA ($K_2=10$) for the XCAT phantom \cite{Segars2008Realistic}. 
The parameters were empirically chosen as $\{\beta,\gamma\} = \{70, 0.07\}$ for DECT-CULTRA and $\{\beta_1, \beta_2, \gamma_1, \gamma_2\} = \{50, 70, 0.03, 0.04\}$ for DECT-ST. 
The pros and cons of the two cases are clearly observed in the result. In particular, compared to DECT-ST, DECT-CULTRA successfully removes the artifacts at the boundaries of the basis materials and also preserves some details. 
However, the edges in the soft tissue-only regions in the water image for DECT-CULTRA looks quite undistinguishable compared to DECT-ST.  
This suggests that the common and cross-material models by themselves may not provide a good trade-off in material decomposition, since they capture different properties of the materials. 
The proposed DECT-MULTRA approach exploiting both a mixed material model and unions of sparsifying transforms can effectively alleviate the drawbacks of these variants. 
Section \ref{Sec:experiment} shows the numerical results and comparisons.

\vspace{-0.02in}
\section{Algorithms}
\label{Sec:algorithm}
This section describes the algorithms for pre-learning the mixed union of transforms from datasets, and for \mbox{optimizing (\ref{Eq:cost_func}).}
\subsection{Algorithm for Training a Mixed Union of Sparsifying Transforms}
\label{Subsec:Learning}
We pre-learn a union of 2D common-material unitary transform matrices $\{\widetilde{\omg}_{1,\widetilde{k}_1}\}_{\widetilde{k}_1=1}^{\widetilde{K}_1}$, and a union of 3D  cross-material \mbox{unitary} transform matrices $\{\widetilde{\omg}_{2,\widetilde{k}_2}\}_{\widetilde{k}_2=1}^{\widetilde{K}_2}$ separately from a dataset of material images.
\begin{figure}[htb]
	\vspace{-0.11in}
	\centering
	\begin{minipage}{0.01\linewidth}
		\centerline{\includegraphics[scale=0.38,trim=10 10 110 10,clip]{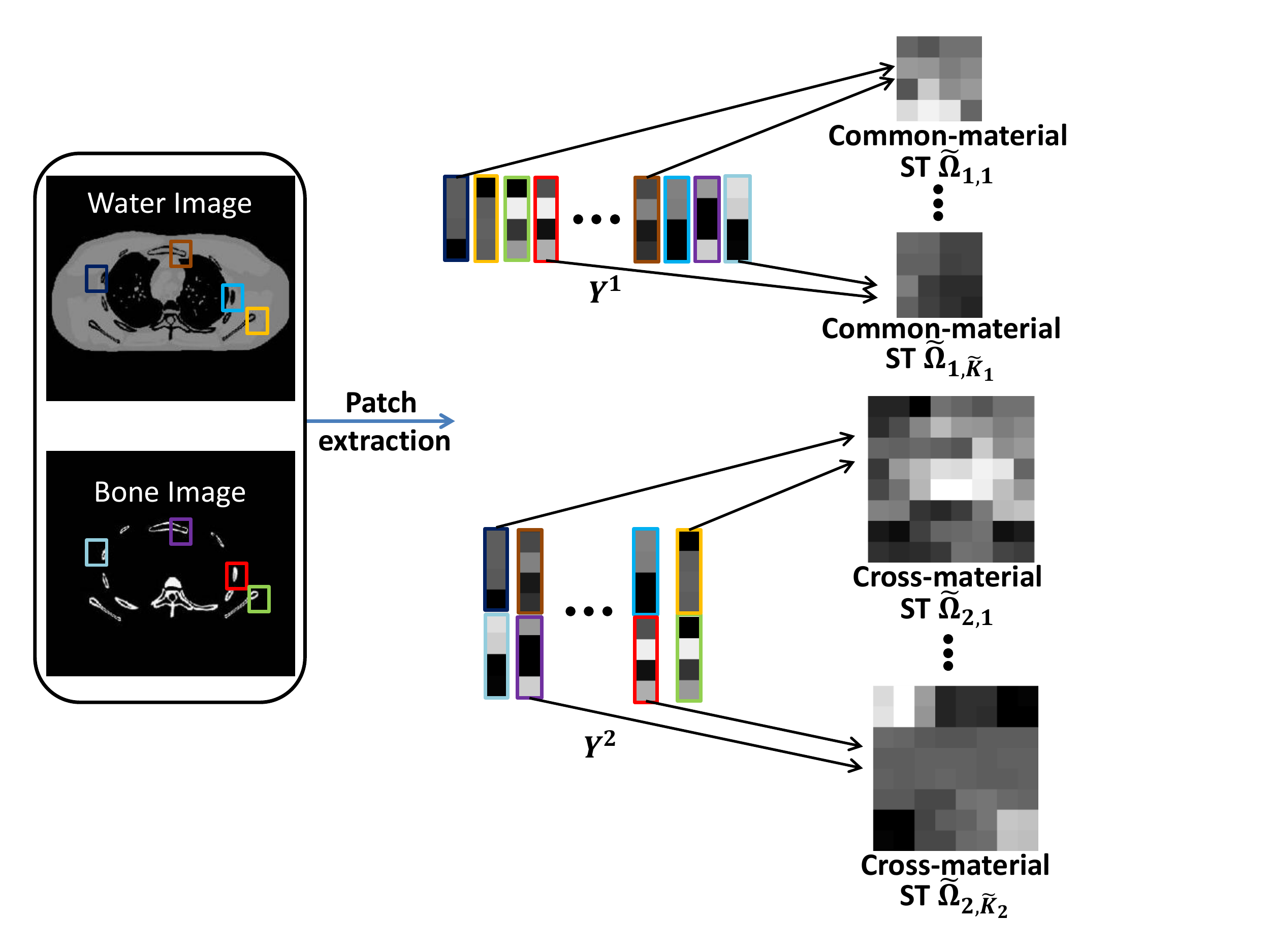}}
		\vspace{-0.05in}
		\centerline{(a)}
	\end{minipage}\\
	\begin{minipage}{0.01\linewidth}
		\centerline{\includegraphics[scale=0.38,trim=5 8 90 8,clip]{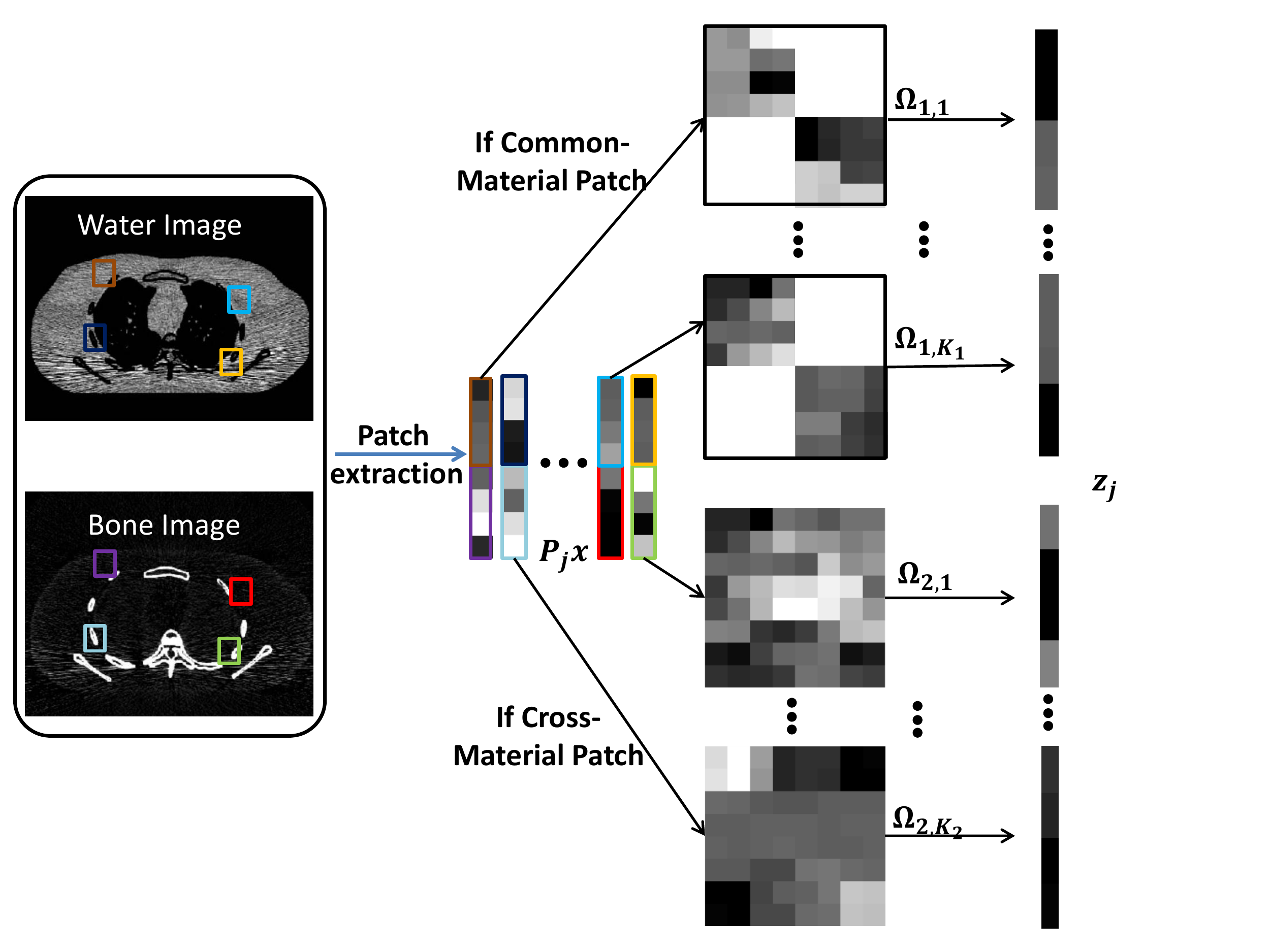}}
		\centerline{(b)}
	\end{minipage}
	\caption{Schematic of the MULTRA model: (a) and (b) illustrate the learning process for MULTRA and the operation of the pre-learned transforms during the decomposition process, respectively. $K_2 = \widetilde{K}_2$.}
	\vspace{-0.1in}
	\label{fig:diagram}
\end{figure} 
For learning the common-material transforms, where each transform $\widetilde{\omg}_{1,\widetilde{k}_1} \in \mathbb R^{m\times m}$ acts on 2D patches, we extract vectorized 2D patches from across all basis materials' images. 
We also extract vectorized 3D or multi-material patches for training the various cross-material transforms $\widetilde{\omg}_{2,\widetilde{k}_2} \in \mathbb R^{2m\times 2m}$, where each patch is formed from the basis material components of an object by stacking 2D patches from the same spatial location in the different basis material images. 
Fig.~\ref{fig:diagram}(a) shows a schematic of this learning model.
When optimizing (\ref{Eq:cost_func}), the $\widetilde{K}_1$ 2D common-material transform matrices $\{\widetilde{\omg}_{1,\widetilde{k}_1}\}$ above are used to independently form the blocks of the block-diagonal matrices $\{\omg_{1,k_1}\}$. In particular, $K_1 = \widetilde{K}_1^2$.
However, $\{{\omg}_{2,k_2}\}$ are identical to $\{\widetilde{\omg}_{2,\widetilde{k}_2}\}$ ($K_2=\widetilde{K}_2$). \footnote{The notation difference is because the training phase works with both 2D and 3D patches, whereas Problem (\ref{Eq:cost_func}) uses a single unified notation for patches (i.e., the multi-material patches $\P_j \x$).} The operation of the pre-learned transforms during the material decomposition process is shown in Fig.~\ref{fig:diagram}(b).


The unions of common and cross material transforms are learned by solving the following problem for $r=1,2$:
\begin{align}
\hspace{-0.06in}
\nonumber \mathop{\arg\min}_{\{\widetilde{\omg}_{r,\widetilde{k}_r}\}} \hspace{-0.03in} \min_{\{C_{\widetilde{k}_r}^r, \Z_{i_r}^r \}} & \sum^{\widetilde{K}_r}_{\widetilde{k}_r=1} \sum_{i_r\in C_{\widetilde{k}_r}^r} \Big\{ \|\widetilde{\omg}_{r,\widetilde{k}_r} \Y_{i_r}^r-\Z_{i_r}^r\|^2_2+\eta_r^2\|\Z_{i_r}^r\|_0 \Big\} \\
& \textup{s.t.} \quad \widetilde{\omg}_{r,\widetilde{k}_r}^T \widetilde{\omg}_{r,\widetilde{k}_r} = \I,  \,\; 1 \leq \widetilde{k}_r \leq \widetilde{K}_r,
\tag{P1}
\label{Eq:tran_update}
\end{align} 
where $\I$ denotes the identity matrix of appropriate size, $\eta_r>0$ is a scalar parameter controlling sparsity, $\Y_{i_1}^1 \in \mathbb R^{m}$ and $\Y_{i_2}^2 \in \mathbb R^{2m}$ denote the $i_1$th and $i_2$th training vectors for $r=1$ and $r=2$, respectively, and $\Z_{i_1}^1 \in \mathbb R^{m}$ and $\Z_{i_2}^2 \in \mathbb R^{2m}$ denote the corresponding sparse coefficient vectors.  
The term $\|\widetilde{\omg}_{r,\widetilde{k}_r} \Y_{i_r}^r-\Z_{i_r}^r\|^2_2$ in (\ref{Eq:tran_update}) is called sparsification error \cite{Ravishankar2015} and captures the deviation of the transformed patches from their sparse approximations. Each patch in (\ref{Eq:tran_update}) is matched to a specific transform, and the goal in (\ref{Eq:tran_update}) is to simultaneously learn the collection of $\widetilde{K}_r$ transforms and cluster the training vectors, and also estimate the sparse coefficient vectors.
We optimize (\ref{Eq:tran_update}) by using the approach in \cite{ravishankar:16:ddl} and alternating between updating $\{C_{\widetilde{k}_r}^r, \Z_{i_r}^r\}$ (sparse coding and clustering step) and $\{\widetilde{\omg}_{r,\widetilde{k}_r}\}$ (transform update step), with efficient updates in each step (involving thresholding or small singular value decompositions).

\vspace{-0.05in}
\subsection{MULTRA Material Image Decomposition Algorithm}
We propose an iterative algorithm for the image-domain DECT material decomposition problem (\ref{Eq:cost_func}) with regularizer (\ref{Eq:cost_func_reg}) that alternates between updating $\x$ (\emph {image update step}) and updating $\{\z_{j}, C_{k_r}^r\}$ (\emph {sparse coding and clustering step}). We exploit such an exact alternating minimization scheme for (\ref{Eq:cost_func}) as it leads to efficient updates, and does not involve additional algorithm parameters.

\subsubsection{Image Update Step}
\label{Sec:Img_update}
Here, we solve for $\x$ in (\ref{Eq:cost_func}) with fixed $\{\z_j, C_{k_r}^r\}$, using the following PWLS sub-problem:
\begin{equation}
\label{Eq:Img_update} \hspace{-0.02in}
\mathop{\arg\min}_{ \x \in \mathbb R^{2 N_p}} \frac{1}{2} \| \y- \A \x\|^2_\W +  \sum_{r=1}^2  \sum_{k_r=1}^{K_r}  \sum_{j\in C_{k_r}^r}\hspace{-0.03in} \beta_r \left\| \omg_{r,k_r}  \P_{j}  \x - \z_{j} \right\|_2^2.
\end{equation}
The exact solution to (\ref{Eq:Img_update}) is obtained efficiently as follows. First, denoting the regularizer component of the cost in (\ref{Eq:Img_update}) by $\R_2(\x)$, its gradient is given as follows:
\begin{equation}
\label{Eq:grad_Img}
\nabla\R_2(\x)=2 \sum_{r=1}^{2} \sum_{k_r=1}^{K_r} \sum_{j\in C_{k_r}^r} \beta_r \P_{j}^T{\omg_{r,k_r}^{T}}({\omg_{r,k_r}}\P_{j}\x-\z_j). 
\end{equation}
Since the transform matrices are all unitary, (\ref{Eq:grad_Img}) is equivalent to
\begin{equation}
\label{Eq:grad_simp_Img}
\nabla\R_2(\x)= 2 \sum_{r=1}^{2} \sum_{k_r=1}^{K_r} \sum_{j\in C_{k_r}^r} \beta_r \P_{j}^T \left( \P_j \x - {\omg_{r,k_r}^{T}} \z_j \right). 
\end{equation} 
The second term in (\ref{Eq:grad_simp_Img}) is independent of $\x$,  and the matrix pre-multiplying $\x$ in the first term in (\ref{Eq:grad_simp_Img}) is diagonal and can in fact be pre-computed. Owing to the structure in $\A$ (it acts independently across pixels) and the diagonal structure of $\W$ and using (\ref{Eq:grad_Img}), the $\x$ update in (\ref{Eq:Img_update}) can be separated into $N_p$ pixel-wise updates.
The update at each pixel $j$ is given as follows:
\begin{equation}
\label{Eq:Img_iter} \hspace{-0.1in}
\hat{\x}_j^T=\mathbf{H}_j^{-1}(\A_0^T \W_j \y_j+2 \M_j \sum_{r=1}^{2} \sum_{k_r=1}^{K_r} \sum_{j\in C_{k_r}^r} \beta_r \P_{j}^T {\omg_{r,k_r}^{T}} \z_j),
\end{equation}
where $\hat{\x}_j = \left(\hat{x}_{1j}, \hat{x}_{2j}\right)$, $\M_j \in \mathbb R^{2\times 2N_p}$ is an operator that extracts elements corresponding to the $j$th \mbox{pixels}, and the $2\times 2$ Hessian matrix $\mathbf{H}_j=\A_0^T \W_j \A_0+2\,\diag\left(\M_j \sum_{r=1}^{2} \sum_{k_r=1}^{K_r} \sum_{j\in C_{k_r}^r} \beta_r \P_j^T \P_j \mathbf{1}\right)$, where $\mathbf{1} \in \mathbb R^{2N_p}$ denotes a column vector of ones.

\subsubsection{Sparse Coding and Clustering Step}
Here, we update $\{\z_{j}, C_{k_r}^r\}$ with fixed $\x$ in (\ref{Eq:cost_func}), using the following sub-problem:
\begin{equation} 
\label{Eq:clusspar}
\mathop{\arg\min} \limits_{ \{ \z_{j}, C_{k_r}^r \} } \sum_{r=1}^2  \sum_{k_r=1}^{K_r}  \sum_{j\in C_{k_r}^r} \beta_r \left\{ \left\| \omg_{r,k_r}  \P_{j}  \x - \z_{j} \right\|_2^2+ \gamma_r^2 \left\|  \z_{j}\right\|_0 \right\} .
\end{equation}
The exact solution to (\ref{Eq:clusspar}) can be obtained efficiently. For a fixed clustering, the optimal sparse code for each patch is obtained by hard-thresholding as  $\z_{j}=H_{\gamma_r}(\omg_{r,k_r}\P_{j}\x)$, where the operator $H_{\gamma_r}(\cdot)$ zeros out vector elements with magnitudes less than $\gamma_r$, and leaves other elements unchanged. Substituting this in (\ref{Eq:clusspar}) yields the following equivalent problem for clustering each patch:
\begin{equation}
\label{Eq:spaclu_update}
\begin{split}
(\hat{r}_j, \hat{k}_j)=\mathop{\arg\min}_{\substack{1\le k_{r} \le K_r \\ r \in \{1,2\}} } \beta_r \Big\{ &\left\| \omg_{r,k_r}  \P_{j}  \x - H_{\gamma_r}(\omg_{r,k_r} \P_{j}\x) \right\|_2^2+ \\& \gamma_r^2 \left\|  H_{\gamma_r}(\omg_{r,k_r} \P_{j}\x)\right\|_0 \Big\}.
\end{split}
\vspace{-0.05in}
\end{equation}
Solving (\ref{Eq:spaclu_update}) requires computing the cost with respect to each transform in the two models to determine the minimum value (or best match). This can be done efficiently as follows. 
For $r=1$, each block of $\omg_{1,k_r}$ can assume any of the $\widetilde{K}_1$ transforms in $\{\widetilde{\omg}_{1,\widetilde{k}_r}\}$. Because the cost for the 3D patch in (\ref{Eq:spaclu_update}) for $r=1$ equals the sum of the corresponding costs for the constituent 2D material patches, the best matching transform for each of those 2D patches is found independently (by searching over the smaller set of $\widetilde{K}_1$ transforms) and the best material-wise transforms are then combined (into a block diagonal matrix) to provide the best matching transform within $r=1$. 
Comparing the smallest cost value for $r=1$ with the smallest value within $r=2$ yields the best matched model and corresponding transform (cluster).
The patches can be optimally clustered in parallel and the optimal sparse codes are then given as  $\hat{\z}_{j}=H_{\gamma_{\hat{r}_j}}(\omg_{{\hat{r}_j},\hat{k}_j} \P_{j}\x)$, $\forall$ $j$.

The proposed alternating minimization algorithm for (\ref{Eq:cost_func}) belongs to the broad class of block coordinate descent (BCD) optimization algorithms, that are guaranteed to decrease the objective function over the iterations. 
Since the objective in (\ref{Eq:cost_func}) is lower bounded, it converges in the proposed algorithm. 
More detailed theoretical convergence results for the iterates (convergence to critical points or partial global minimizers, etc.) can also be shown for DECT-MULTRA similar to the results shown in recent work \cite{ravishankar:16:ddl} (cf. Theorems 1 and 2 in \cite{ravishankar:16:ddl}) for related BCD schemes.
Algorithm \ref{alg:1} describes the proposed iterative scheme for optimizing Problem (\ref{Eq:cost_func}). 
 \begin{algorithm}
 	\caption{DECT-MULTRA Algorithm for (\ref{Eq:cost_func})}  
 	\begin{algorithmic}[1] 
 		\REQUIRE  initial material image $\hat{\x}^{(0)}$, pre-learned $\{\omg_{r,k_r}\}$, parameters $\beta_r$ and $\gamma_r$ for $r=1,2$, number of iterations $I$.
 		\ENSURE decomposed material images $\hat{\x}^{(I)}$.\\	    
 		\FOR{$i = 0, 1, 2,\cdots, I-1 $}
 		\STATE \textbf{(1) Image update}: With $\{\hat{\z}_{j}^{(i)}, {\hat{C}_{k_r}^{r^{(i)}}} \}$ fixed,
 		\STATE compute $\hat{\x}_j^{(i+1)}$ at each pixel $j$ according to (\ref{Eq:Img_iter}).
 		\STATE \textbf{(2) Sparse Coding and Clustering}: with $\hat{\x}^{(i+1)}$ fixed, update the  cluster assignment $(\hat{r}_j^{(i+1)}, \hat{k}_j^{(i+1)})$ for each patch using (\ref{Eq:spaclu_update}), and the updated sparse codes are $\hat{\z}_{j}^{(i+1)}=H_{\gamma_{\hat{r}_j^{(i+1)}}}(\omg_{{\hat{r}_j^{(i+1)}},\hat{k}_j^{(i+1)}} \P_{j}\hat{\x}^{(i+1)})$ $\forall$ $j$.
 		\ENDFOR	
 		\RETURN $\hat{\x}^{(I)}$
 	\end{algorithmic}
 	 	\label{alg:1}
 \end{algorithm}
\vspace{-0.2in} 

\subsection{Computational Cost}
The computational cost per outer iteration of the proposed algorithm for (\ref{Eq:cost_func}) scales as $O(m^2 (\sqrt{K_1} + K_2) N_p)$, and is dominated by matrix-vector multiplications in the sparse coding and clustering step. Importantly, being an image-domain decomposition scheme, the proposed algorithm does not involve expensive forward and back-projections.
\vspace{-0.06in}

\section{Results}
\label{Sec:experiment}
We employed both numerical simulations with phantoms and clinical DECT data to evaluate the proposed \mbox{methods}, namely DECT-MULTRA, DECT-CULTRA, and DECT-ST. This section describes the experiments evaluating the performance of the proposed methods in comparison with competing methods. 
Additional experimental results of DECT-MULTRA and other methods are provided in the supplement. \footnote{Supplementary material is available in the supplementary files/multimedia	tab.}
A link to software to reproduce our results will be provided at \url{https://web.eecs.umich.edu/~fessler/}.
\vspace{-0.1in}
\subsection{Methods for Comparison}
\begin{enumerate}
	\item \textbf{Direct Matrix Inversion} \cite{niu2014iterative}: solving (\ref{Eq:cost_func}) by matrix inversion, i.e., without regularization.
	\item \textbf{DECT-EP} \cite{xue:2017:statistical}: optimizes (\ref{Eq:cost_func}) with an edge-preserving regularizer, which is defined as $\R(\x)= \sum_{l=1}^{2} \beta_l \R_l (\x_l) $, where the regularizer for the $l$th material is $\R_l (\x_l) = {\sum_{k=1}^{K}\psi_l([\mathbf{C}\x_l]_k)}$,  where $K=N_p N_{lj}$, with $N_{lj}$ denoting the number of neighbors for each pixel $x_{lj}$,  $\mathbf{C} \in \mathbb R^{K\times N_p}$ is the 2D finite difference matrix and  $\psi_l (t) \triangleq  \frac{\delta_l^2}{3} \left(\sqrt{1+3(t/\delta_l) ^2}-1 \right)$ \cite{xue:2017:statistical}, where $\delta_l$ is the edge-preserving parameter for the $l$th material.
	\item \textbf{DECT-TDL} \cite{zhang:15:tbd}: DECT image-domain decomposition with regularization based on a learned tensor dictionary that is trained by K-CPD (an extension of the K-SVD algorithm \cite{aharon:06:ksa} to incorporate tensor models). During the image decomposition process, the tensor sparse codes are updated using the multilinear orthogonal matching pursuit (MOMP) method \cite{K-CPD}.
\end{enumerate}	 

All the cross-compared methods in the image-domain have the same data-fidelity term but differ in the regularizer. 
In particular, the direct matrix inversion method and DECT-EP are recent non-adaptive (i.e., not involving learning) methods. They are quite distinct from the proposed transform learning-based methods, which rely on a learned sparsification operator $\omg$. 
The proposed DECT-ST and DECT-CULTRA are simpler forms of DECT-MULTRA, while the regularizer for the recent DECT-TDL exploits a tensor synthesis dictionary model and is quite different from the proposed method.
These methods are recent works in the image-domain decomposition literature and thus form an important subset of methods to compare with.

\subsection{Training the Image Models for DECT}
We pre-train dictionaries for DECT-TDL and transforms for the proposed DECT-ST, DECT-CULTRA (see Section \ref{Sec:II-B}), and DECT-MULTRA. 
We first chose five training slices (\mbox{different} from the test slices in our experiments) of the XCAT phantom \cite{Segars2008Realistic} and seperated each slice into water and bone density images 
according to the table of linear attenuation coefficients for organs provided for the XCAT phantom. We grouped water, fat, muscle, and blood into the water density image, and spine bone and rib bone  into the bone density image. 
For DECT-MULTRA, we pre-learned a union of common-material transforms ($\widetilde{K}_1=15$) from $8\times 8$ patches with a patch stride of $1\times 1$, extracted separately from the five slices of water images and bone images. 
A union of cross-material transforms ($\widetilde{K}_2=10$) was also learned from $8\times 8 \times 2$ overlapping multi-material patches with a spatial (2D) patch stride of $1 \times 1$ that were extracted from five stacked water+bone (3D) images. 
The sparsity parameter $\eta_r$ was set as $0.21$ and $0.17$ for the common-material ($r=1$) and cross-material ($r=2$) models, respectively. 
For DECT-ST, we pre-learned two different transform matrices for water and bone from $8 \times 8$ overlapping patches (with patch stride $1 \times 1$) extracted from the five water and five bone training images, respectively. 
The training parameters $\{\lambda,\,\eta\}$ for DECT-ST \cite{li2018image} were empirically set as $\{5.28\times 10^8,\,0.12\}$ and $\{9.74\times 10^7,\,0.15\}$ for water and bone, respectively.
For DECT-CULTRA, we used the same collection of cross-material transforms as learned for the DECT-MULTRA method.
We ran $2000$ iterations of all the transform learning methods to ensure convergence. 

For DECT-TDL, we pre-learned a tensor dictionary from $8\times 8\times 2$ overlapping multi-material patches with a spatial (2D) patch stride of $1 \times 1$, extracted from the five stacked water+bone images (the same training slices as used for DECT-MULTRA and DECT-CULTRA). We used a maximum patch-wise sparsity level of $50$ along with a error tolerance of $1$ during sparse coding. We used the above learned transforms and dictionaries in all experiments

\subsection{XCAT Phantom Results and Analysis}
\label{Sec:XCAT}
\subsubsection{Framework and Data}
We evaluated the proposed DECT-MULTRA, DECT-CULTRA, and DECT-ST methods for image-domain material decomposition of three test slices of the XCAT phantom \cite{Segars2008Realistic}. We compared the image quality of the decomposed material images obtained with the proposed methods with those for Direct Matrix Inversion, DECT-EP, and DECT-TDL. 
	\begin{figure}[h]
		\centering
		\begin{minipage}[H]{0.3\linewidth}  
			\centering  
			\centerline{Water}
			\centerline{\includegraphics[scale=0.38]{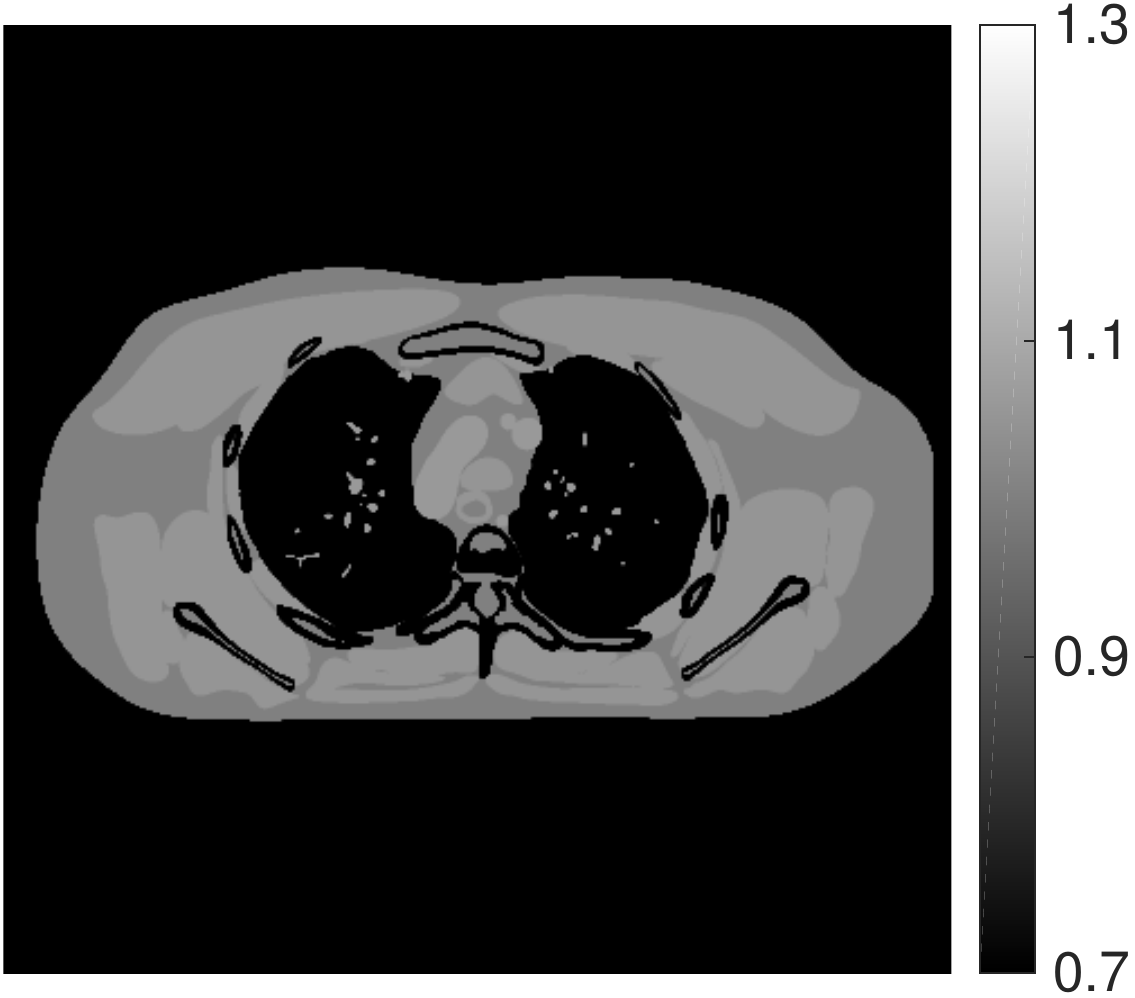}} 	 
		\end{minipage}    \hspace{0.6in} \vspace{0.05in}
		\begin{minipage}[H]{0.3\linewidth}  
			\centering 
			\centerline{Bone}  
			\centerline{\includegraphics[scale=0.38]{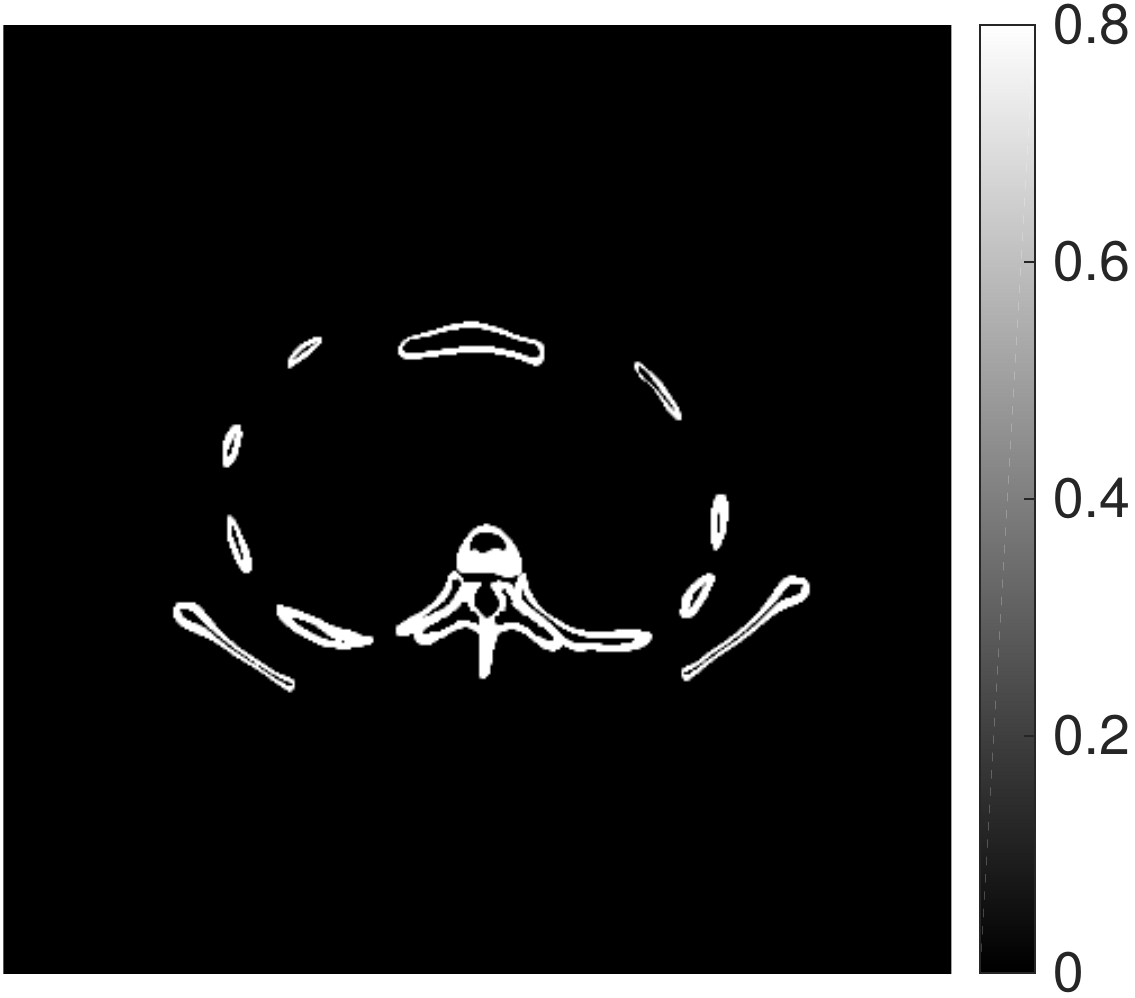}}  		 
		\end{minipage}	\\
		\begin{minipage}[H]{0.3\linewidth}  
			\centering
			\centerline{140kVp}   
			\centerline{\includegraphics[scale=0.378]{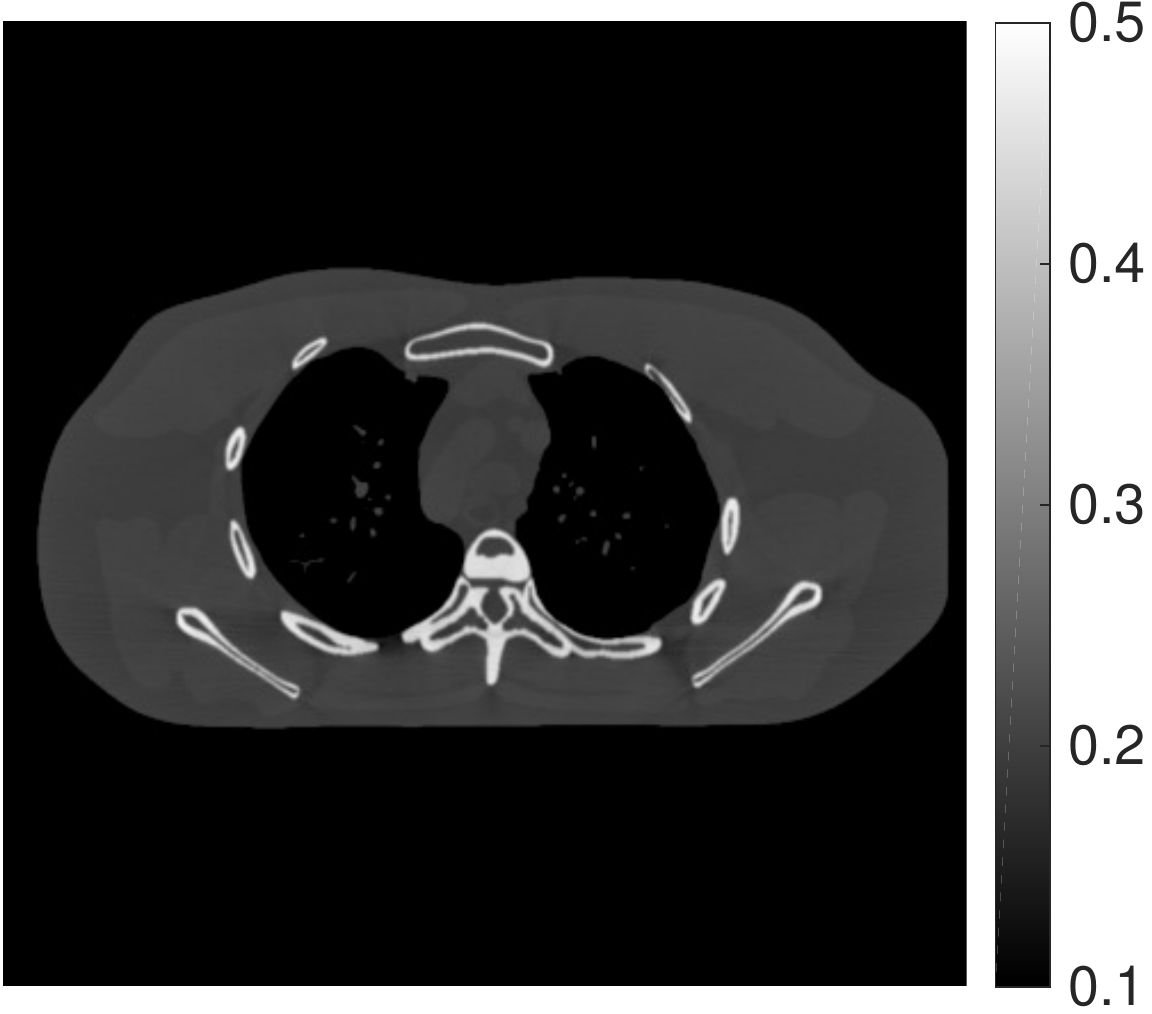}}  		 
		\end{minipage}    \hspace{0.6in}
		\begin{minipage}[H]{0.3\linewidth}  
			\centering 
			\centerline{80kVp}  
			\centerline{\includegraphics[scale=0.378]{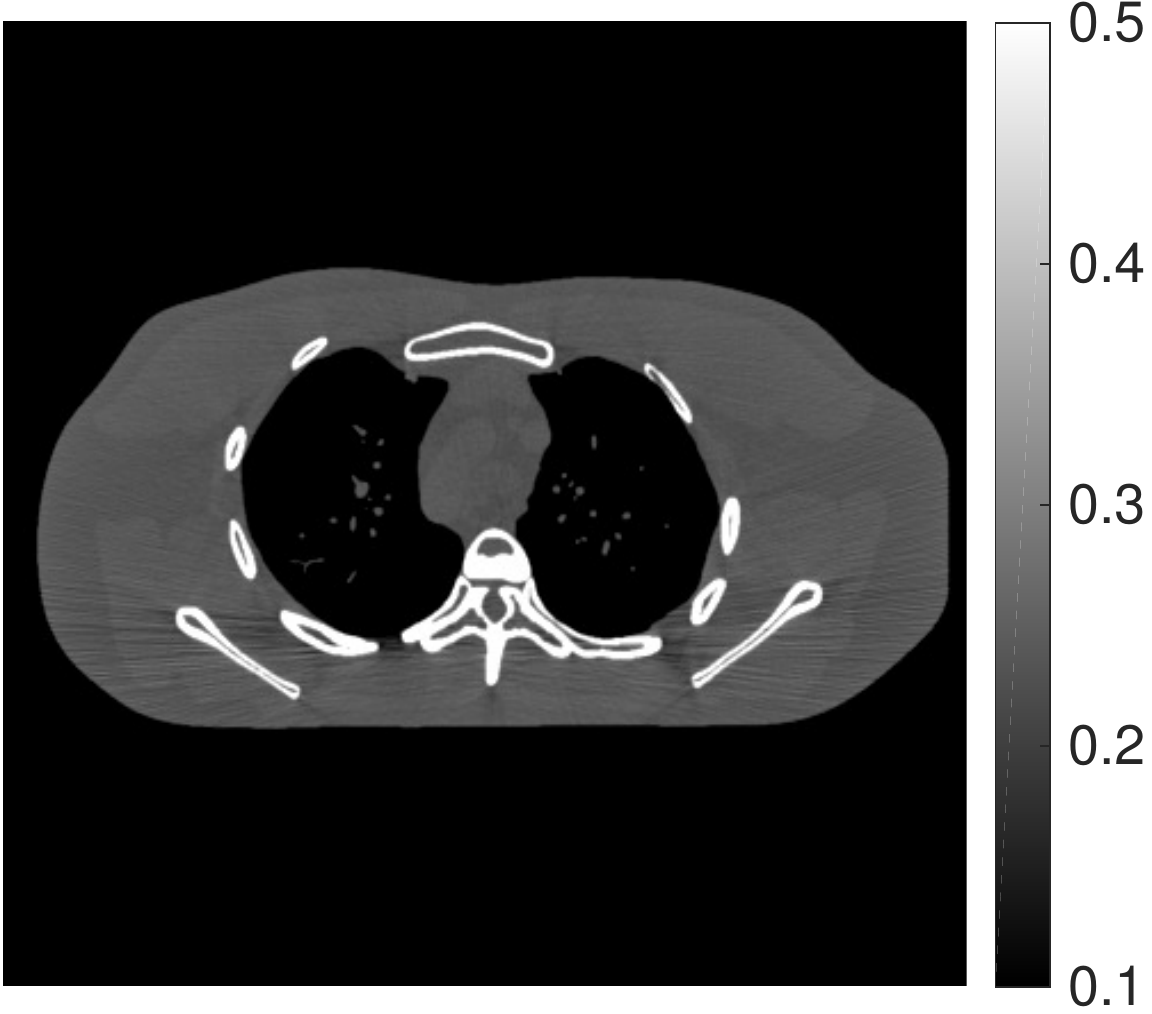}}  		 
		\end{minipage}
		\caption{The first row shows the true water and bone material images (cropped) for a test slice. The second row shows the corresponding attenuation images (cropped) at high and low energies, respectively.}
		\label{Fig:highlow}	
		\vspace{-0.18in}
	\end{figure} 
	

Fig.~\ref{Fig:highlow} shows the true water and bone density images for a test slice (central slice of the XCAT phatom). The true density images are of size $1024 \times 1024$, with the pixel size being $0.49 \times 0.49$~mm$^2$. 
We generated noisy (Poisson noise) sinograms of test slices of the XCAT phantom of size $888 \times 984$ (radial samples $\times$ angular views) using GE LightSpeed X-ray CT fan-beam system geometry corresponding to a poly-energetic source at 140kVp and 80kVp with $1\times 10^6$ and $1.86\times 10^5$ incident photons per ray, respectively. 
We used filtered back projection (FBP) to reconstruct $512\times 512$ high and low energy attenuation images (Fig.~\ref{Fig:highlow} displays them for the central test slice) with a pixel size of $0.98\times0.98$~mm$^2$, which are used as inputs for the image-domain material decomposition methods. Note that although each pixel of the XCAT phantom has only one material, our proposed method is quite general and capable of handling mixed materials in image pixels, which is demonstrated for the clinical data case in Section~\ref{subsec:clinical}.

To evaluate the performance of various methods quantitatively,  we computed the Root Mean Square Error (RMSE) for the decomposed material images in a region of interest (ROI). The ROI was a circular (around the center) region that removed all the black background area that was not interesting. For a decomposed material density $\hat{\x}_l$, the RMSE in density (g/cm$^3$) is defined as $\sqrt{ \sum_{j=1}^{N_{ROI}} (\hat{x}_{lj}-x_{lj}^\star)^2/{N_{ROI}}}$,  where $x_{lj}^\star$ denotes the downsampled\footnote{The $1024 \times 1024$ true density images were downsampled to size $512 \times 512$ by linear averaging.} true density of the $l$th material at the $j$th pixel location and $N_{ROI}$ is the number of pixels in a ROI. 

\begin{figure*}[htb] 
	\begin{tikzpicture}
	[spy using outlines={rectangle,red,magnification=2,width=16mm, height =12mm, connect spies}]				
	\node {\includegraphics[width=0.233\textwidth]{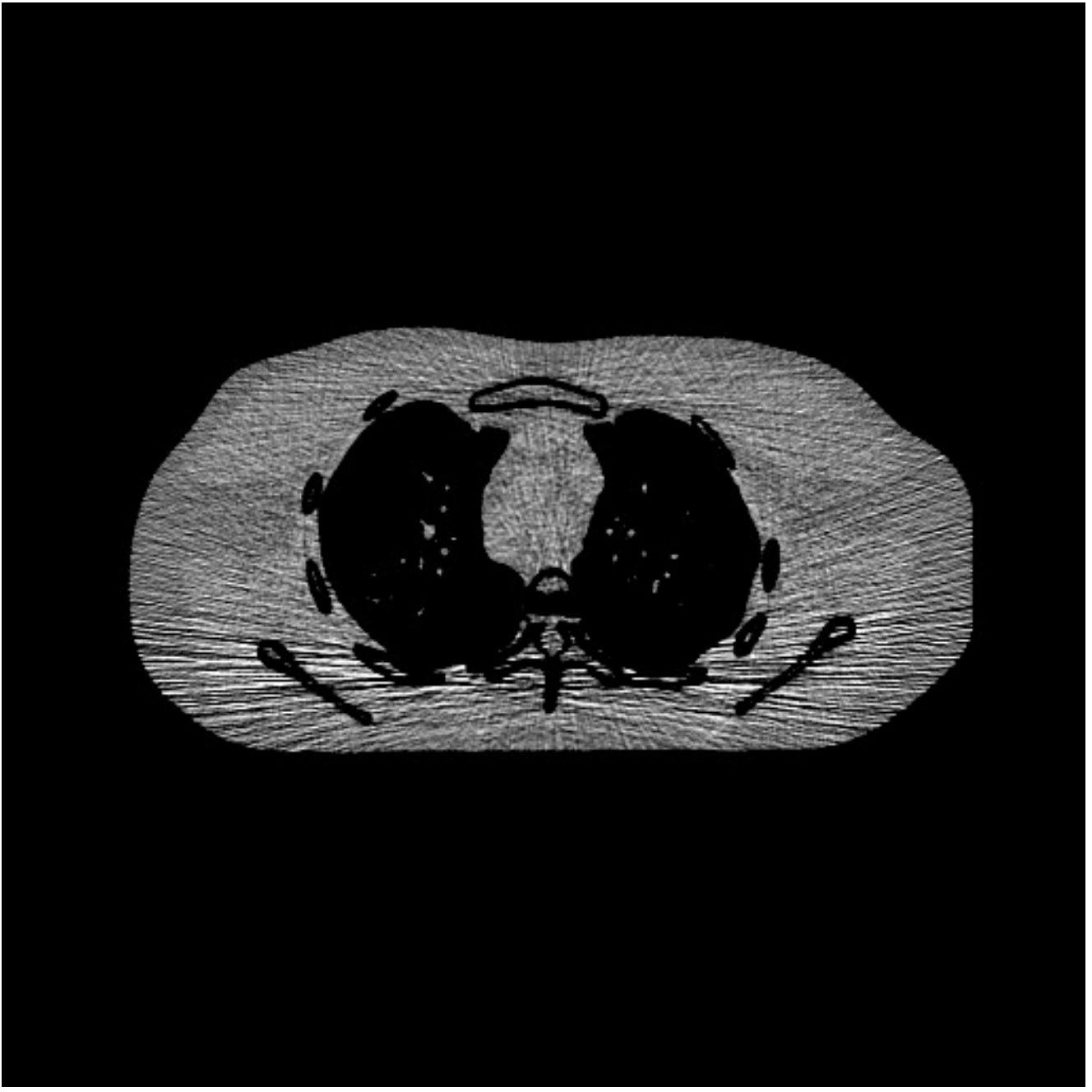}	};	
	\spy on (0.4,-0.5) in node [right] at (0.5,-1.45);
	\spy on (-1.1,0.05) in node [right] at (-2.0,1.5);
	\end{tikzpicture}
	\begin{tikzpicture}
	[spy using outlines={rectangle,red,magnification=2,width=16mm, height =12mm, connect spies}]				
	\node {\includegraphics[width=0.233\textwidth]{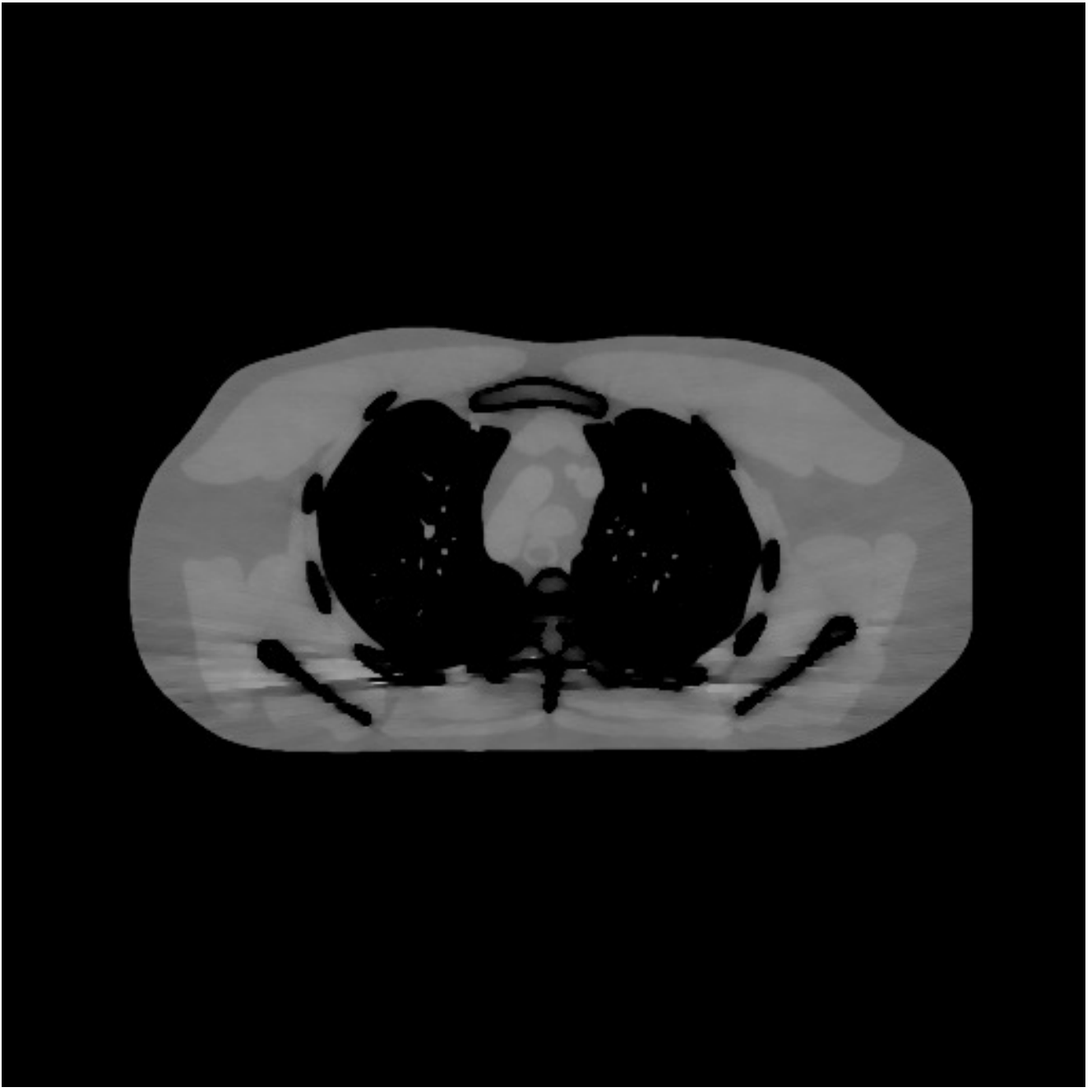}	};	
	\spy on (0.4,-0.5) in node [right] at (0.5,-1.45);
	\spy on (-1.1,0.05) in node [right] at (-2.0,1.5);
	\end{tikzpicture}
	\begin{tikzpicture}
	[spy using outlines={rectangle,red,magnification=2,width=16mm, height =12mm, connect spies}]				
	\node {\includegraphics[width=0.233\textwidth]{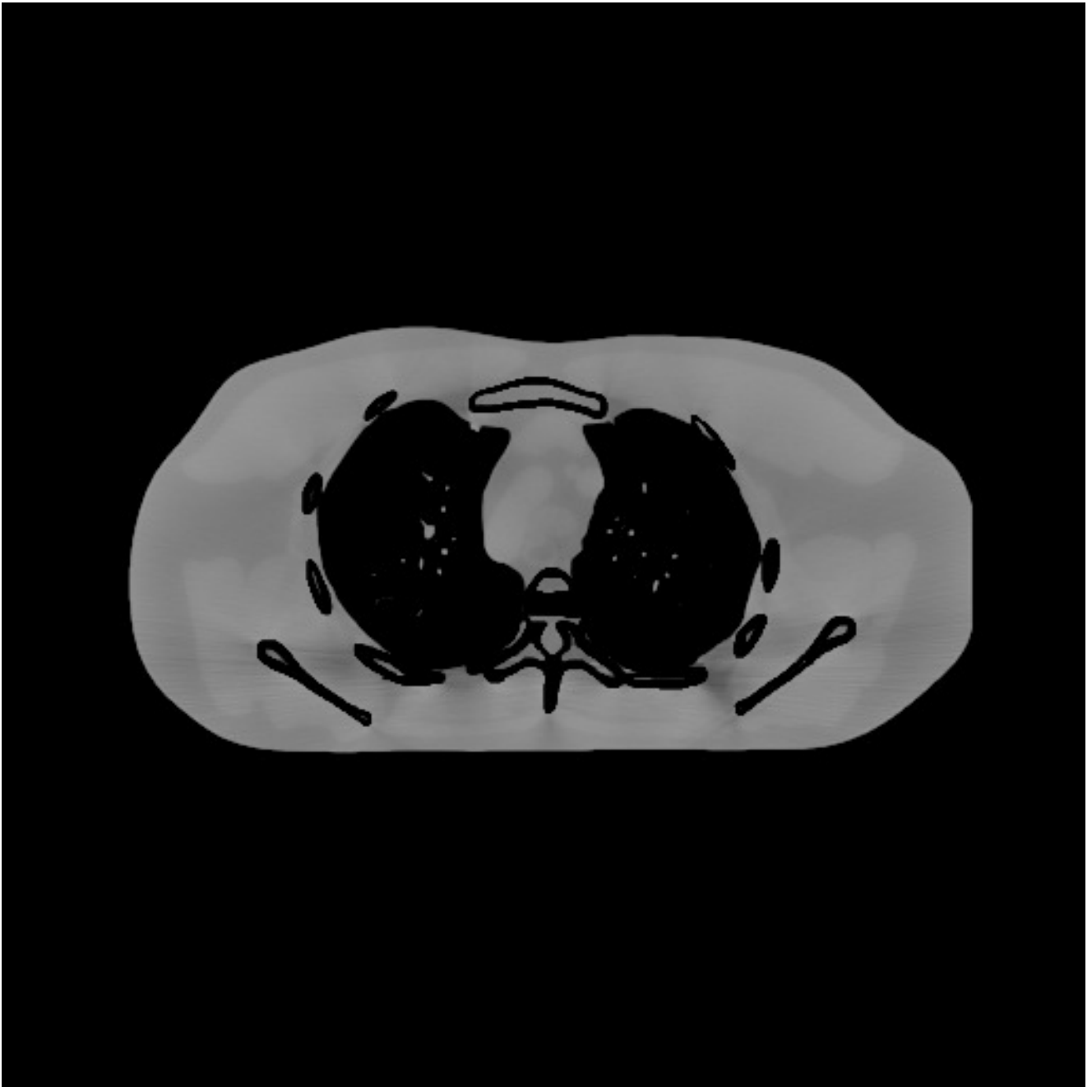}	};
	\spy on (0.4,-0.5) in node [right] at (0.5,-1.45);
	\spy on (-1.1,0.05) in node [right] at (-2.0,1.5);
	\end{tikzpicture}	
	\begin{tikzpicture}
	[spy using outlines={rectangle,red,magnification=2,width=16mm, height =12mm, connect spies}]				
	\node {\includegraphics[width=0.233\textwidth]{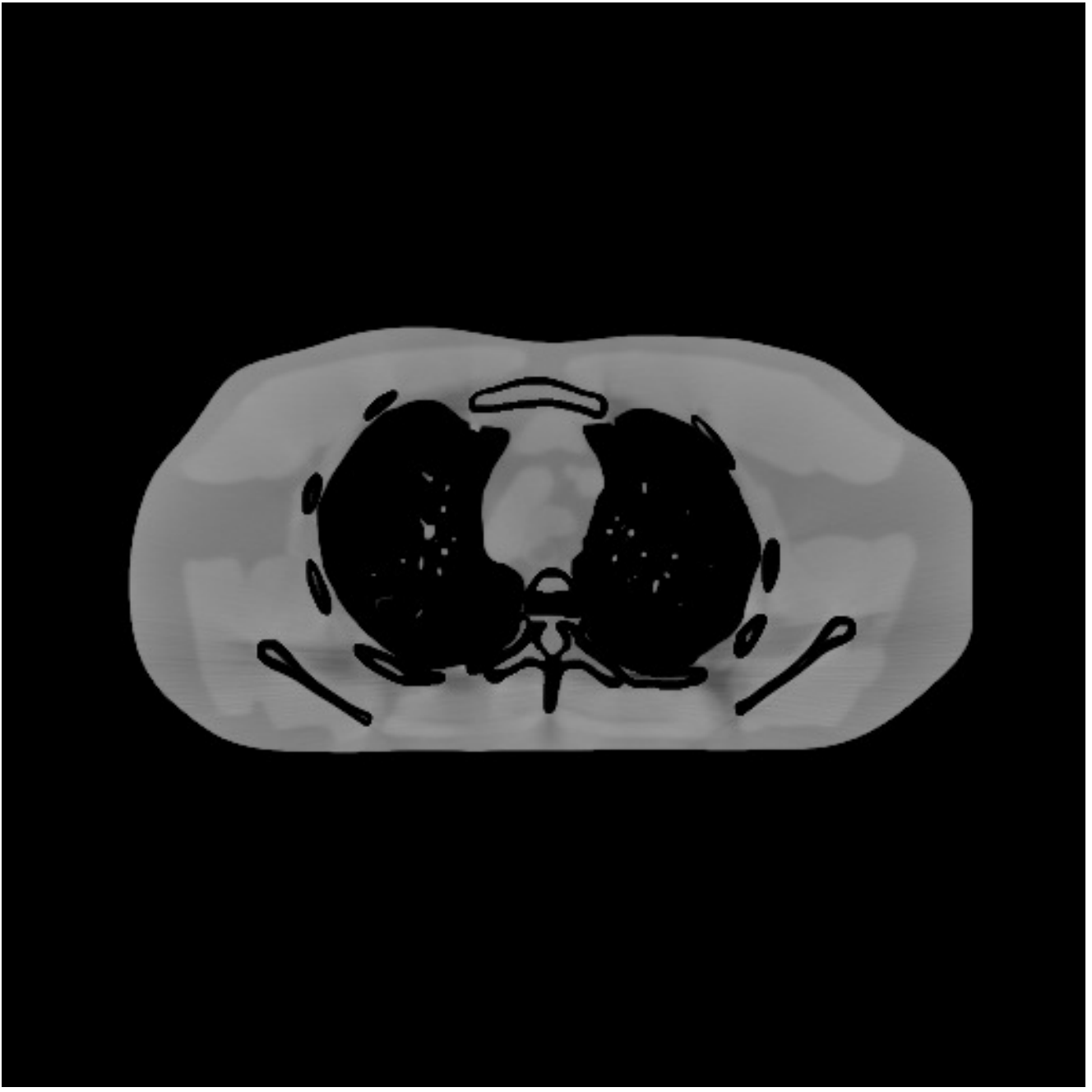}	};
	\spy on (0.4,-0.5) in node [right] at (0.5,-1.45);
	\spy on (-1.1,0.05) in node [right] at (-2.0,1.5);	
	\end{tikzpicture}   \\ \vspace{-0.2in}
	
	\begin{tikzpicture}
	[spy using outlines={rectangle,red,magnification=4,width=16mm, height =12mm, connect spies}]				
	\node {\includegraphics[width=0.233\textwidth]{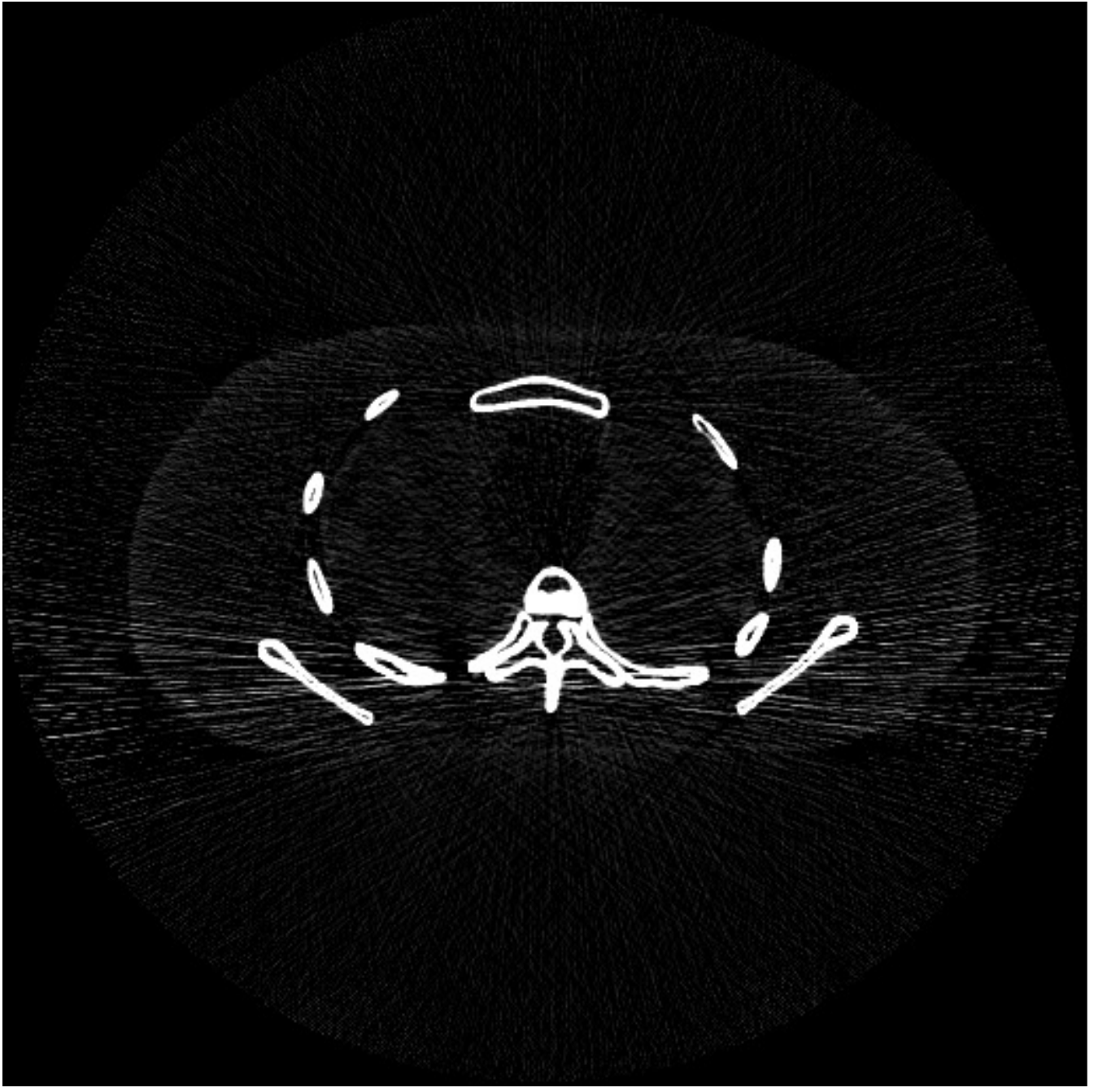}	};	
	\spy on (0.45,-0.5) in node [right] at (0.55,-1.45);
	\spy [width=16mm, height =12mm, magnification=3] on (-0.55,-0.45) in node [right] at (-2.1,-1.45);					
	\end{tikzpicture}
	\begin{tikzpicture}
	[spy using outlines={rectangle,red,magnification=4,width=16mm, height =12mm, connect spies}]				
	\node {\includegraphics[width=0.233\textwidth]{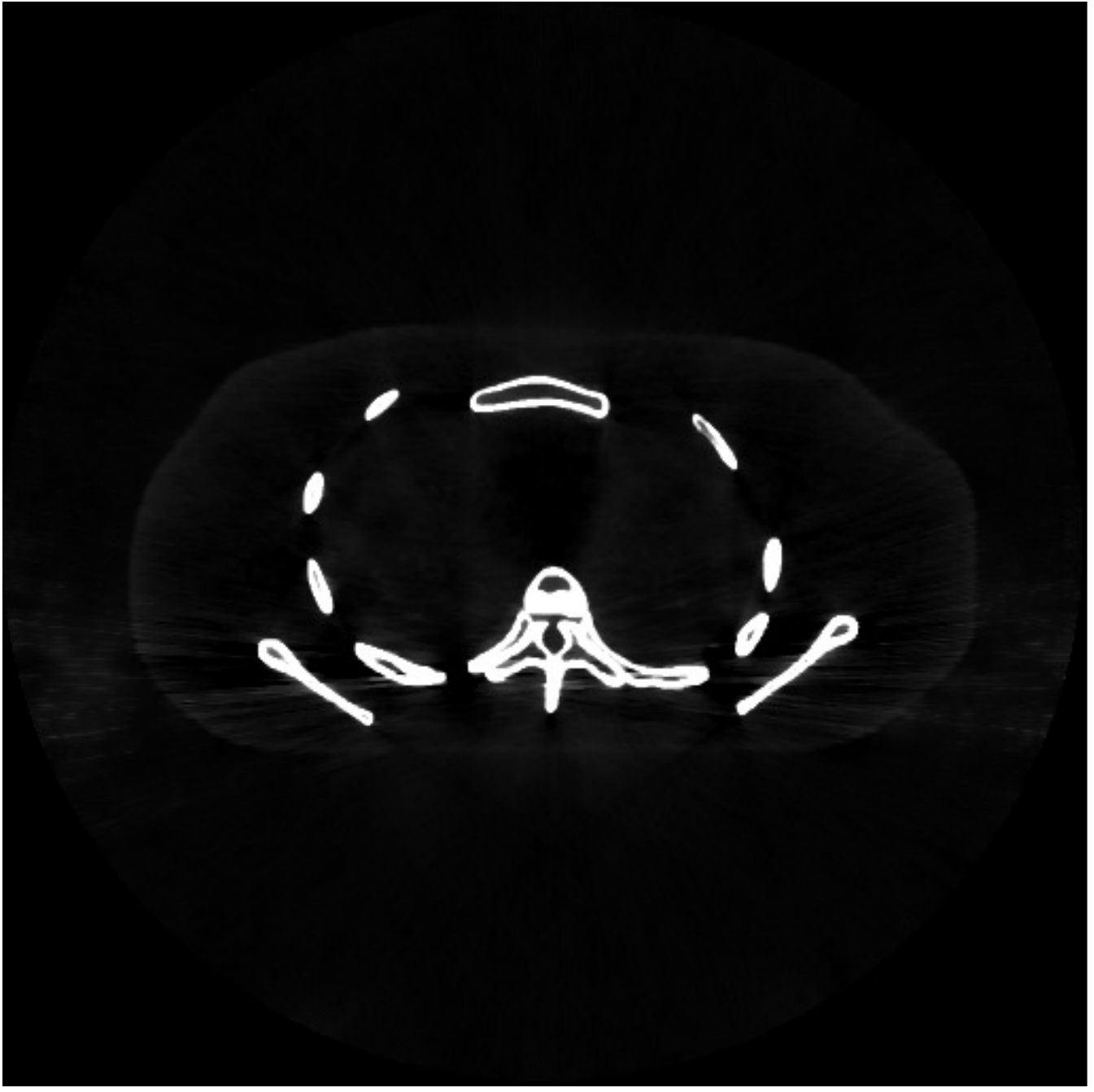}	};	
	\spy on (0.45,-0.5) in node [right] at (0.55,-1.45);
	\spy [width=16mm, height =12mm, magnification=3] on (-0.55,-0.45) in node [right] at (-2.1,-1.45);				
	\end{tikzpicture}	
	\begin{tikzpicture}
	[spy using outlines={rectangle,red,magnification=4,width=16mm, height =12mm, connect spies}]				
	\node {\includegraphics[width=0.233\textwidth]{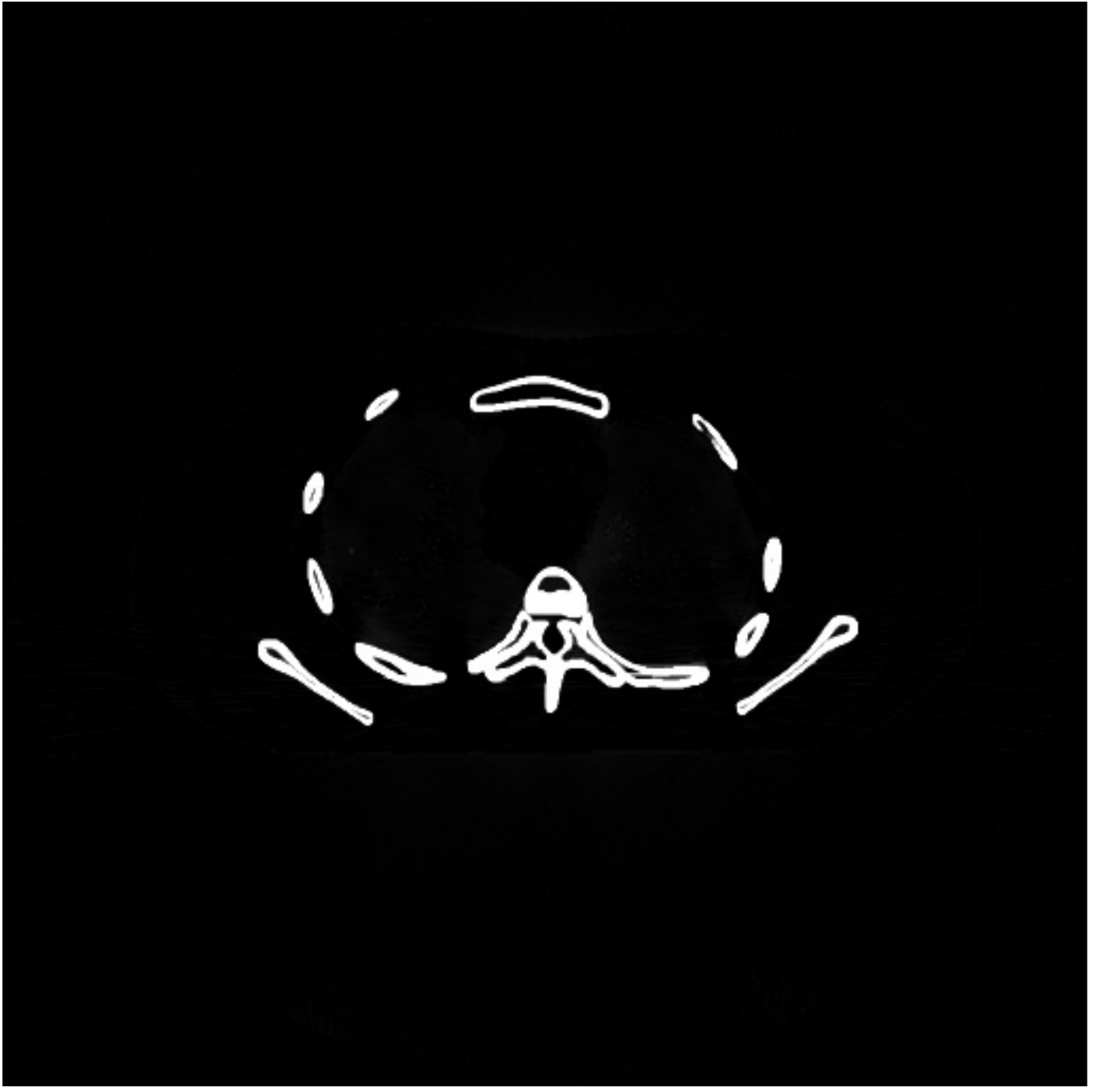}	};
	\spy on (0.45,-0.5) in node [right] at (0.55,-1.45);
	\spy [width=16mm, height =12mm, magnification=3] on (-0.55,-0.45) in node [right] at (-2.1,-1.45);					
	\end{tikzpicture}
	\begin{tikzpicture}
	[spy using outlines={rectangle,red,magnification=4,width=16mm, height =12mm, connect spies}]				
	\node {\includegraphics[width=0.233\textwidth]{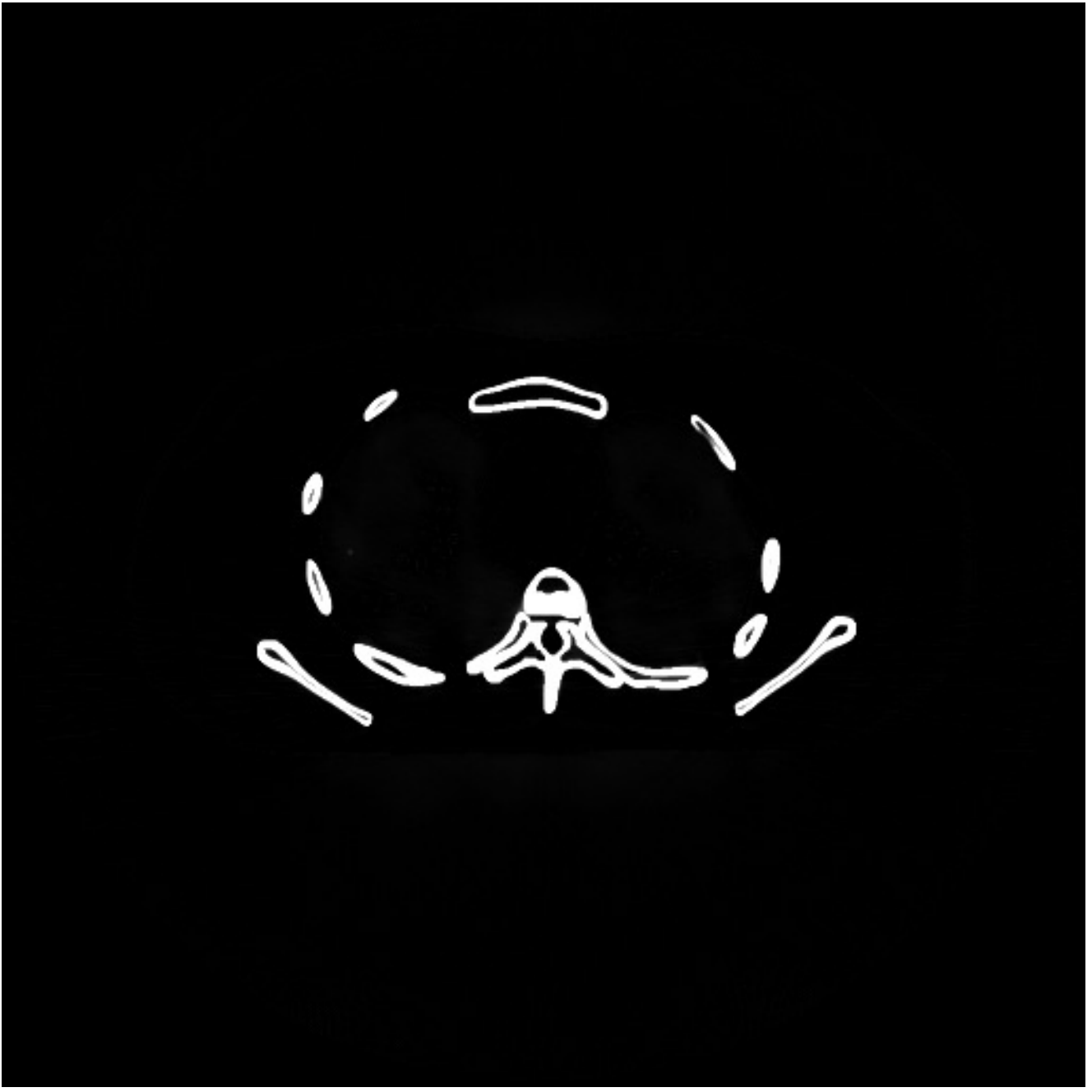}	};
	\spy on (0.45,-0.5) in node [right] at (0.55,-1.45);
	\spy [width=16mm, height =12mm, magnification=3] on (-0.55,-0.45) in node [right] at (-2.1,-1.45);				
	\end{tikzpicture} 	 	
	\caption{Left to right: basis material images decomposed by Direct Matrix Inversion, DECT-EP, DECT-TDL, and DECT-MULTRA.  The top and bottom rows show the water and bone images with display windows [0.7  1.3]\,g/cm$^3$ and [0 0.8]\,g/cm$^3$, respectively. }		
	\label{Fig:comp_error_result}
	\vspace{-0.13in}
\end{figure*}

\subsubsection{Decomposition Results and Comparisons}
We first obtained basis material images from the attenuation images at high and low energies using the Direct Matrix Inversion method, which were then used to initialize the DECT-EP method (that uses a convex regularizer). 
We ran sufficient ($500$) iterations of the DECT-EP algorithm, with \mbox{parameters} $\{\beta_1\,,\beta_2\}$ and $\{\delta_1,\,\delta_2\}$ being $\{2^8,\,2^{8.5}\}$ and $\{0.01,\,0.02\}$~g/cm$^3$, respectively. 
For DECT-ST, DECT-CULTRA, DECT-TDL and DECT-MULTRA, we initialized the algorithms with the DECT-EP decompositions. 
For DECT-TDL, we set a maximum sparsity level of $50$ during sparse coding along with an error tolerance of $0.3$, and set the regularization parameter as $30$.
For the DECT-ST scheme, the parameters $\{\beta_1,\,\beta_2$,\,$\gamma_1,\,\gamma_2\}$ were set as $\{50,\,70,\,0.03,\,0.04\}$. 
For DECT-CULTRA (${K}_2=10$), the parameters $\beta$ and $\gamma$ were set as $70$ and $0.07$, respectively. 
For DECT-MULTRA ($\sqrt{K_1}=15, {K}_2=10$), the parameters $\{\beta_1,\,\beta_2,\,\gamma_1,\,\gamma_2\}$ were set as $\{50,\,50,\,0.13,\,0.09\}$. 
We empirically selected these optimal parameter combinations for the various methods to achieve the best image quality and decomposition accuracy in our experiments. 

Table \ref{Tab: RMSE} shows the RMSE values of material images decomposed by various methods for different test slices. 
DECT-MULTRA clearly achieves the smallest RMSE values in Table~\ref{Tab: RMSE}, followed by DECT-CULTRA, and then the RMSE values increase gradually from DECT-TDL, DECT-ST, DECT-EP to Direct Matrix Inversion. To capture the rich features of basis material images,  DECT-CULTRA uses a union of learned cross-material sparsifying transforms, DECT-TDL uses a pre-learned overcomplete tensor-based dictionary, and DECT-MULTRA uses a mixed union of learned transforms. 
These three methods achieve smaller RMSE  than DECT-ST that uses only two square sparsifying transforms to sparsify the two basis materials.
Moreover, compared to DECT-TDL and DECT-CULTRA that exploit only a cross-material model, DECT-MULTRA learns both common-material properties and cross-material dependencies, which enables it to outperform the former methods. 

\begin{table}[h]
	\centering
	\caption{RMSE of decomposed images of basis \mbox{materials} for Direct Matrix Inversion, DECT-EP, DECT-ST, DECT-TDL, DECT-CULTRA $(K_2=10)$, and DECT-MULTRA $(\sqrt{K_1}=15, K_2=10)$, respectively, for multiple slices of the XCAT phantom. The unit for RMSE is $10^{-3}$\,g/cm$^3$. }
	\begin{tabular}{p{0.4cm}<{\centering} p{0.4cm}<{\centering} p{1cm}<{\centering} p{0.61cm}<{\centering} p{0.61cm}<{\centering} p{0.63cm}<{\centering} p{1cm}<{\centering} p{1cm}<{\centering}}
		\hline\hline
		\multicolumn{2}{c}{Method}  & \tabincell{c}{Direct \\Inversion} & \tabincell{c}{DECT \\EP} & \tabincell{c}{DECT \\ST} & \tabincell{c}{DECT \\TDL} &\tabincell{c}{DECT \\CULTRA}& \tabincell{c}{DECT \\ MULTRA}  \\
		\hline
		\multirow{2}{*}{\tabincell{c}{Slice \\61}} &  Water & 72.8 & 60.9 & 51.3 & 44.8 & 43.1 & \textbf{42.8} \\ \cline{2-8}
		&  Bone  & 68.4 & 60.2 & 51.6 & 44.9 & 44.1 & \textbf{43.9} \\ 
		\hline
		\multirow{2}{*}{\tabincell{c}{Slice \\77}} &  Water & 92.4 & 65.9 & 55.6 & 41.8 & 39.0 & \textbf{38.7} \\ \cline{2-8}
		&  Bone  & 89.0 & 72.2 & 61.8 & 50.8 & 50.0 & \textbf{49.8} \\  
		\hline
		\multirow{2}{*}{\tabincell{c}{Slice \\150}} &  Water & 116.7 & 69.1 & 61.7 & 43.5 & 40.8 & \textbf{38.6}  \\  \cline{2-8}
		&  Bone  & 110.8 & 76.7 & 67.0 & 53.4 & 52.3 &  \textbf{50.8} \\	  \hline	\hline				   
	\end{tabular} 
	\label{Tab: RMSE}
	\vspace{-0.15in}		
\end{table}

Fig.~\ref{Fig:comp_error_result} shows the material density images decomposed by the Direct Matrix Inversion method, DECT-EP, DECT-TDL, and DECT-MULTRA for a test slice (Slice 77). 
DECT-EP reduces the severe streak artifacts and noise observed in the decomposed water and bone images obtained by Direct Matrix Inversion. 
Compared to DECT-EP, DECT-MULTRA and DECT-TDL further reduce the artifacts and improve edge details at the boundaries of different materials. 
However, compared to DECT-EP, the DECT-TDL result suffers from poor soft-tissue contrast in the water image.
DECT-MULTRA that exploits both common and cross material learned models removes artifacts while improving image features and sharpness of soft-tissue edges, which is clearly noticeable in the zoom-ins of the water and bone images.
\textcolor{black}{The total runtime for the 500 iterations (using unoptimized Matlab code on a machine with two 2.70 GHz 12-core Intel Xeon E5-2697 v2 processors) was 181.9 minutes for DECT-TDL, whereas it was only 72.1 minutes for DECT-MULTRA. Unlike DECT-TDL that involves expensive and approximate sparse coding, DECT-MULTRA performs cheap and closed-form sparse coding and clustering leading to low runtimes.} 
Additional comparisons between decomposition error images are included in the supplement.
  
\subsubsection{Clustering and Convergence Behavior of DECT-MULTRA}
\begin{figure*}[htb]
	\centering
	\includegraphics[width=0.22\textwidth]{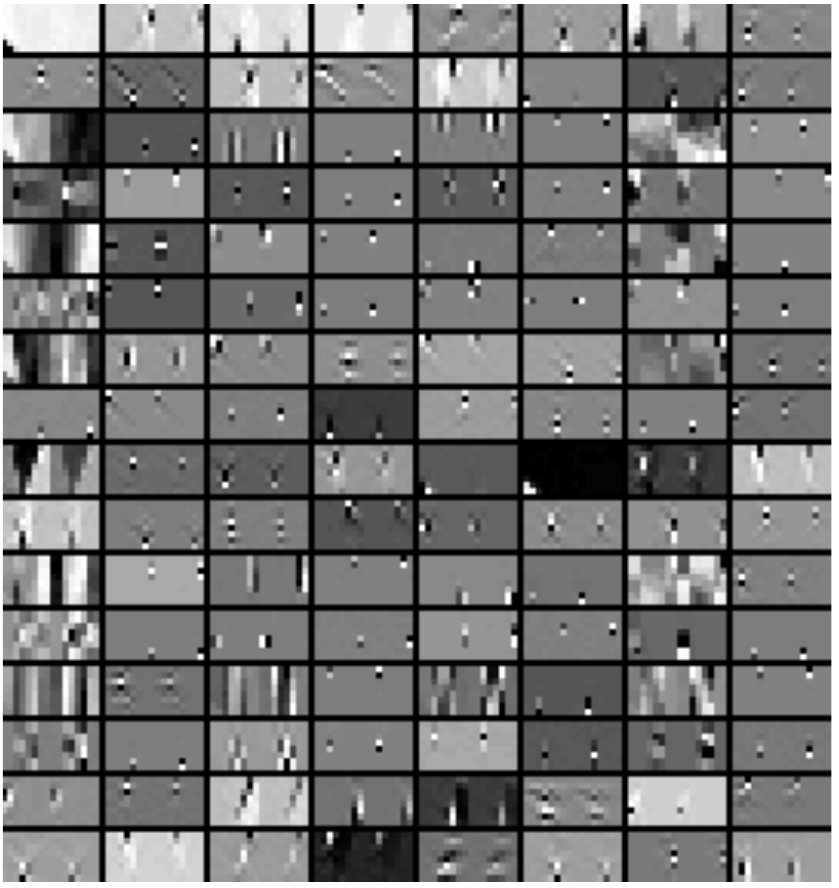}
	\includegraphics[width=0.22\textwidth]{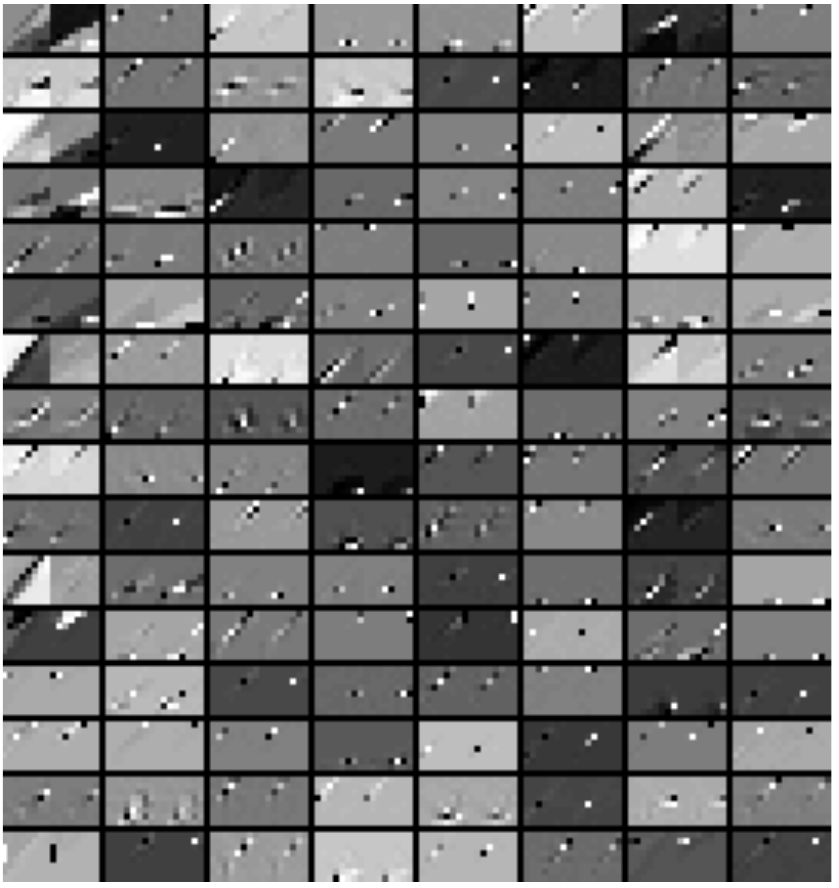}
	\includegraphics[width=0.22\textwidth]{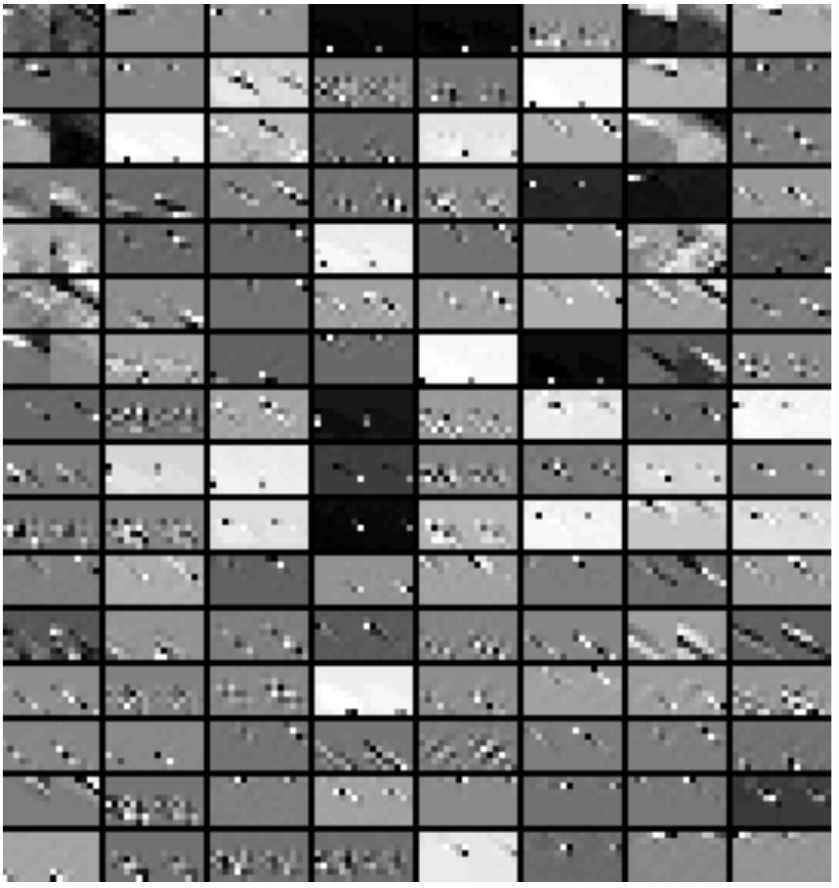}
	\includegraphics[width=0.22\textwidth]{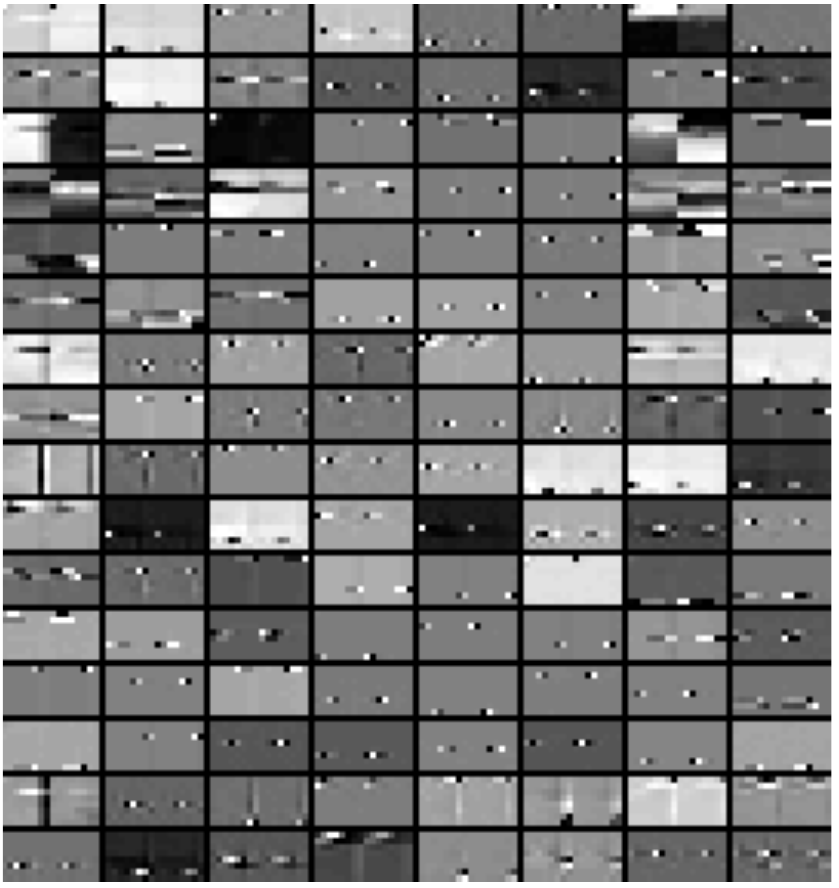}  \\
	\vspace{0.02in}
	\begin{overpic}[width=0.22\textwidth]{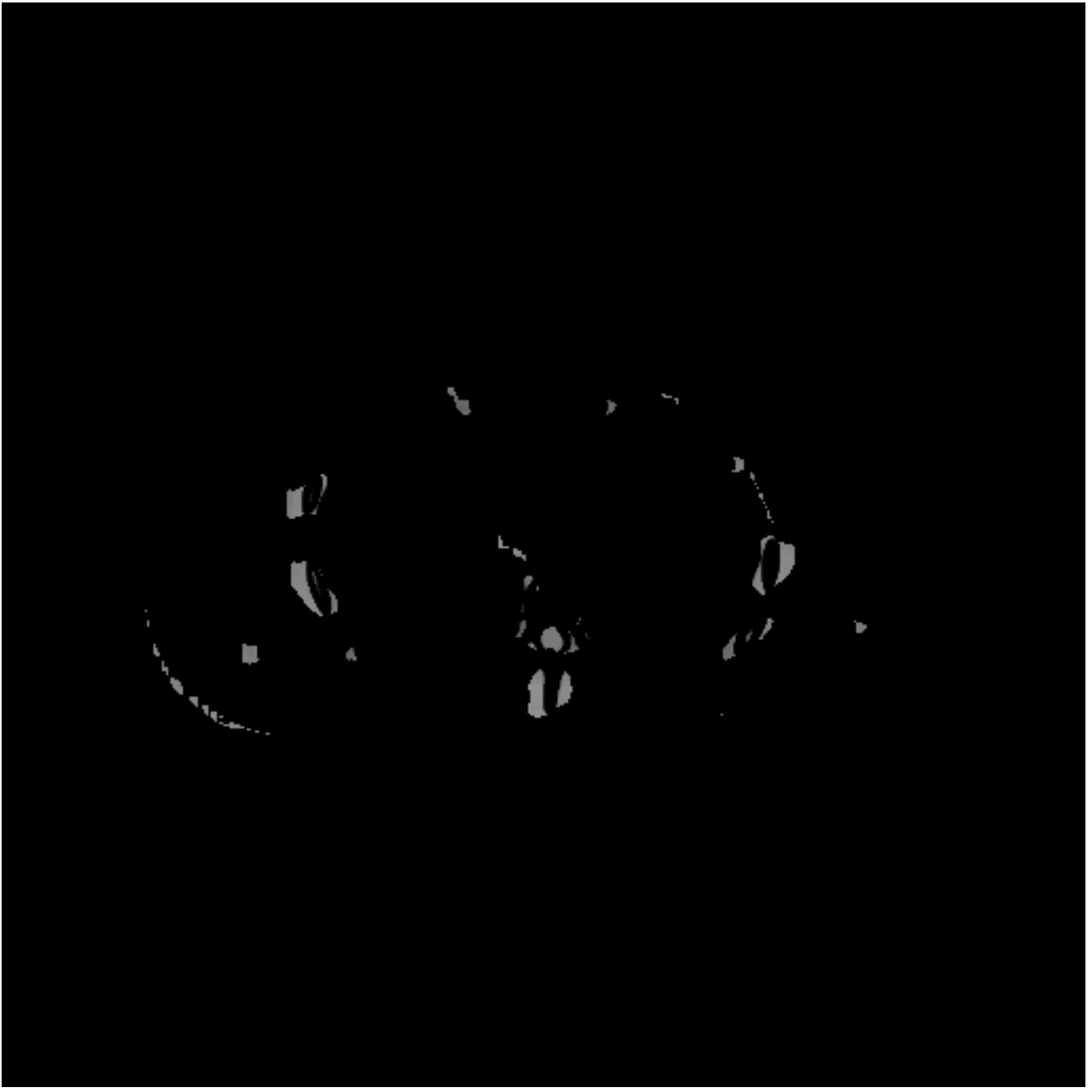}
		\put(30,85){ \color{white}{\bf \large{Class 4}}}     
	\end{overpic}
	\begin{overpic}[width=0.22\textwidth]{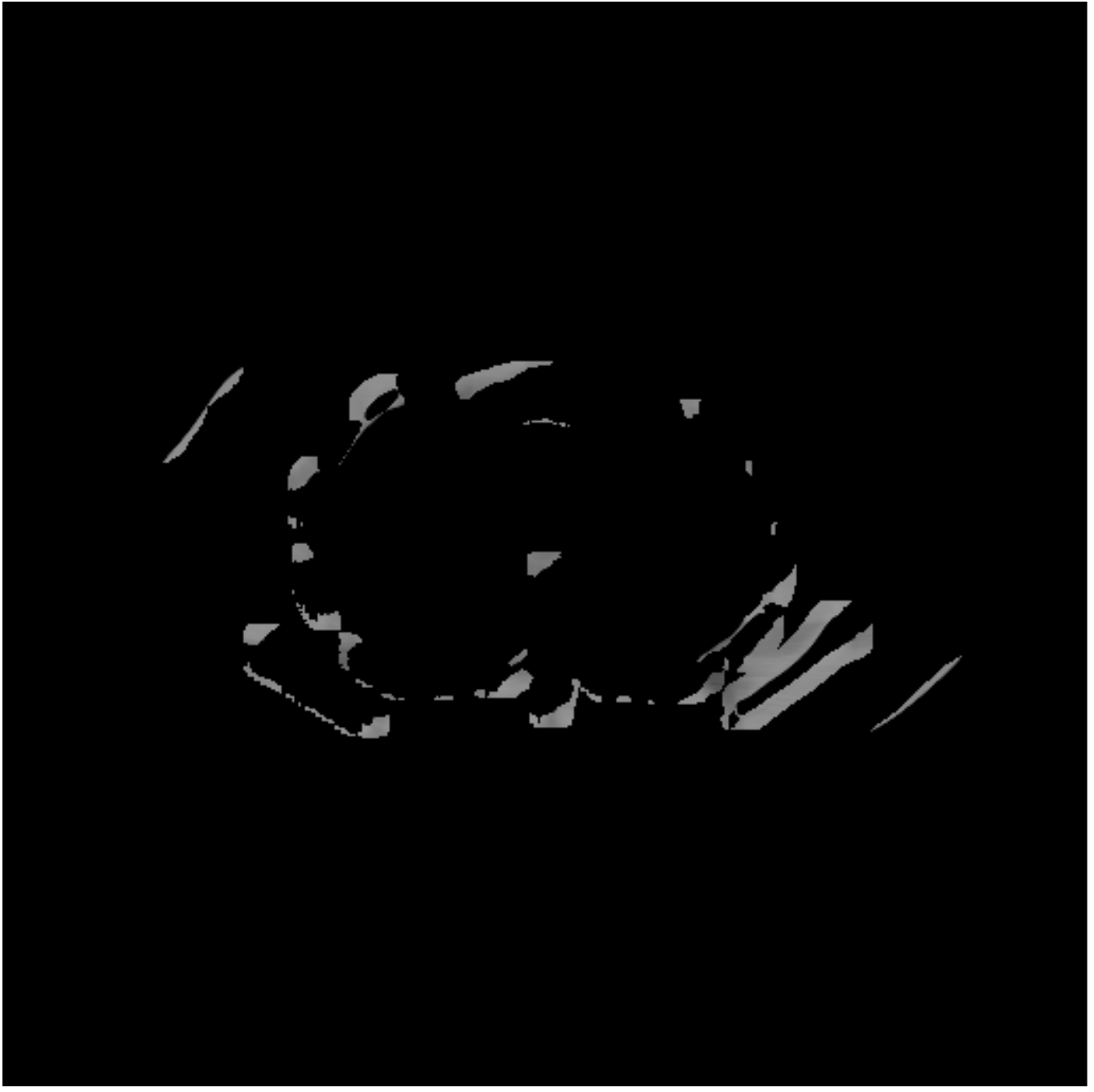}
		\put(30,85){ \color{white}{\bf \large{Class 5}}}     
	\end{overpic}
	\begin{overpic}[width=0.22\textwidth]{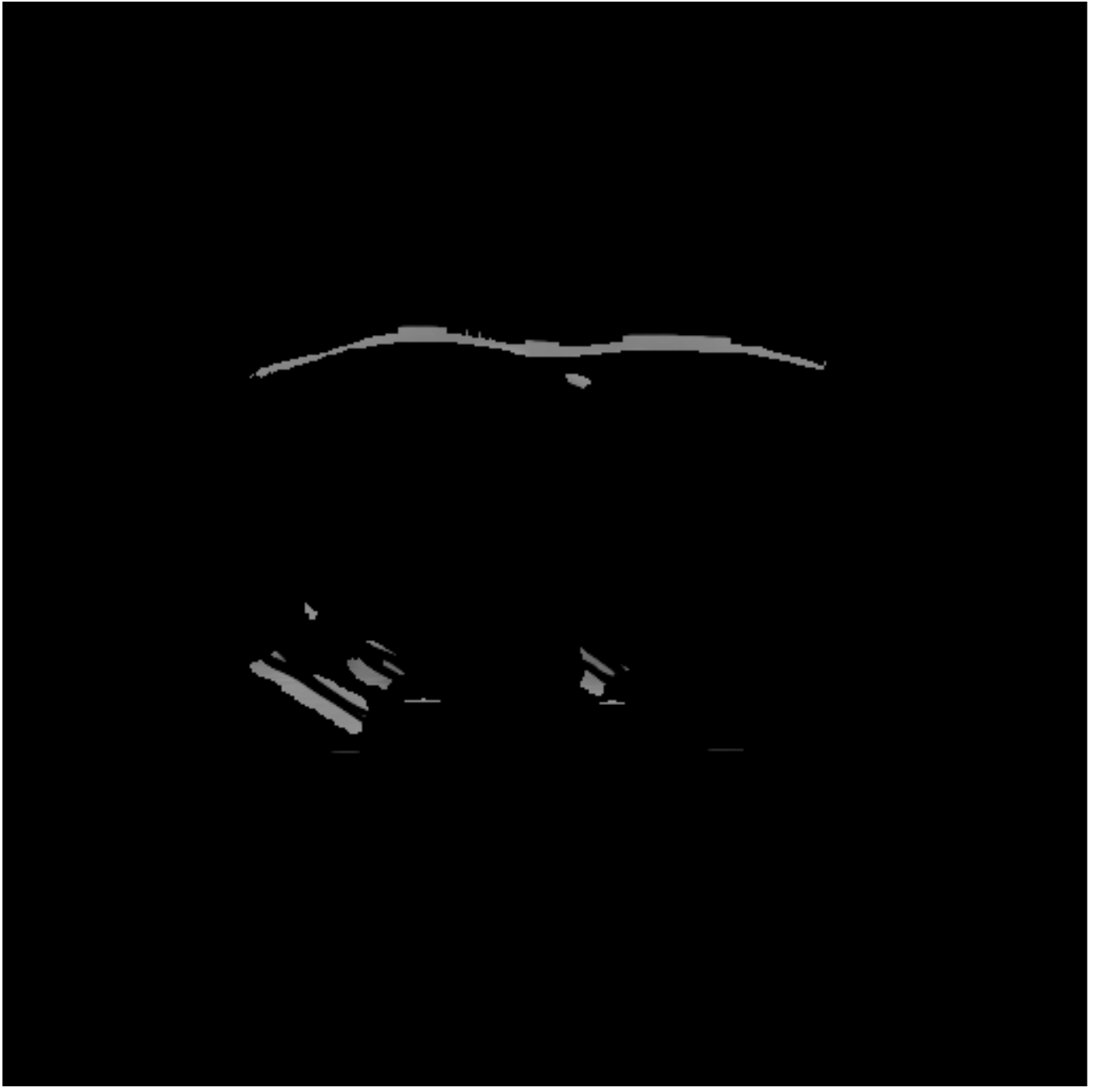}
		\put(30,85){ \color{white}{\bf \large{Class 7}}}     
	\end{overpic}
	\begin{overpic}[width=0.22\textwidth]{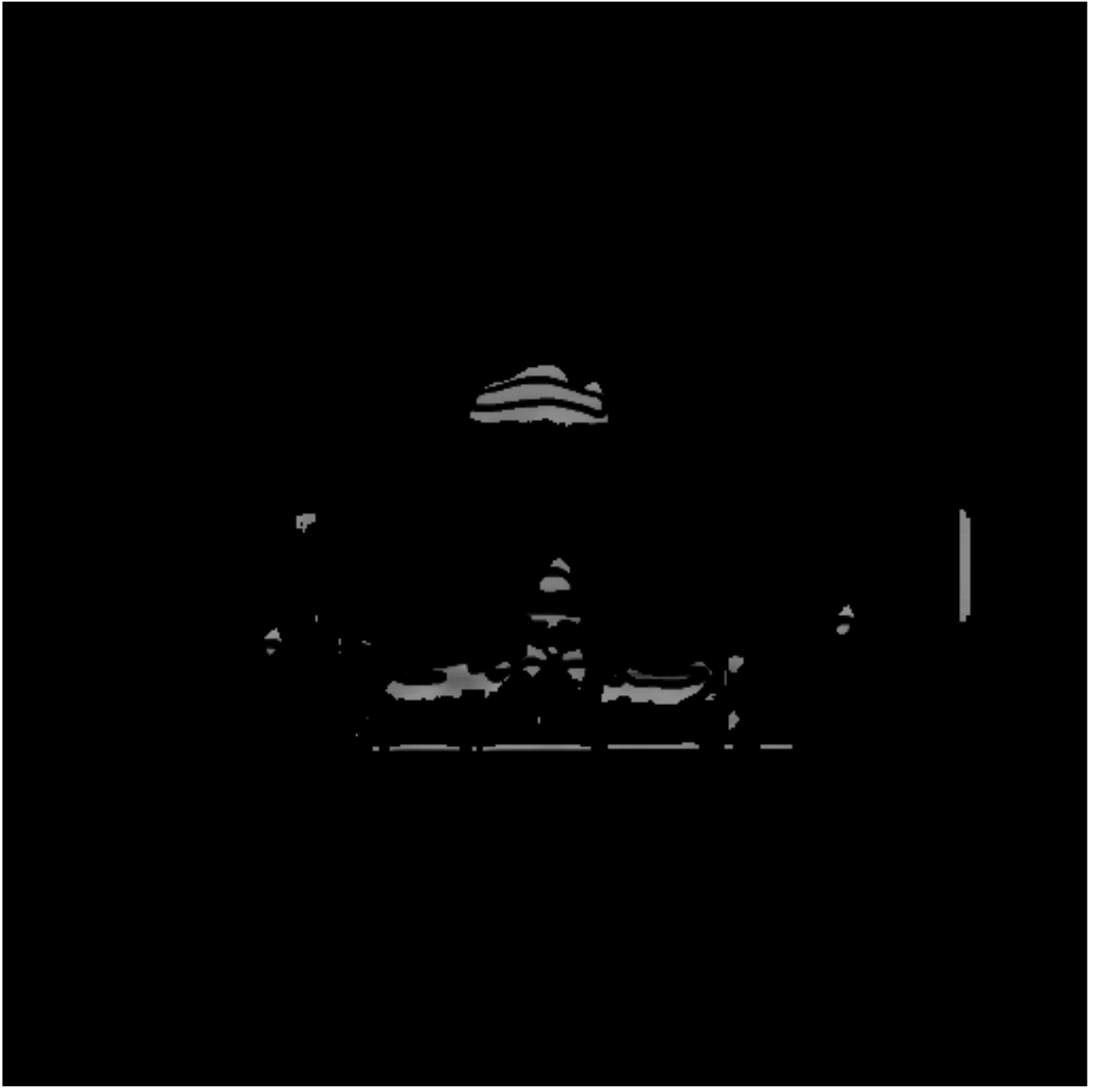}
		\put(30,85){ \color{white}{\bf \large{Class 8}}}     
	\end{overpic}				
	\includegraphics[width=0.22\textwidth]{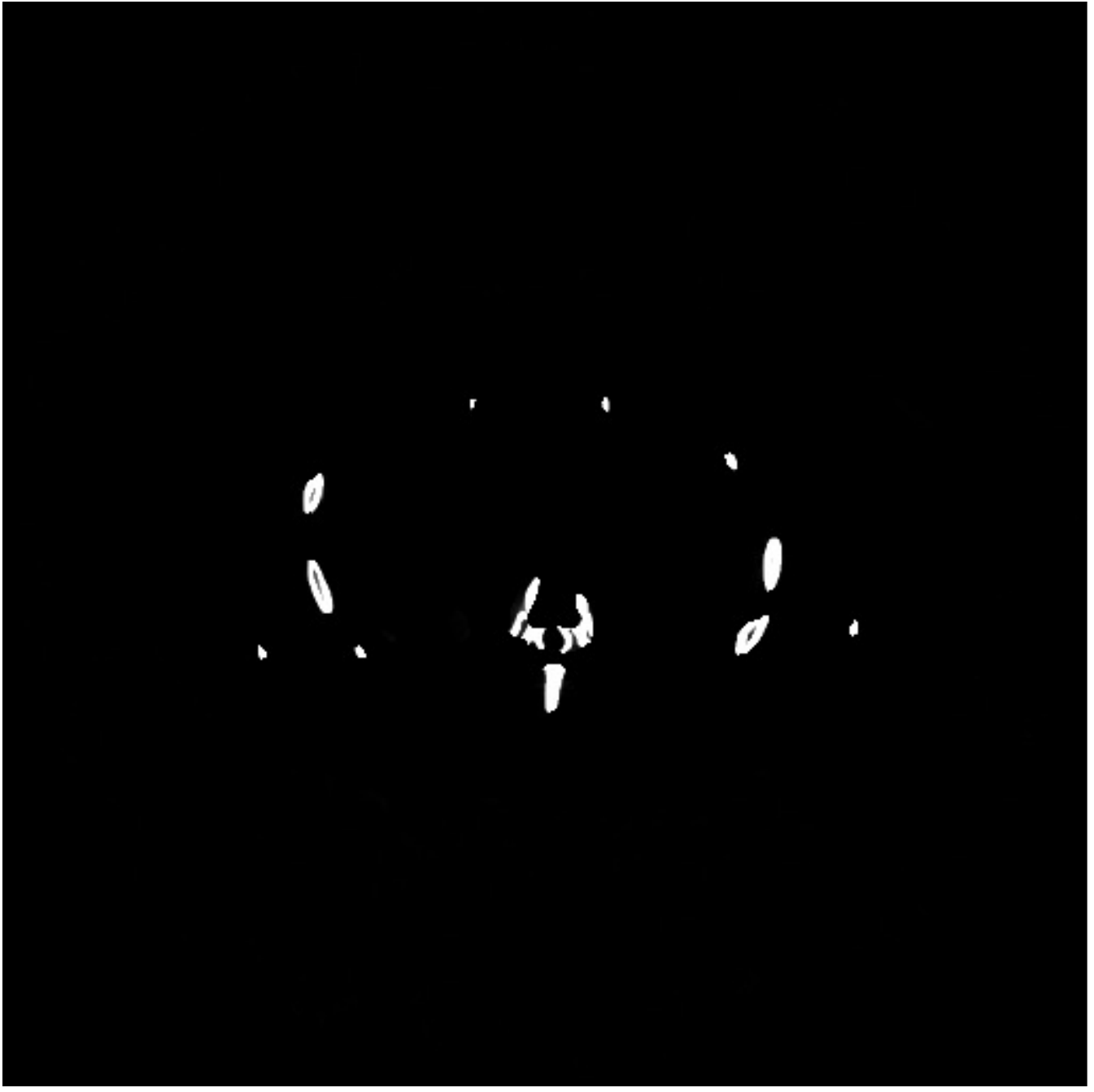}
	\includegraphics[width=0.22\textwidth]{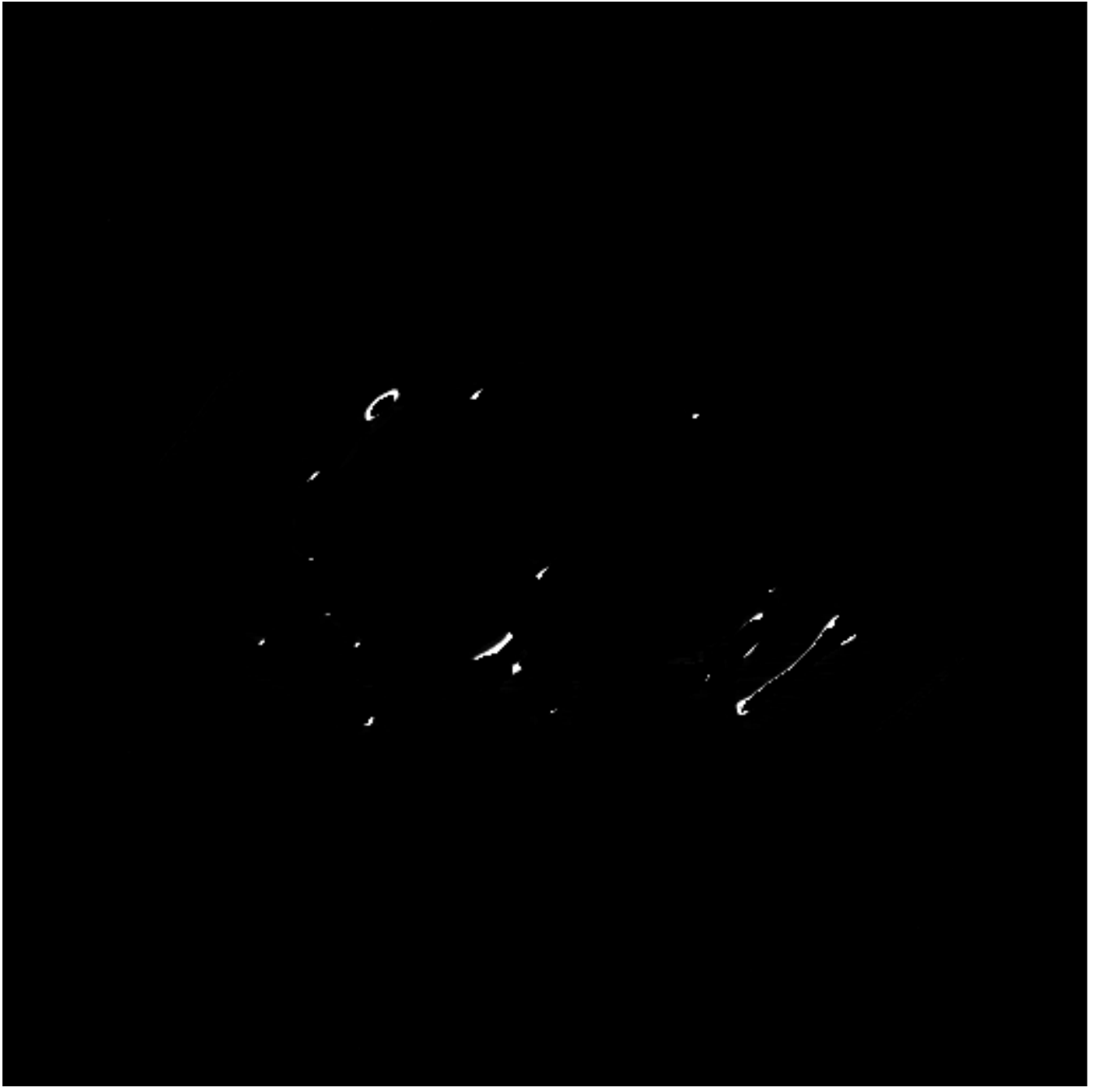}
	\includegraphics[width=0.22\textwidth]{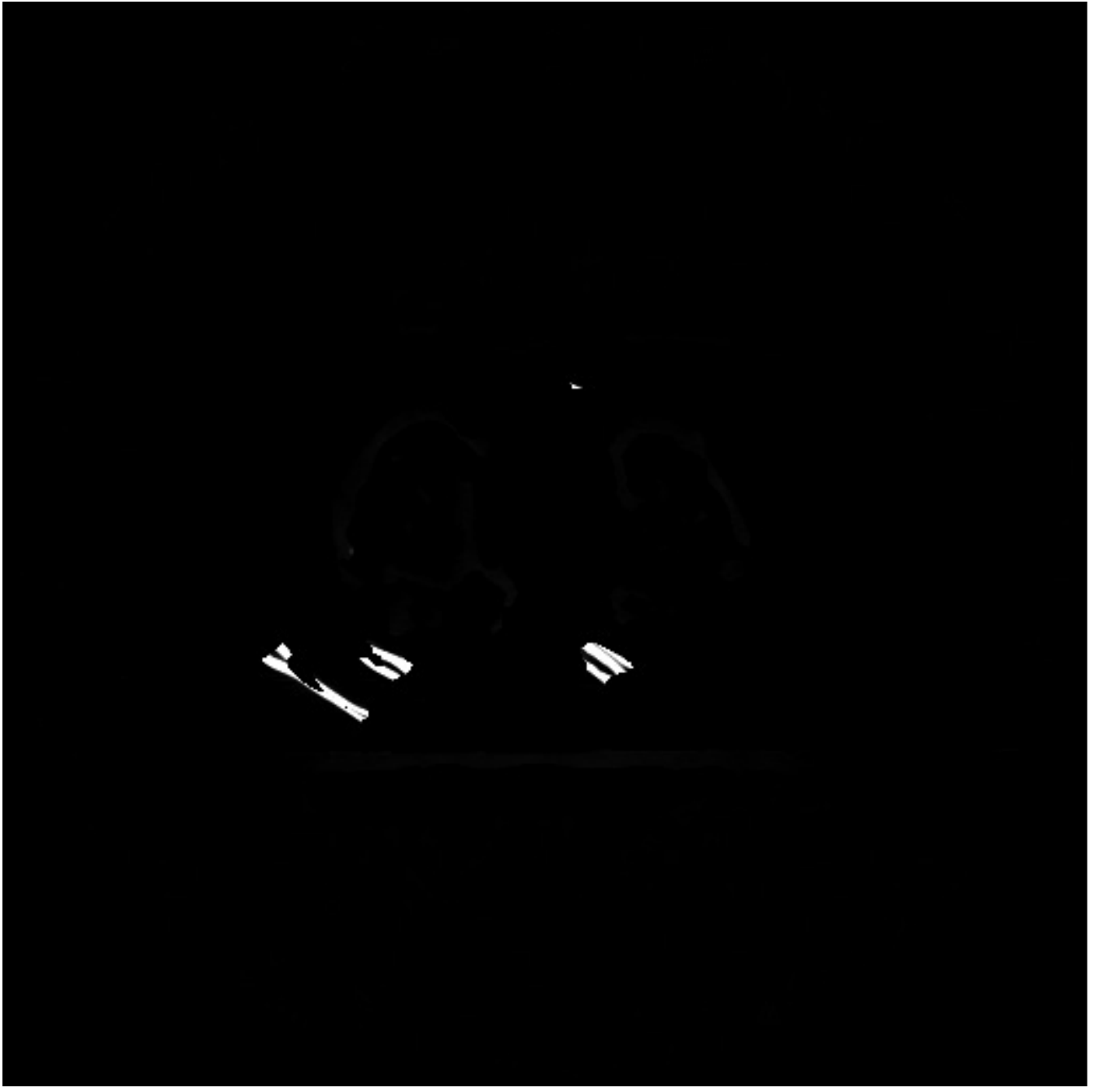}
	\includegraphics[width=0.22\textwidth]{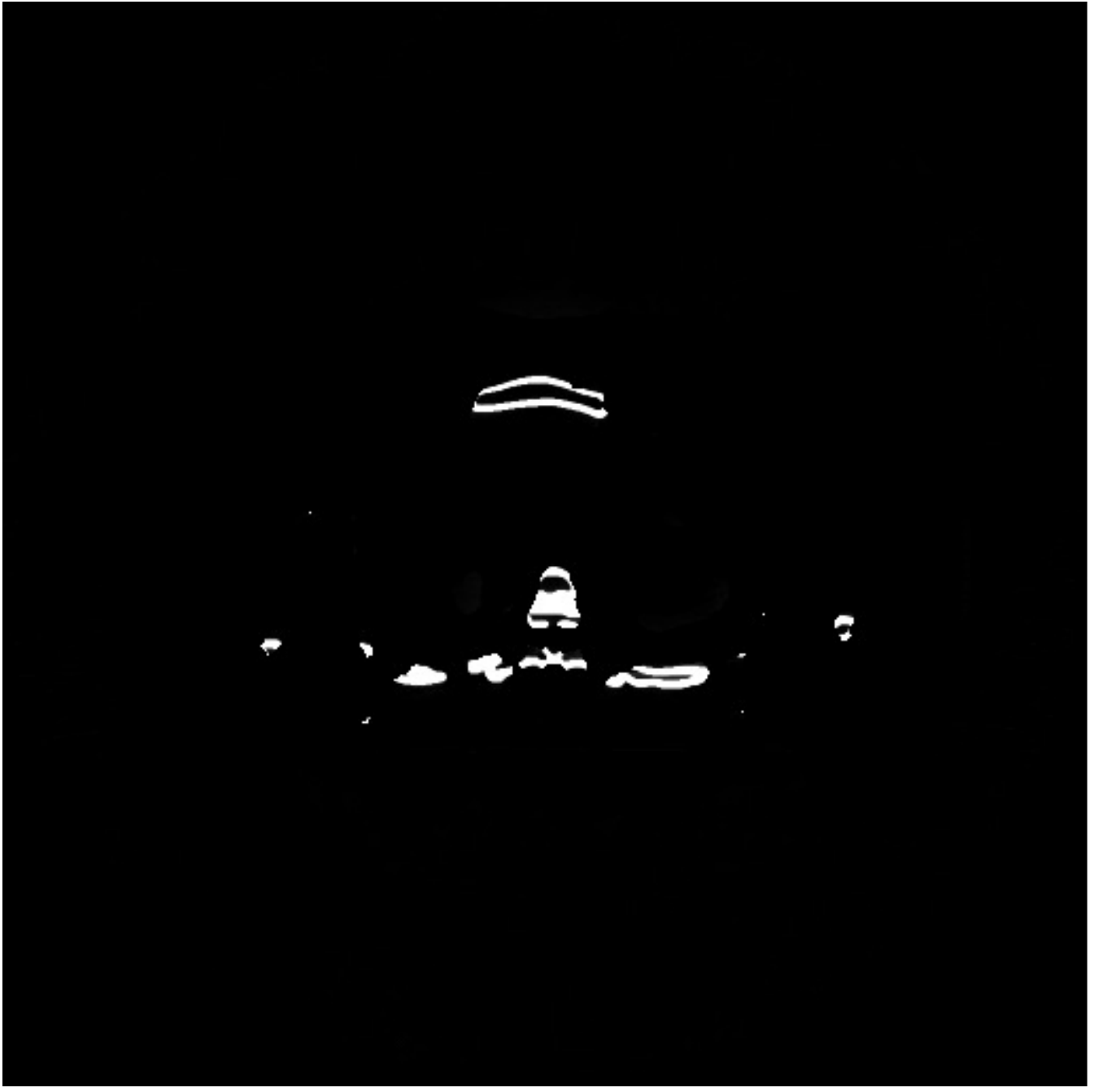}
	\caption{Pixel-level clustering in slice 77 with the DECT-MULTRA cross-material model: The top row shows the individual cross-material transforms for classes $4$, $5$, $7$, and $8$, with the transform rows shown as $8 \times 16$ patches. The middle and bottom rows show the corresponding water and bone pixels (using estimated densities in the decomposition) grouped into each class, with display windows [0.7  1.3]\,g/cm$^3$ and [0 0.8]\,g/cm$^3$, respectively. }
	\label{Fig:Clustering}
\end{figure*}

To better illustrate the effect of the learned models in DECT-MULTRA, Fig.~\ref{Fig:Clustering} shows examples of pixel-level clustering of water and bone pixels in Slice 77 for the cross-material model ($K_1=225,~K_2=10$). 
Since each pixel in the stacked water+bone result of DECT-MULTRA belongs to many overlapping (3D) patches, it is clustered into either the common or cross material model and further to a specific transform (class) in the model, by a majority vote among the (already clustered) patches overlapping it. 
The clustering results show that the cross-material model effectively captures pixels straddling the boundary regions of materials. The clustering captures various oriented edges, with Class 4 contains many vertical edges; classes 5 and 7 contain many oriented edges (e.g., at 45-degree and 135-degree orientations); and class 8 contains mostly horizontal edges. Fig.~\ref{Fig:Clustering} also illustrates the cross-material transforms for each class, which show various directional and gradient-like features that jointly sparsify the water and bone dependencies. Additional clustering results for the common-material model and the corresponding transforms are shown in the supplement.

Fig.~\ref{Fig:Objective} shows that the objective function of DECT-MULTRA monotonically decreases and converges quickly  over iterations for decomposing the central slice of the XCAT phantom. Fig.~\ref{Fig:RMSE} shows the convergence of the RMSE of water and bone images for DECT-ST, DECT-TDL, DECT-CULTRA and DECT-MULTRA over their iterations. 
DECT-MULTRA achieves the lowest RMSE and also converges faster than the other methods.

\subsubsection{Study of Different Weights for the Common-material and Cross-material Models}
	The common-material and cross-material parts of the DECT-MULTRA regularizer in (\ref{Eq:cost_func_reg}) receive weight $\beta_1$ and $\beta_2$, respectively. 
	Here, we tuned $\beta_1$ and $\beta_2$ to study the effects of the common-material and cross-material models.   
	Fig.~\ref{fig:weight_comp} shows the water and bone material images decomposed by DECT-MULTRA using different weights for the common-material and cross-material models.
\begin{figure}[h]
	\centering
	\includegraphics[scale=0.47]{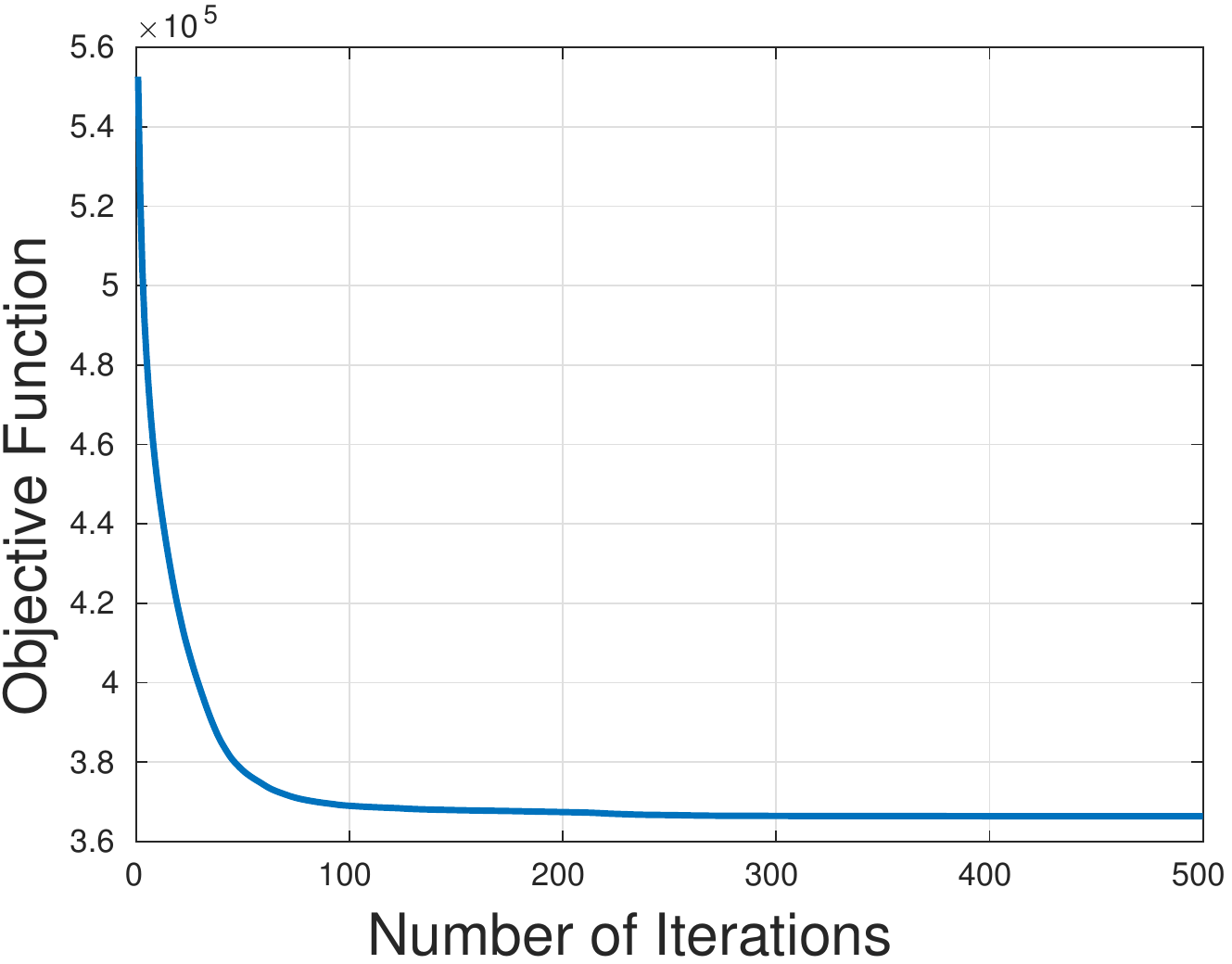}
	\caption{Objective function in (\ref{Eq:cost_func}) plotted over the iterations of the DECT-MULTRA algorithm when decomposing the central slice (Slice 77) of the XCAT phantom.}
	\label{Fig:Objective}
	\vspace{-0.3in}
\end{figure}
	A smaller weight $\beta_1$ for the common-material model and a  correspondingly larger weight $\beta_2$ for the cross-material model leads to lower contrast and streak artifacts in water images, but can remove artifacts near boundaries of water and bone materials (first column of Fig.~\ref{fig:weight_comp}). A larger $\beta_1$ and smaller $\beta_2$ can improve the contrast and eliminate artifacts in water images, but leads to artifacts in mixed material regions (forth and fifth columns of Fig.~\ref{fig:weight_comp}).	
	 Table \ref{Tab:rmse_comp} shows the RMSE values of material images decomposed by DECT-MULTRA with different weights for the common-material and cross-material models in (\ref{Eq:cost_func}).  The equal weights case $\{\beta_1= 50, \beta_2=50\}$ and the weight combination $\{\beta_1= 25, \beta_2=75\}$  achieved lower RMSE values compared to the other weight combinations. 
	 Fig.~15 in the supplement shows the water and bone images obtained by DECT-MULTRA with $\beta_1 = 99$ and $\beta_2 =1$, which have high RMSE. Thus,
	 we can infer that in the MULTRA model, both the cross-material and common-material components are important, but the cross-material model is slightly more important than the common-material model.

\begin{figure}[htb]
	\vspace{-0.05in}
	\centering
	\includegraphics[scale=0.45]{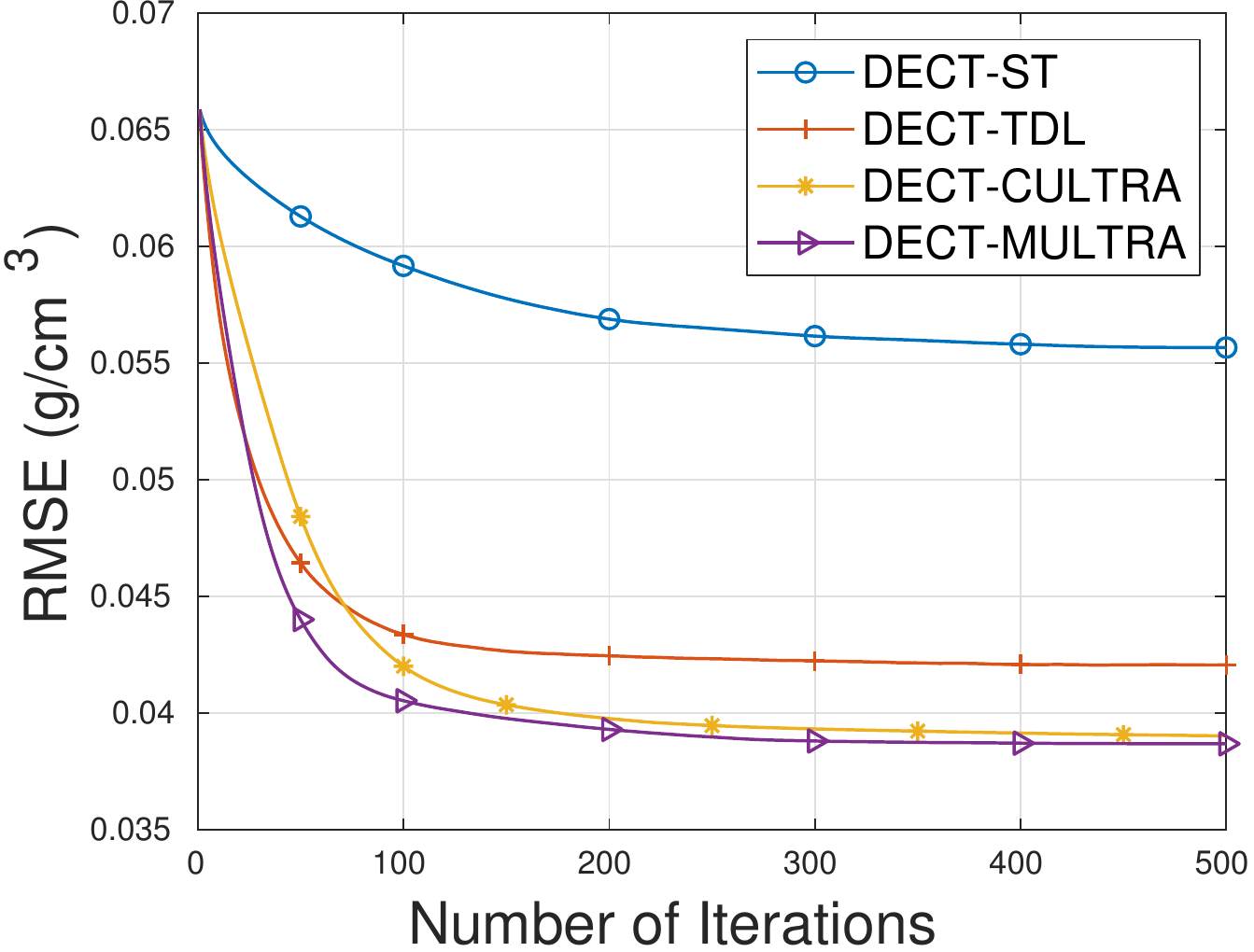} \centerline{(a)}\\
	\includegraphics[scale=0.45]{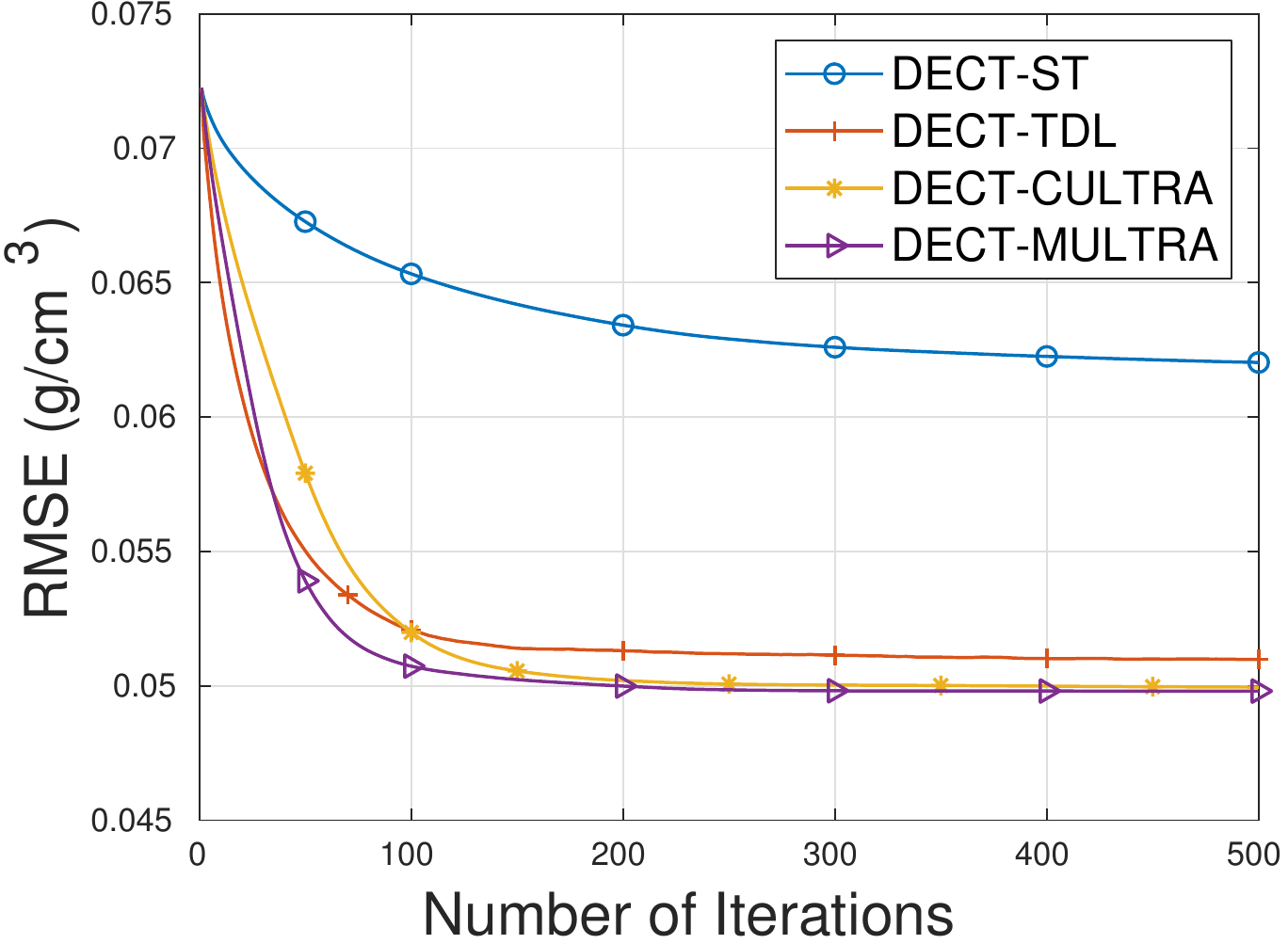} \centerline{(b)}
	\caption{RMSE for the central slice of the XCAT phantom plotted over the iterations of the DECT-TDL, DECT-ST, DECT-CULTRA, and DECT-MULTRA algorithms. (a) and (b) show the RMSE of water and bone images, respectively.}
	\label{Fig:RMSE}
	\vspace{-0.2in}
\end{figure}

\textcolor{blue}{
	\vspace{-0.05in}	
	\begin{table}[htb]
		\centering  
		\begin{tabular}{cccccc} 
			\hline \hline 
			$\{\beta_1\,,\beta_2\}$ & $\{1,\,99\}$ & $\{25,\,75\}$ & $\{50,\,50\}$  & $\{75,\,25\}$ &$\{90,\,10 \}$ \\
			\hline
			Water & 39.4 & 38.3 & 38.7  & 39.3 & 41.9\\
			\hline
			Bone & 50.4 & 49.8 & 49.8  & 50.1 & 52.0\\ 
			\hline \hline 
		\end{tabular}  
		\caption{RMSE values of decomposed basis material images for different choices of weights for the common-material and cross-material model terms in (\ref{Eq:cost_func}). The unit of RMSE is $10^{-3}$\,g/cm$^3$. } 
		\label{Tab:rmse_comp}	
		\vspace{-0.2in}  
	\end{table}	 
}

	\begin{figure*}[bth]
		\centering
		\begin{tikzpicture}
		[spy using outlines={rectangle,red,magnification=2,width=15mm, height =10mm, connect spies}]				
		\node {\includegraphics[width=0.18\textwidth]{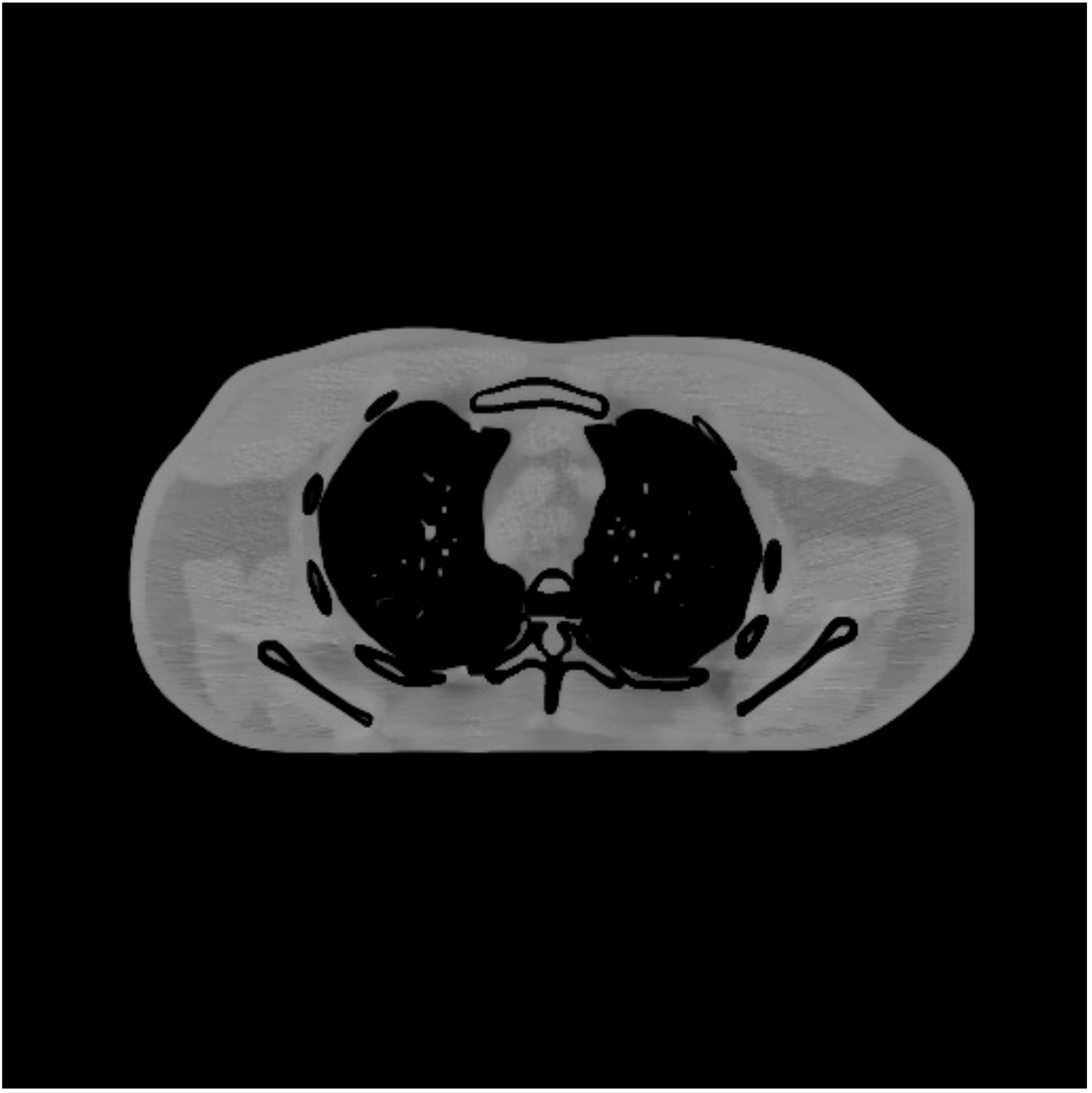}};
		\spy on (0.4,-0.5) in node [right] at (0.05,-1.15);	
		\spy on (-0.5,-0.5) in node [right] at (-1.6,-1.15);
		\spy on (0.85,0.45) in node [right] at (0.05,1.15);	
		\draw[line width=1pt, thin, -latex, red] (1.1,0.6) -- node[auto] {} (0.67,0.56);
		\draw[line width=1pt, thin, -latex, red] (0.45,-0.65) -- node[auto] {} (0.3,-0.43);
		\draw[line width=1pt, thin, -latex, red] (-0.45,-0.65) -- node[auto] {} (-0.43,-0.43);					
		\end{tikzpicture}
		\begin{tikzpicture}
		[spy using outlines={rectangle,red,magnification=2,width=15mm, height =10mm, connect spies}]				
		\node {\includegraphics[width=0.18\textwidth]{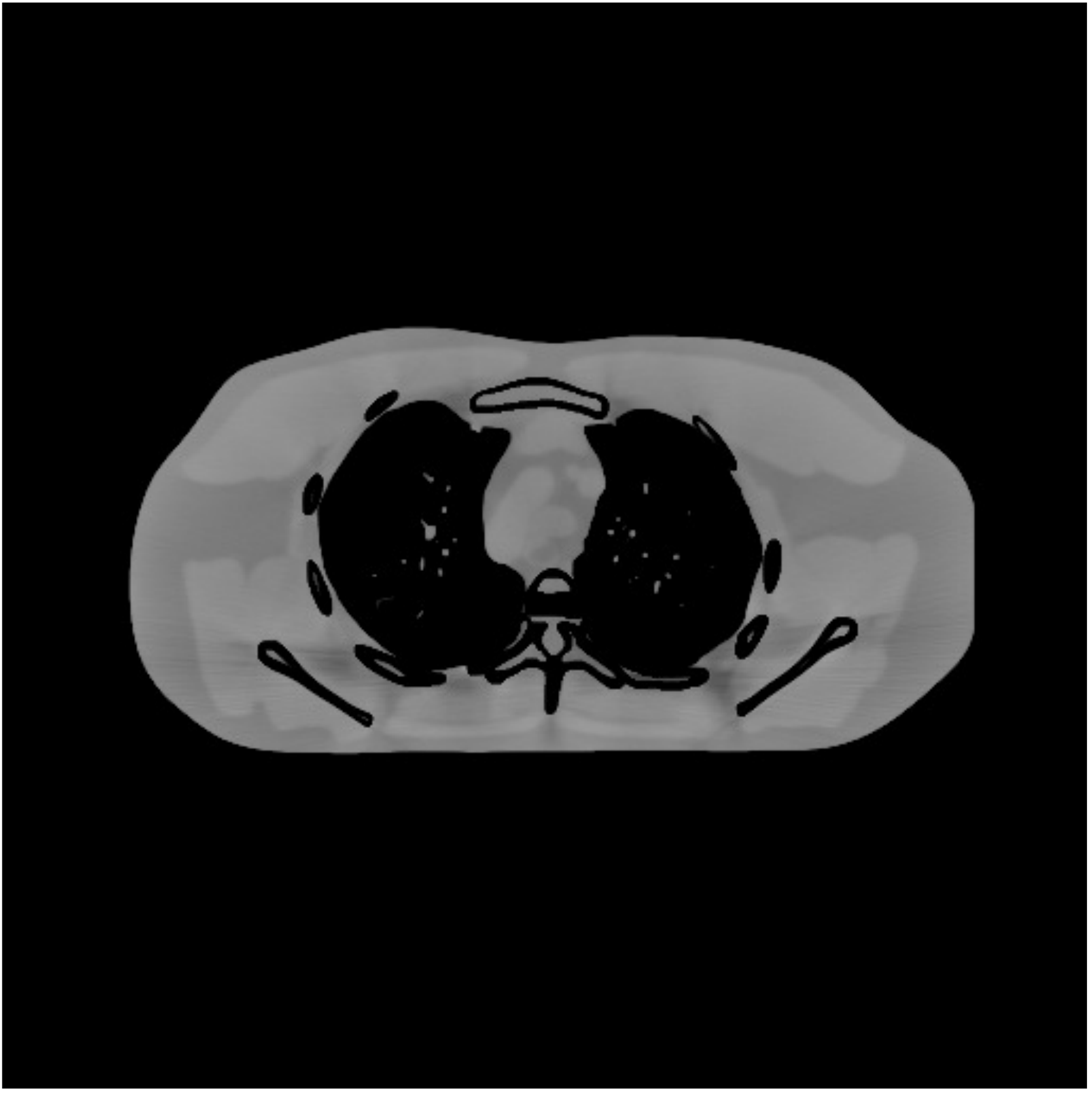}};
		\spy on (0.4,-0.5) in node [right] at (0.05,-1.15);
		\spy on (-0.5,-0.5) in node [right] at (-1.6,-1.15);
		\spy on (0.85,0.45) in node [right] at (0.05,1.15);	
		\draw[line width=1pt, thin, -latex, red] (1.1,0.6) -- node[auto] {} (0.67,0.56);
		\draw[line width=1pt, thin, -latex, red] (0.45,-0.65) -- node[auto] {} (0.3,-0.43);
		\draw[line width=1pt, thin, -latex, red] (-0.45,-0.65) -- node[auto] {} (-0.43,-0.43);						
		\end{tikzpicture}
		\begin{tikzpicture}
		[spy using outlines={rectangle,red,magnification=2,width=15mm, height =10mm, connect spies}]				
		\node {\includegraphics[width=0.18\textwidth]{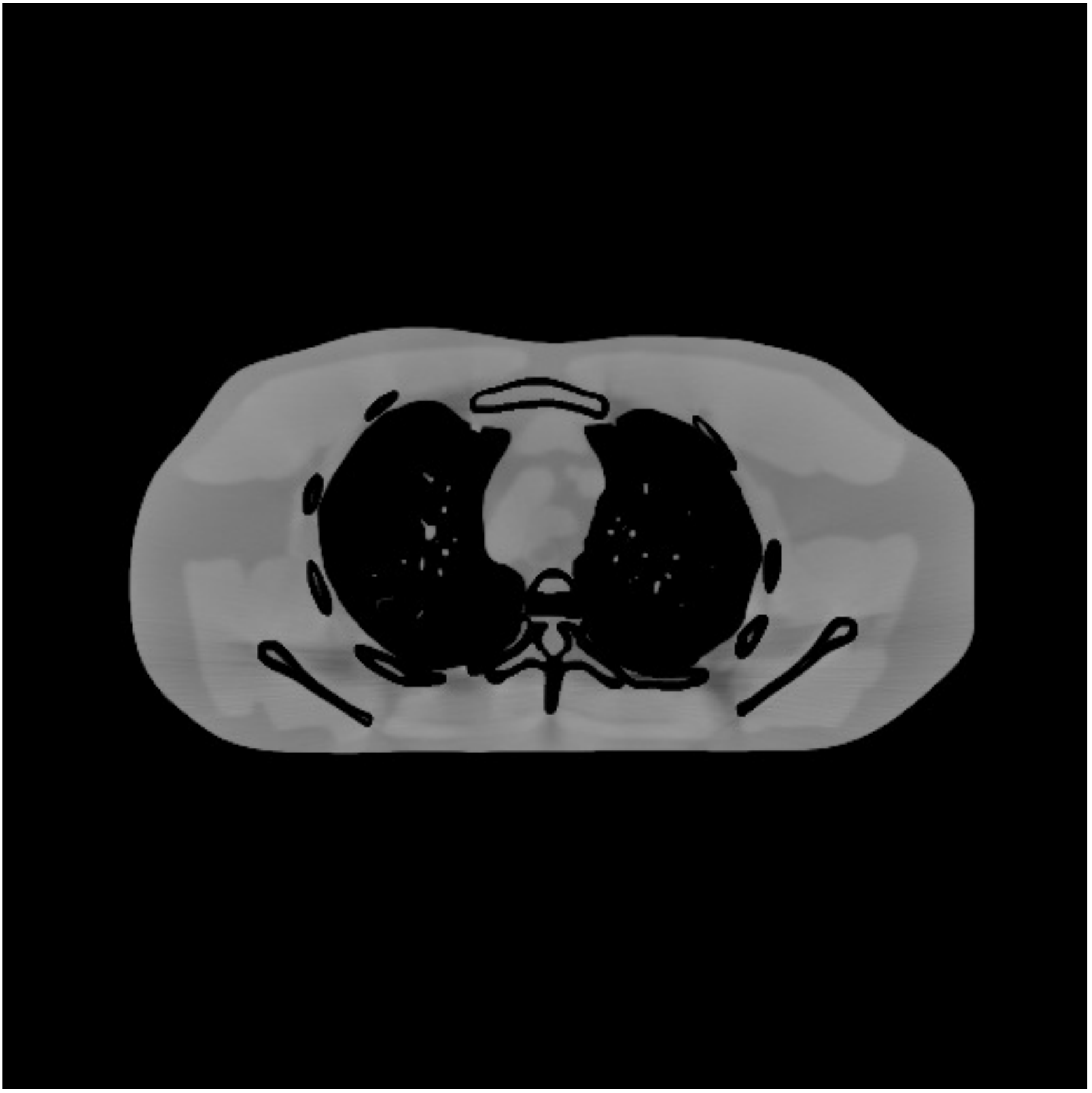}};
		\spy on (0.4,-0.5) in node [right] at (0.05,-1.15);
		\spy on (-0.5,-0.5) in node [right] at (-1.6,-1.15);
		\spy on (0.85,0.45) in node [right] at (0.05,1.15);	
		\draw[line width=1pt, thin, -latex, red] (1.1,0.6) -- node[auto] {} (0.67,0.56);
		\draw[line width=1pt, thin, -latex, red] (0.45,-0.65) -- node[auto] {} (0.3,-0.43);
		\draw[line width=1pt, thin, -latex, red] (-0.45,-0.65) -- node[auto] {} (-0.43,-0.43);							
		\end{tikzpicture}
		\begin{tikzpicture}
		[spy using outlines={rectangle,red,magnification=2,width=15mm, height =10mm, connect spies}]				
		\node {\includegraphics[width=0.18\textwidth]{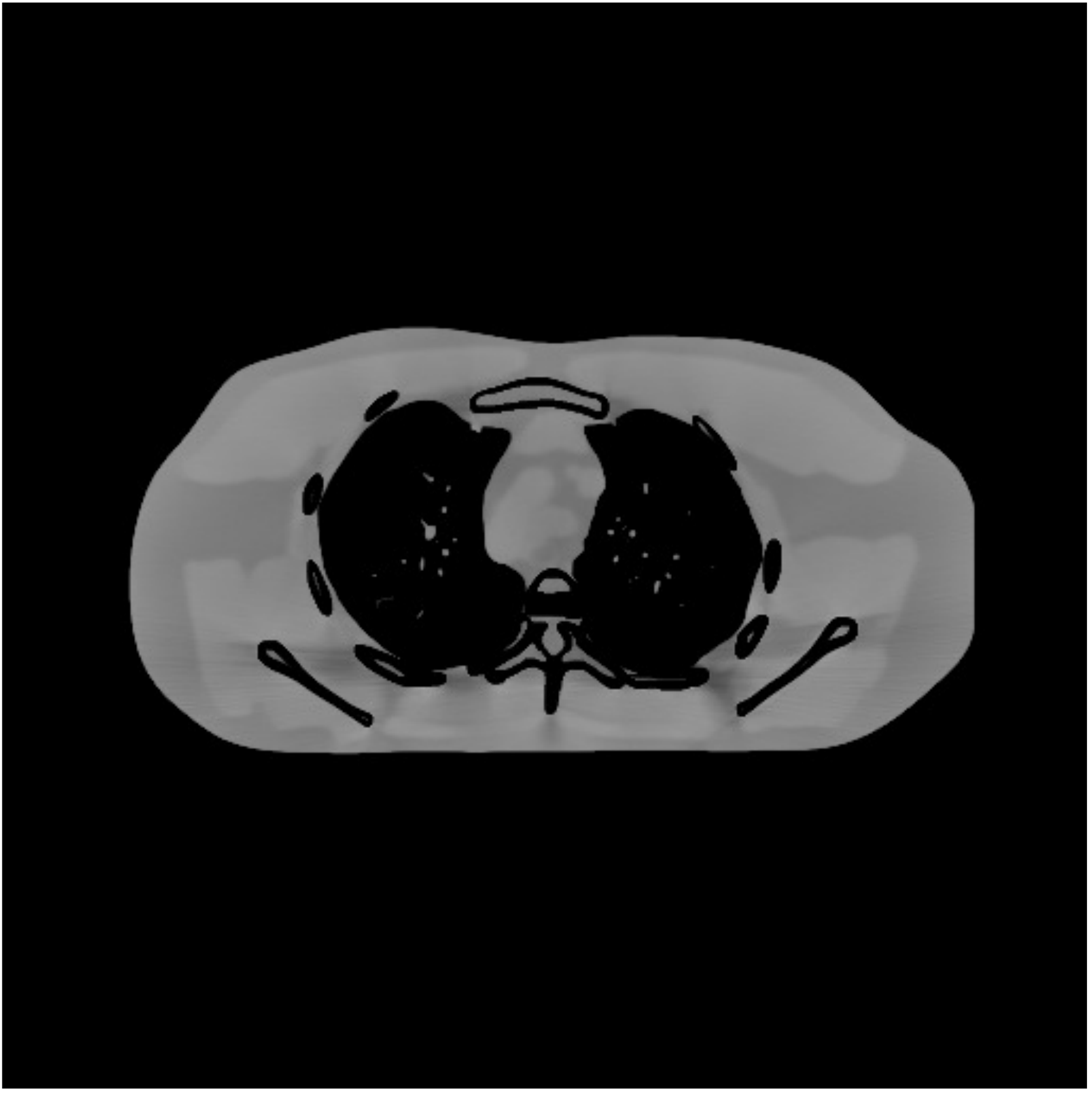}};
		\spy on (0.4,-0.5) in node [right] at (0.05,-1.15);
		\spy on (-0.5,-0.5) in node [right] at (-1.6,-1.15);
		\spy on (0.85,0.45) in node [right] at (0.05,1.15);	
		\draw[line width=1pt, thin, -latex, red] (1.1,0.6) -- node[auto] {} (0.67,0.56);
		\draw[line width=1pt, thin, -latex, red] (0.45,-0.65) -- node[auto] {} (0.3,-0.43);
		\draw[line width=1pt, thin, -latex, red] (-0.45,-0.65) -- node[auto] {} (-0.43,-0.43);							
		\end{tikzpicture}
		\begin{tikzpicture}
		[spy using outlines={rectangle,red,magnification=2,width=15mm, height =10mm, connect spies}]				
		\node {\includegraphics[width=0.18\textwidth]{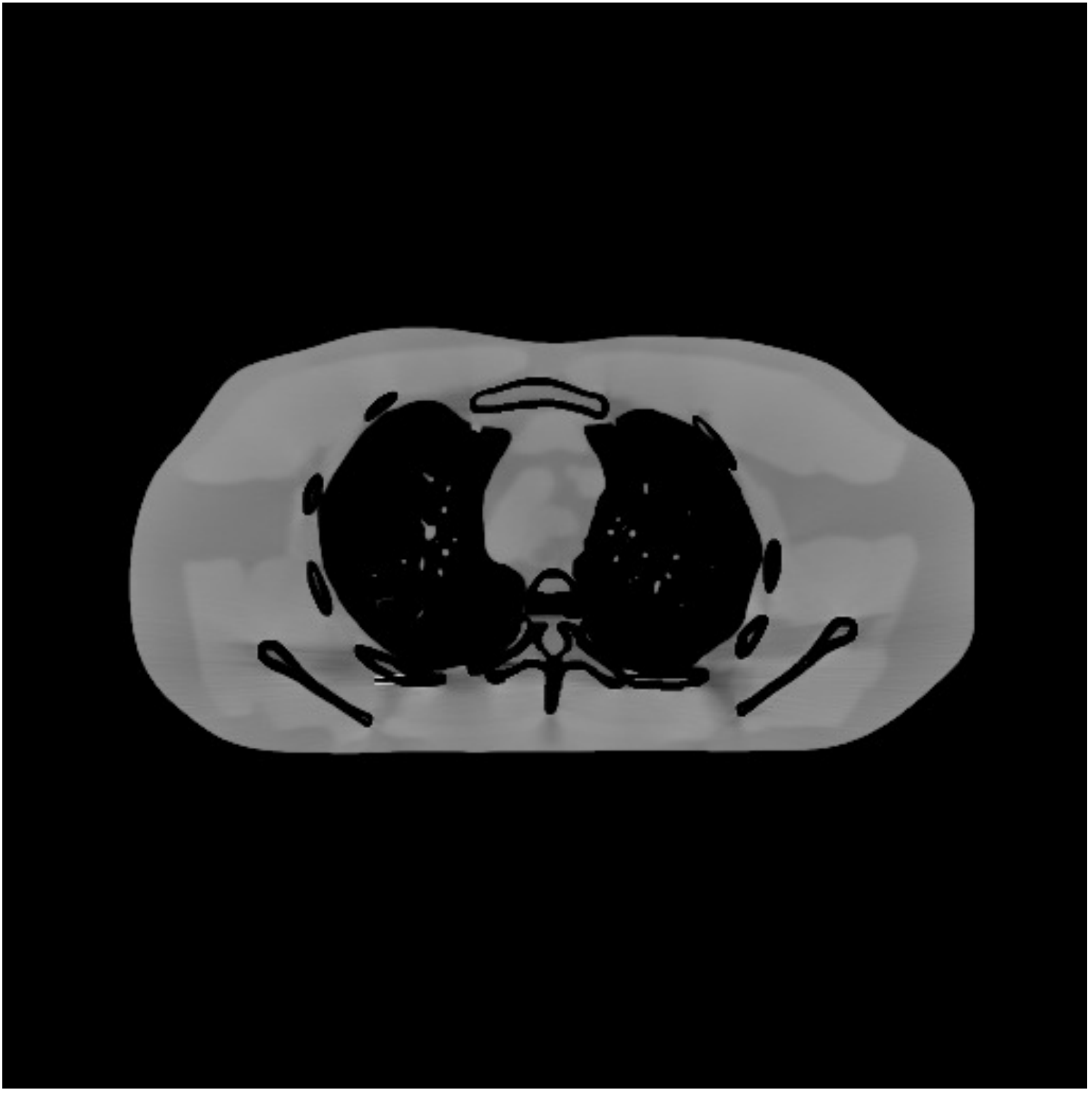}};
		\spy on (0.4,-0.5) in node [right] at (0.05,-1.15);
		\spy on (-0.5,-0.5) in node [right] at (-1.6,-1.15);
		\spy on (0.85,0.45) in node [right] at (0.05,1.15);	
		\draw[line width=1pt, thin, -latex, red] (1.1,0.6) -- node[auto] {} (0.67,0.56);
		\draw[line width=1pt, thin, -latex, red] (0.45,-0.65) -- node[auto] {} (0.3,-0.43);
		\draw[line width=1pt, thin, -latex, red] (-0.45,-0.65) -- node[auto] {} (-0.43,-0.43);							
		\end{tikzpicture}		\\
		\begin{tikzpicture}
		[spy using outlines={rectangle,red,magnification=2,width=15mm, height =10mm, connect spies}]				
		\node {\includegraphics[width=0.18\textwidth]{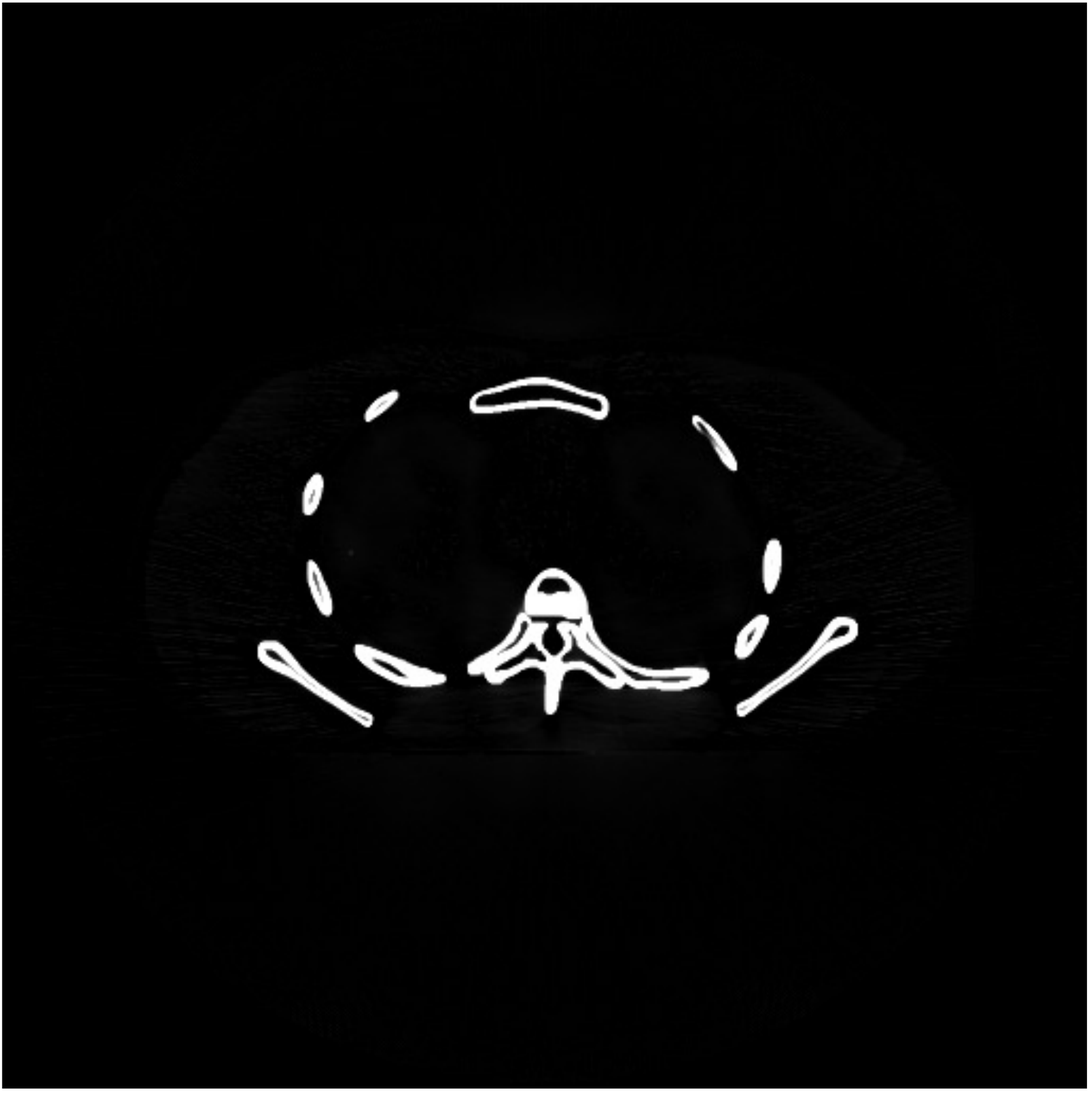}};
		\spy on (0.4,-0.5) in node [right] at (0.05,-1.15);	
		\spy on (-0.5,-0.5) in node [right] at (-1.6,-1.15);
		\draw[line width=1pt, thin, -latex, red] (0.45,-0.65) -- node[auto] {} (0.3,-0.43);					
		\end{tikzpicture}
		\begin{tikzpicture}
		[spy using outlines={rectangle,red,magnification=2,width=15mm, height =10mm, connect spies}]				
		\node {\includegraphics[width=0.18\textwidth]{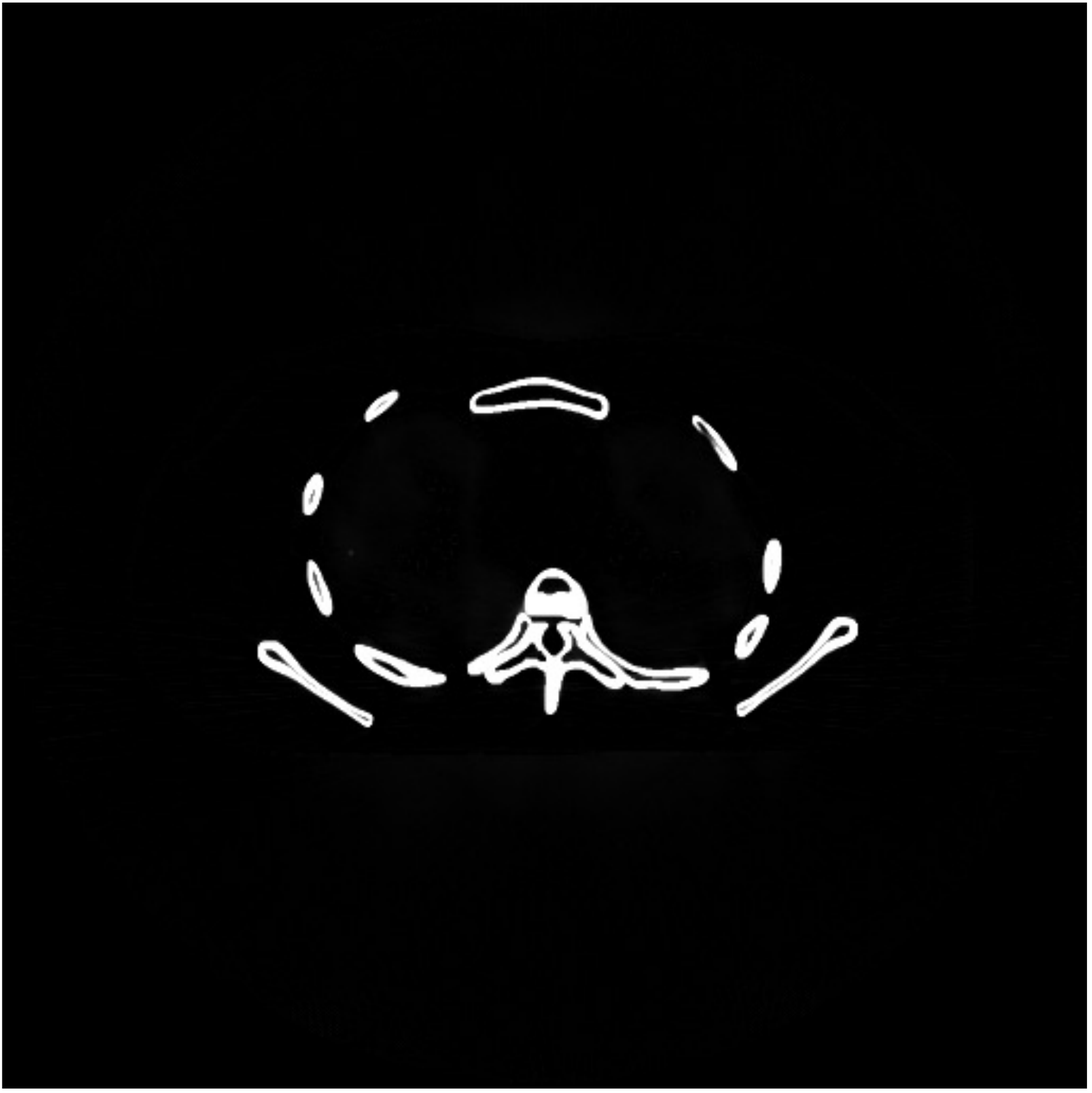}};
		\spy on (0.4,-0.5) in node [right] at (0.05,-1.15);		
		\spy on (-0.5,-0.5) in node [right] at (-1.6,-1.15);
		\draw[line width=1pt, thin, -latex, red] (0.45,-0.65) -- node[auto] {} (0.3,-0.43);				
		\end{tikzpicture}
		\begin{tikzpicture}
		[spy using outlines={rectangle,red,magnification=2,width=15mm, height =10mm, connect spies}]				
		\node {\includegraphics[width=0.18\textwidth]{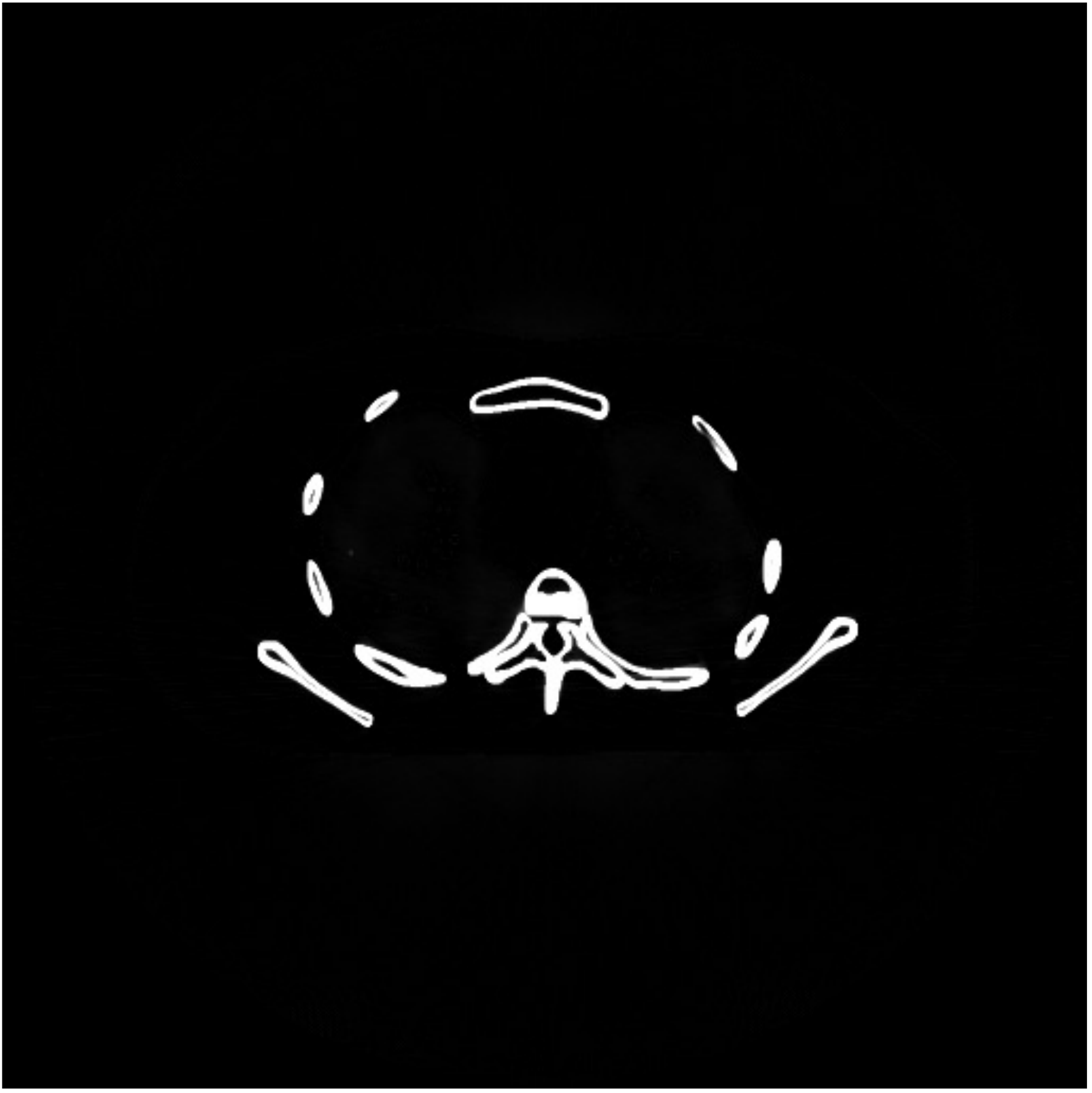}};
		\spy on (0.4,-0.5) in node [right] at (0.05,-1.15);	
		\spy on (-0.5,-0.5) in node [right] at (-1.6,-1.15);
		\draw[line width=1pt, thin, -latex, red] (0.45,-0.65) -- node[auto] {} (0.3,-0.43);						
		\end{tikzpicture}
		\begin{tikzpicture}
		[spy using outlines={rectangle,red,magnification=2,width=15mm, height =10mm, connect spies}]				
		\node {\includegraphics[width=0.18\textwidth]{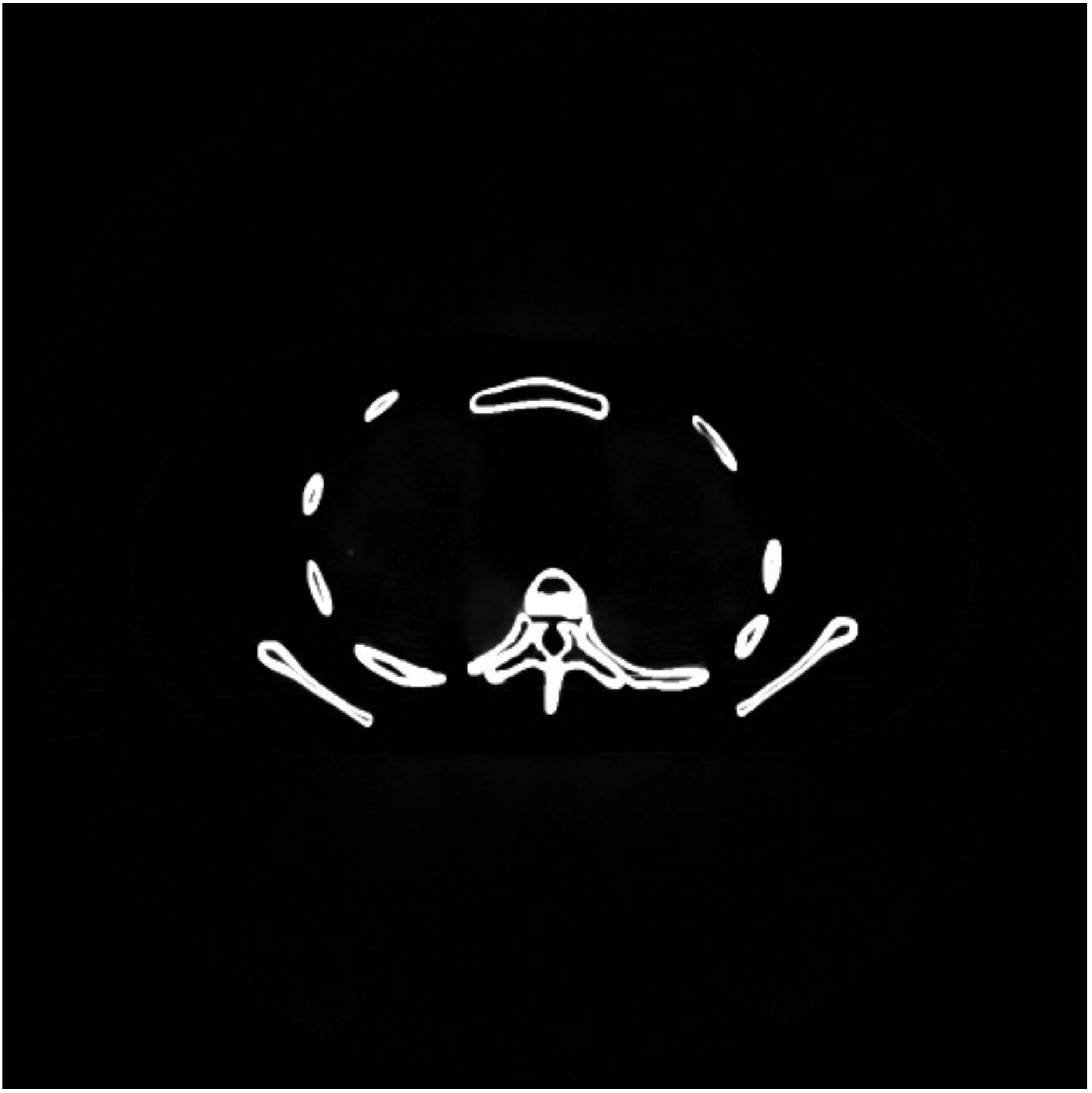}};
		\spy on (0.4,-0.5) in node [right] at (0.05,-1.15);	
		\spy on (-0.5,-0.5) in node [right] at (-1.6,-1.15);
		\draw[line width=1pt, thin, -latex, red] (0.45,-0.65) -- node[auto] {} (0.3,-0.43);						
		\end{tikzpicture}
		\begin{tikzpicture}
		[spy using outlines={rectangle,red,magnification=2,width=15mm, height =10mm, connect spies}]				
		\node {\includegraphics[width=0.18\textwidth]{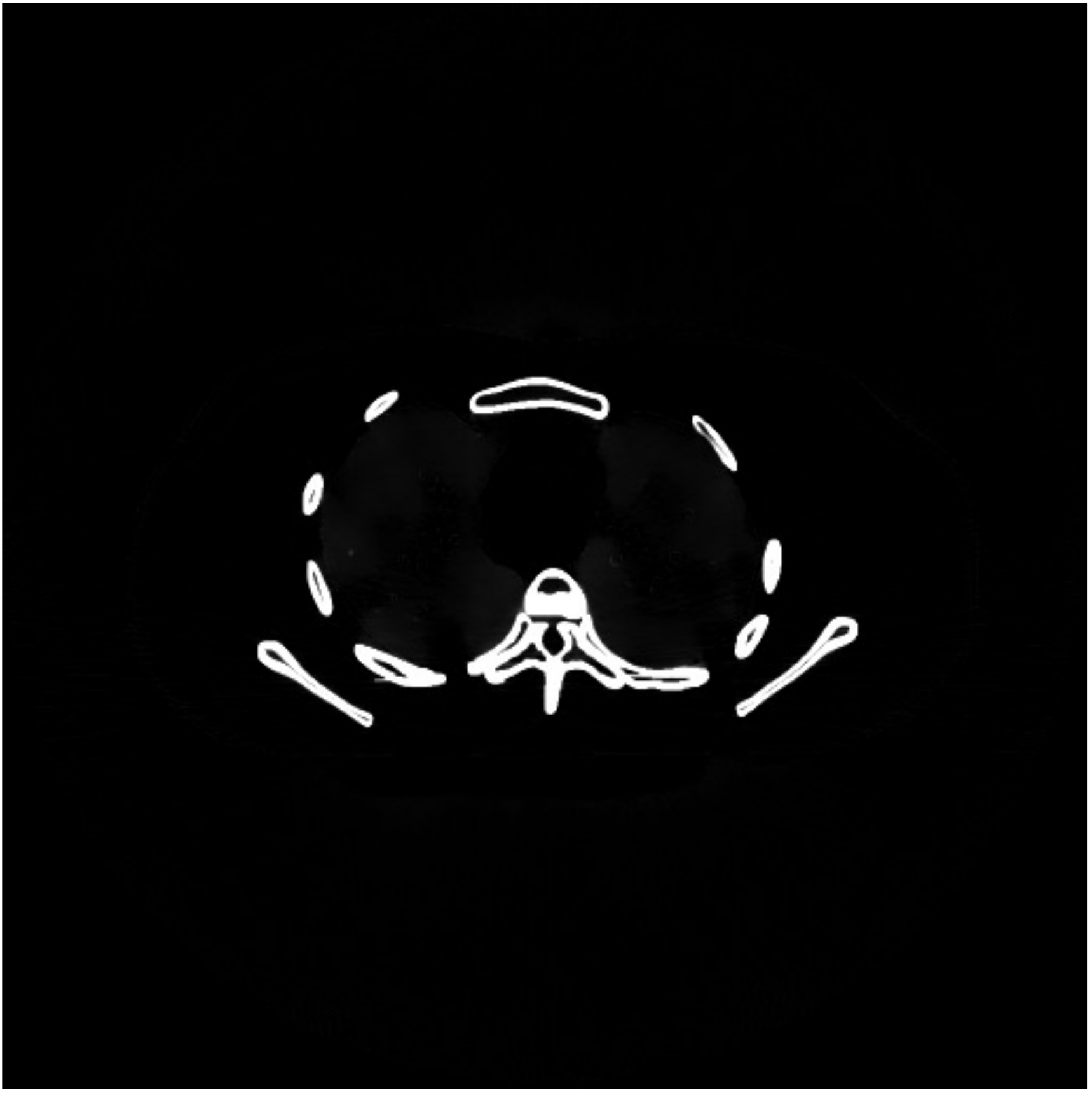}};
		\spy on (0.4,-0.5) in node [right] at (0.05,-1.15);	
		\spy on (-0.5,-0.5) in node [right] at (-1.6,-1.15);
		\draw[line width=1pt, thin, -latex, red] (0.45,-0.65) -- node[auto] {} (0.3,-0.43);						
		\end{tikzpicture}
		\caption{Material image decomposition results for DECT-MULTRA with different weights for the common-material and cross-material models. From left to right: $\{\beta_1\,,\beta_2\}$ are set as $\{1,\,99\}$, $\{25,\,75\}$, $\{50,\,50\}$, $\{75,\,25\}$, and $\{90,\,10\}$. The top and bottom rows show the water and bone images with display windows [0.7  1.3]\,g/cm$^3$ and [0 0.8]\,g/cm$^3$, respectively. }
		\label{fig:weight_comp}
	\end{figure*}
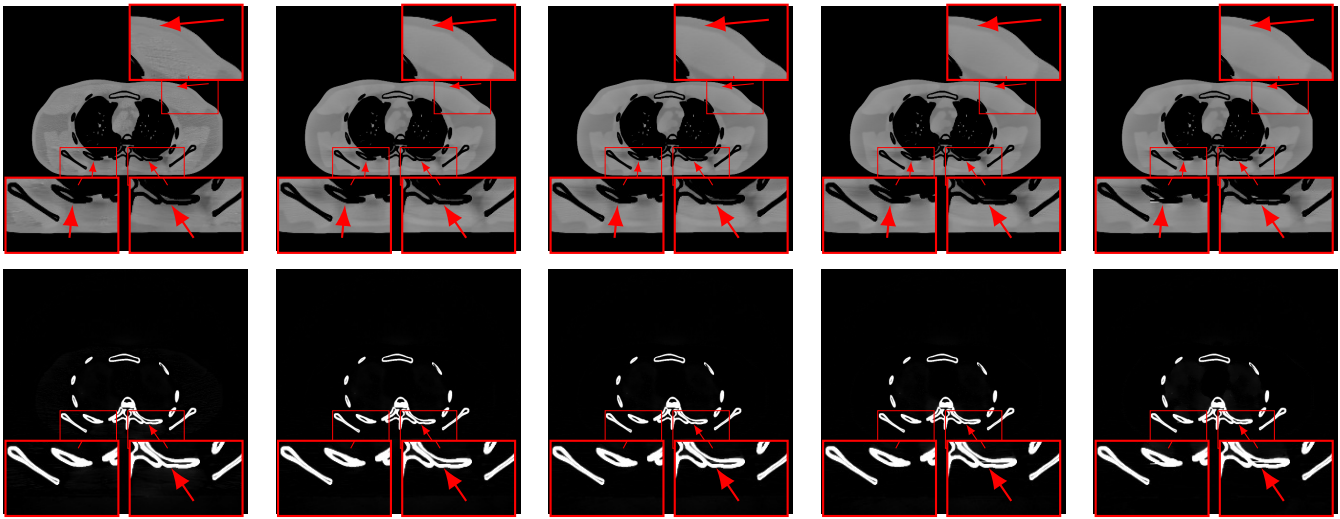

\subsection{Clinical Data Study}
\label{subsec:clinical}
\subsubsection{Framework and Data}
\begin{figure}[htb]
	\centering
	\begin{tikzpicture}
		[spy using outlines={rectangle,red,magnification=2,width=7mm, height =12mm, connect spies}]	
		\spy on (0.03,0.75) in node [right] at (0.6,-1.);
		\spy [width=16mm, height =6mm] on (-0.29,-0.1) in node [right] at (-2.0,-1.3);			
		\node {\includegraphics[scale=0.28]{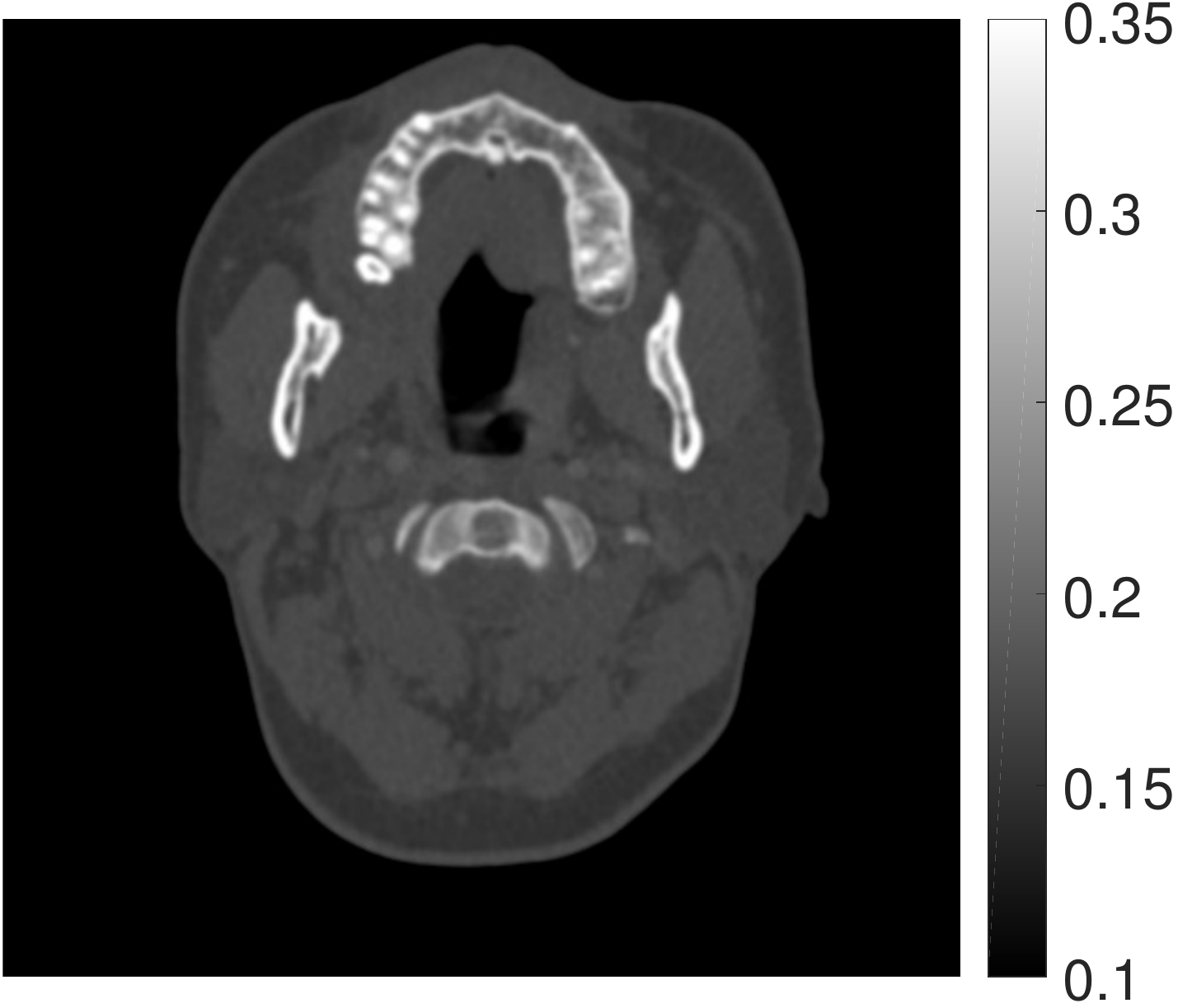}	};				
		\draw[line width=1pt, thin, -latex, red] (0.2,0.57) -- node[auto] {} (0.005,0.55);
	\end{tikzpicture}
	\begin{tikzpicture}
	[spy using outlines={rectangle,red,magnification=2,width=7mm, height =12mm, connect spies}]	
		\spy on (0.03,0.75) in node [right] at (0.6,-1.);
		\spy [width=16mm, height =6mm] on (-0.29,-0.1) in node [right] at (-2.0,-1.3);			
	\node {\includegraphics[scale=0.28]{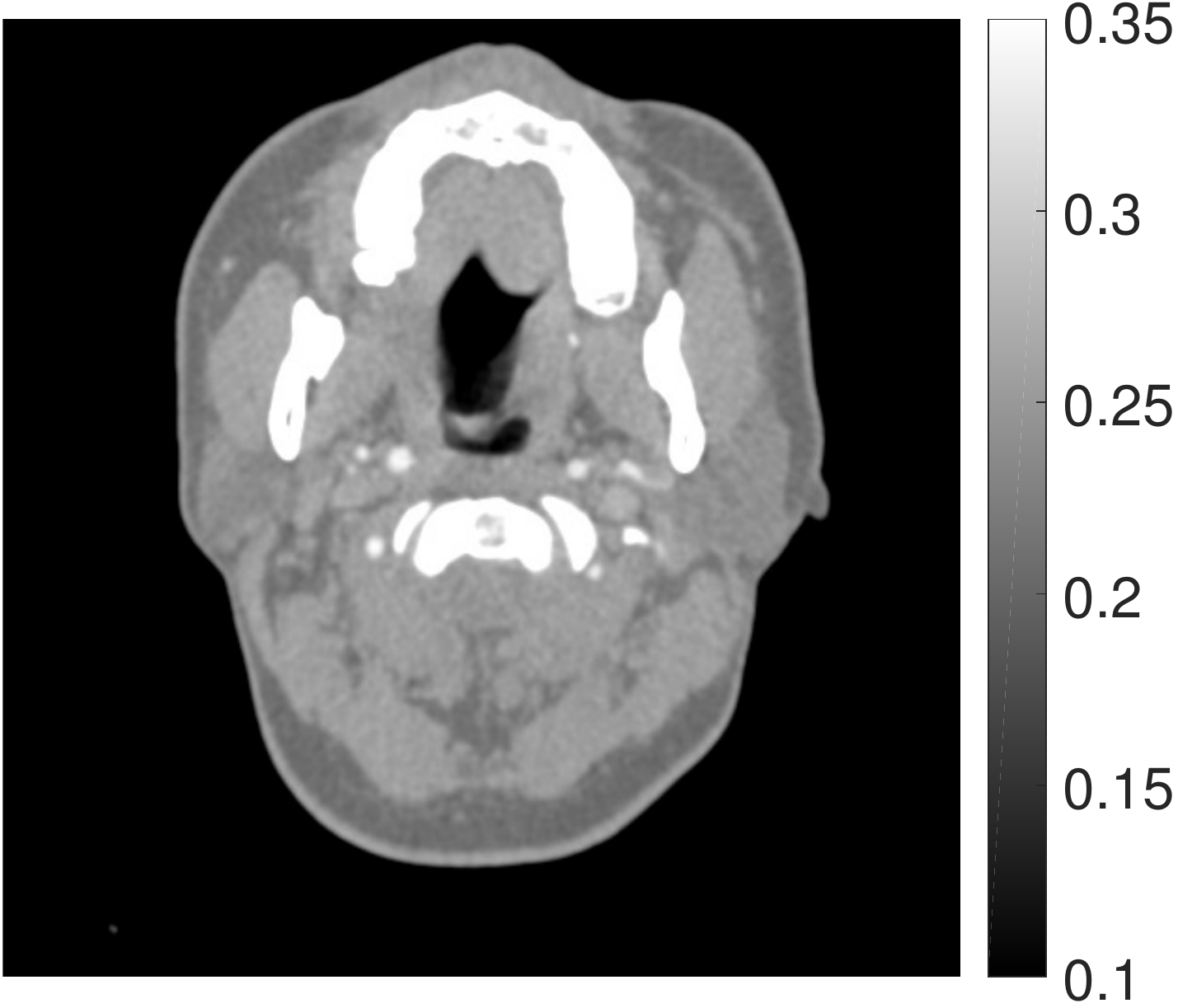}	};				
		\draw[line width=1pt, thin, -latex, red] (0.2,0.57) -- node[auto] {} (0.005,0.55);
	\end{tikzpicture}
	\caption{Head CT images at 140~kVp (left) and 80~kVp (right). The display window is [0.1 0.35]\,cm$^{-1}$. }
	\label{Fig:head_atten}
	\vspace{-0.1in}
\end{figure}

We evaluated the proposed methods using clinical DECT head data.  
The patient data was collected  by a Siemens SOMATOM Definition flash CT scanner using dual-energy CT imaging protocols with dual-source at 80kVp and 140kVp for dual-energy data acquisition. Table \ref{Tab: protocols} lists the protocols of the patient data acquisition. 
Fig.~\ref{Fig:head_atten} shows the head CT images at 140~kVp and 80~kVp. The filtered back projection method was used to reconstruct these attenuation maps. 

\begin{table}[htb]
	\centering
	\caption{Data acquisition parameters applied in head data acquisition.}
	\begin{tabular}{p{3.5cm}<{\centering} p{2cm}<{\centering} p{2cm}<{\centering} }
		\hline\hline
		Scanner &  High-Energy image  & Low-Energy image \\  
		\hline
		Peak voltage (kVp)  & 140 & 80\\
		\hline
		X-ray Tube Current (mA) & 364 &	 648 \\
		\hline
		Exposure Time (s) & \multicolumn{2}{c}{0.285}  \\				   
		\hline
		Current-exposure Time Product (mAs) & \tabincell{c}{103.7} & \tabincell{c}{184.7}\\
		\hline
		Helical Pitch & \multicolumn{2}{c}{0.7} \\
		\hline
		Gantry Rotation Speed (circle/second) & \multicolumn{2}{c}{\tabincell{c}{0.28}} \\
		\hline  \hline
	\end{tabular} 
	\label{Tab: protocols}	
	\vspace{-0.1in}	
\end{table}

\subsubsection{Decomposition Results}
 We initialized the DECT-EP method with the results obtained by Direct Matrix Inversion.  
 We ran the DECT-EP algorithm for 500 iterations and used its results to initialize DECT-TDL and DECT-MULTRA. 
 For DECT-EP, we chose the parameters $\{\beta_1,\,\delta_1,\,\beta_2,\,\delta_2\}$ as $\{2^{10.5},\,0.008,\,2^{11},\,0.015\}$.  
For DECT-TDL, we set the maximum sparsity level during sparse coding to $60$ along with an error tolerance of $0.2$, and the regularization parameter was 40. 
We set the DECT-MULTRA parameters $\{\beta_1,\,\beta_2,\,\gamma_2\}$ as $\{180,\,180,\,0.018\}$. Moreover, for $r=1$, we used different sparsity regularization parameters $0.006$ and $0.03$ for the water and bone components of $\z_j$ in (\ref{Eq:cost_func_reg}), which provided better image quality.

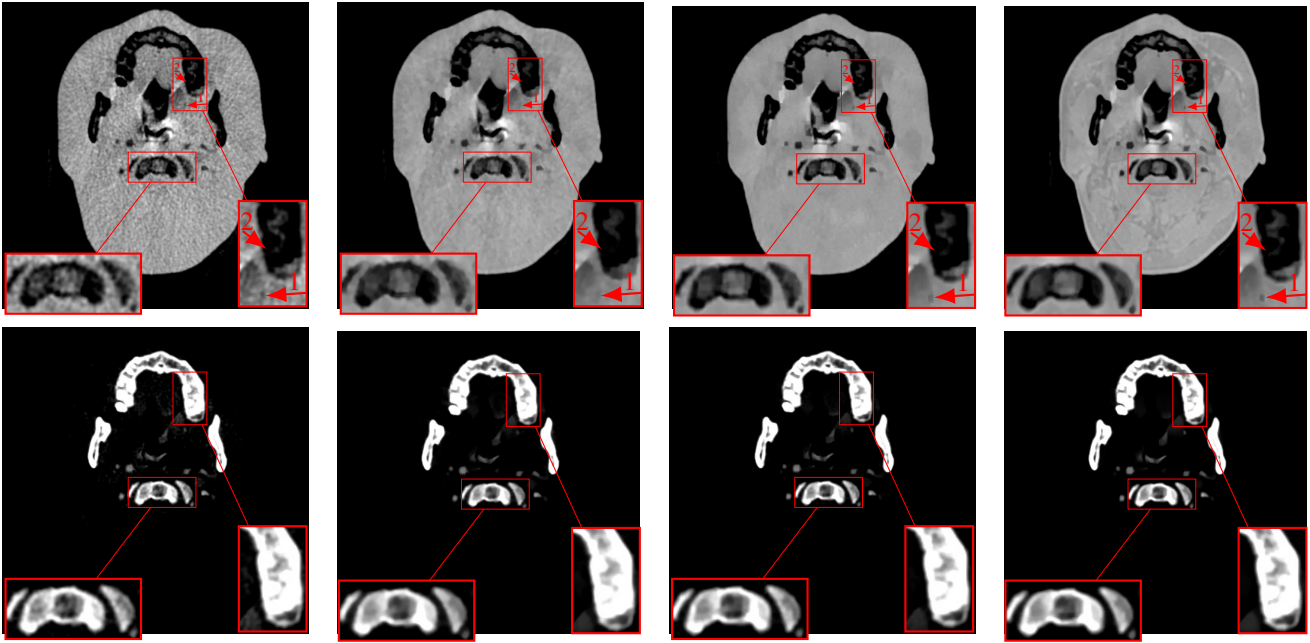
\begin{figure*}[htb]
	\centering
	\begin{tikzpicture}
	[spy using outlines={rectangle,red,magnification=2,width=9mm, height =14mm, connect spies}]
	\spy on (0.45,0.95) in node [right] at (1.1,-1.3);
	\spy [width=18mm, height =8mm] on (0.08,-0.15) in node [right] at (-2.0,-1.7);			
	\node {\includegraphics[scale=0.3]{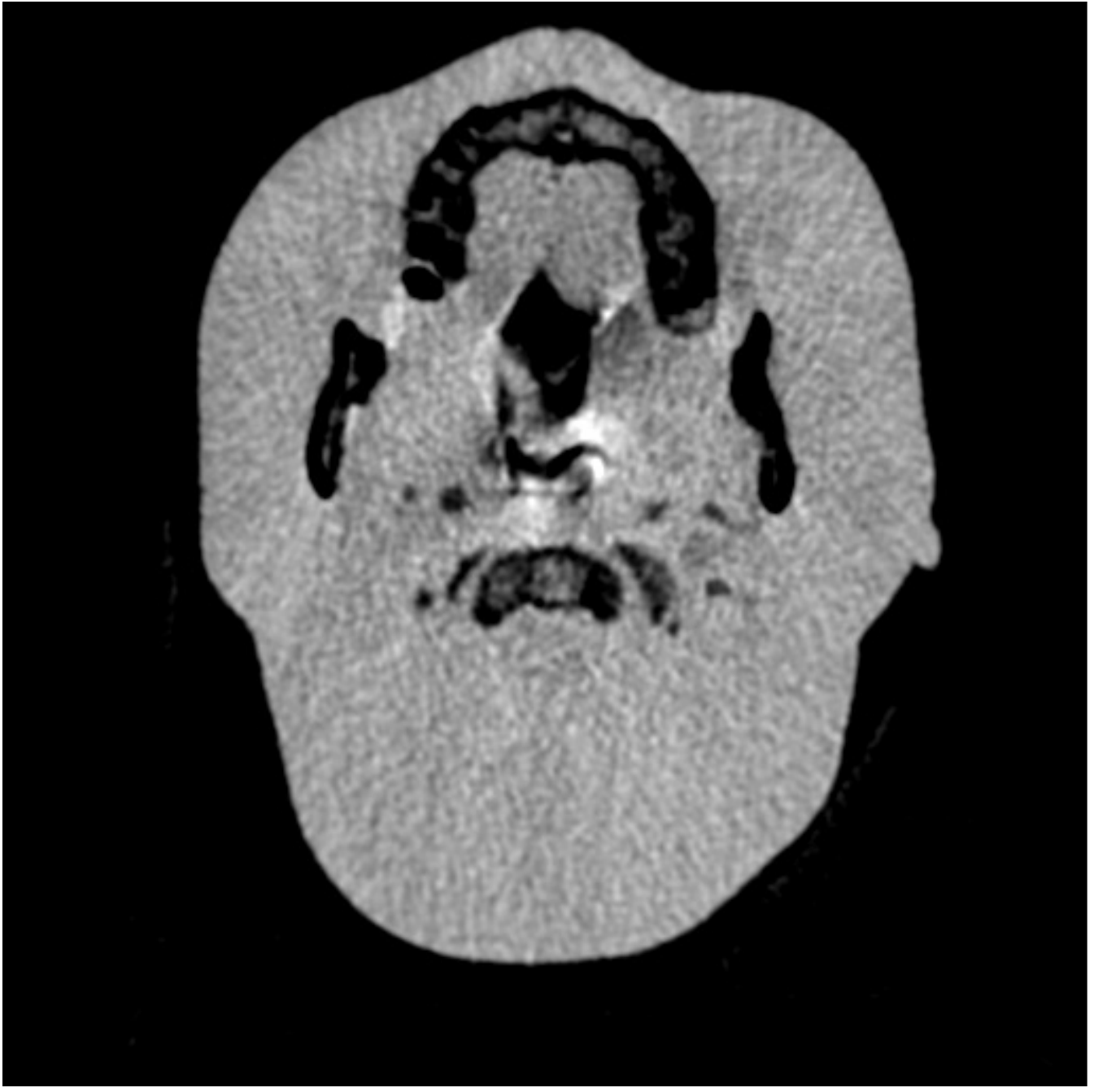}	};	
	\draw[line width=1pt, thin, -latex, red] (0.69,0.69) -- node[xshift=0.05cm,yshift=0.09cm] {\tiny{1}} (0.4,0.67);
	\draw[line width=1pt, thin, -latex, red] (0.25,1.1) -- node[xshift=-0.05cm,yshift=0.13cm] {\tiny{2}} (0.41,0.98);				
	\end{tikzpicture}   
	\begin{tikzpicture}
	[spy using outlines={rectangle,red,magnification=2,width=9mm, height =14mm, connect spies}]				
	\node {\includegraphics[scale=0.3]{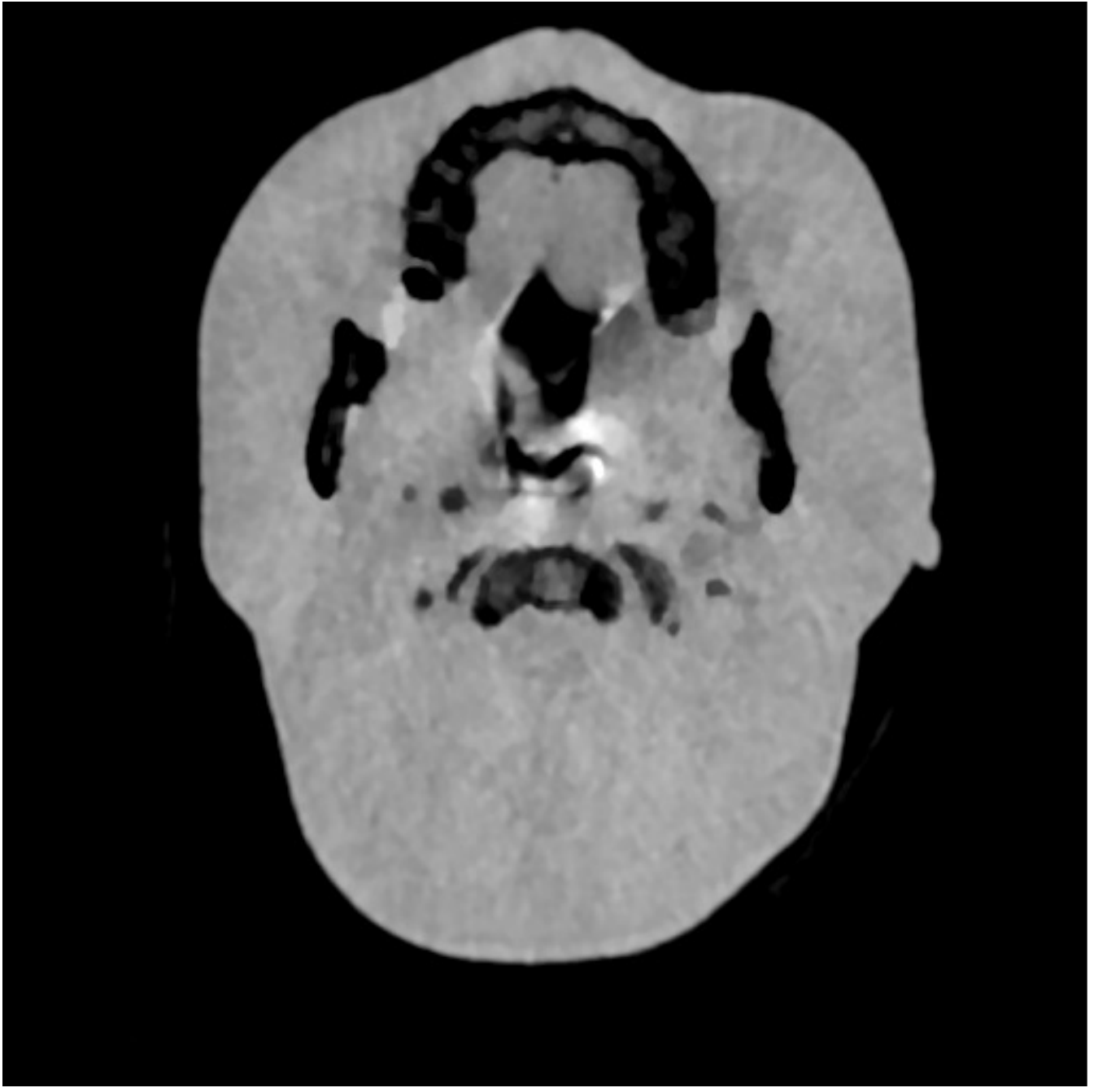}	};				
	\spy on (0.45,0.95) in node [right] at (1.1,-1.3);
	\spy [width=18mm, height =8mm] on (0.08,-0.15) in node [right] at (-2.0,-1.7);
	\draw[line width=1pt, thin, -latex, red] (0.69,0.69) -- node[xshift=0.05cm,yshift=0.09cm] {\tiny{1}} (0.4,0.67);
	\draw[line width=1pt, thin, -latex, red] (0.25,1.1) -- node[xshift=-0.05cm,yshift=0.13cm] {\tiny{2}} (0.41,0.98);
	\end{tikzpicture}   
	\begin{tikzpicture}
	[spy using outlines={rectangle,red,magnification=2,width=9mm, height =14mm, connect spies}]				
	\node {\includegraphics[scale=0.3]{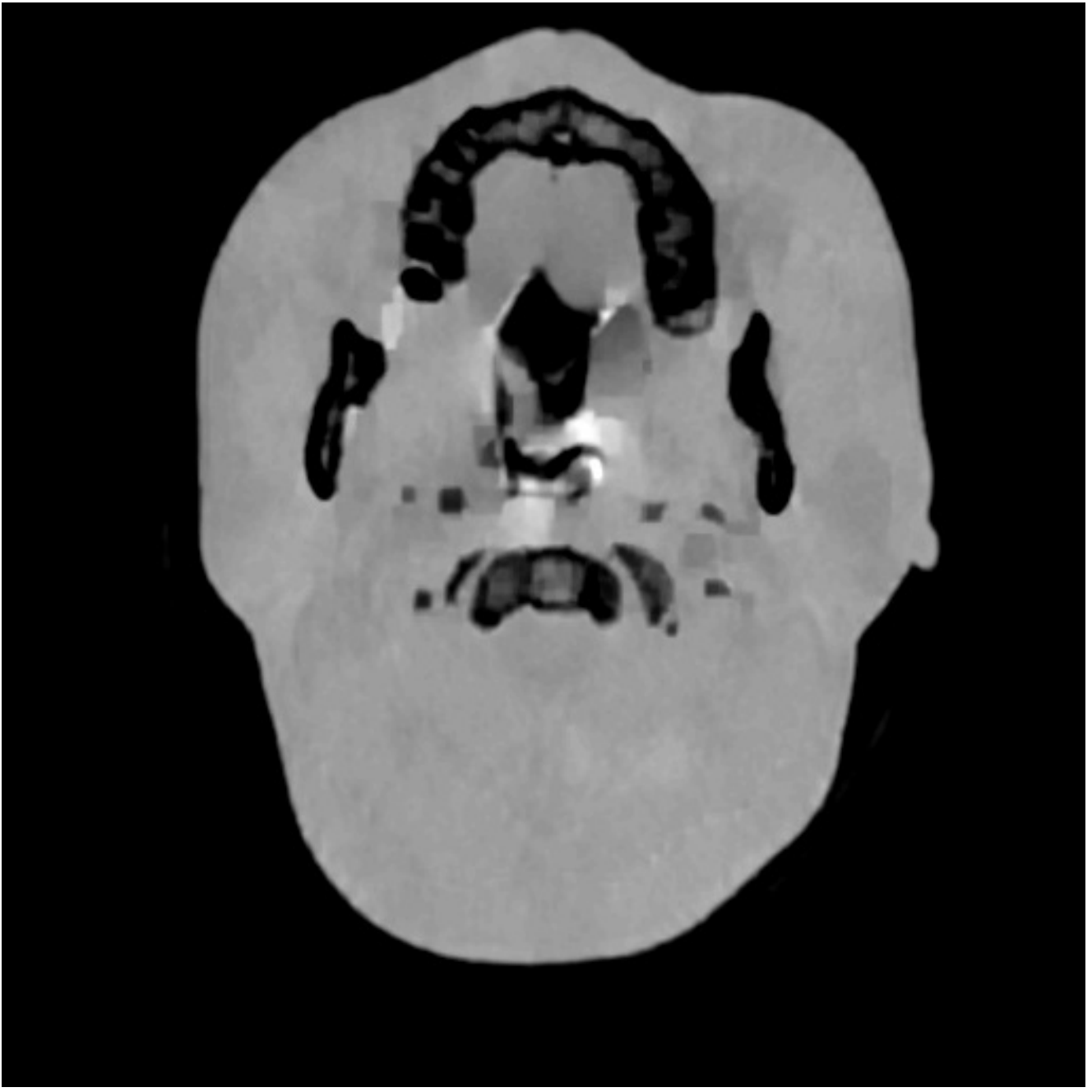}	};				
	\spy on (0.45,0.95) in node [right] at (1.1,-1.3);
	\spy [width=18mm, height =8mm] on (0.08,-0.15) in node [right] at (-2.0,-1.7);
	\draw[line width=1pt, thin, -latex, red] (0.69,0.69) -- node[xshift=0.05cm,yshift=0.09cm] {\tiny{1}} (0.4,0.67);
	\draw[line width=1pt, thin, -latex, red] (0.25,1.1) -- node[xshift=-0.05cm,yshift=0.13cm] {\tiny{2}} (0.41,0.98);
	\end{tikzpicture} 
	\begin{tikzpicture}
	[spy using outlines={rectangle,red,magnification=2,width=9mm, height =14mm, connect spies}]				
	\node {\includegraphics[scale=0.3]{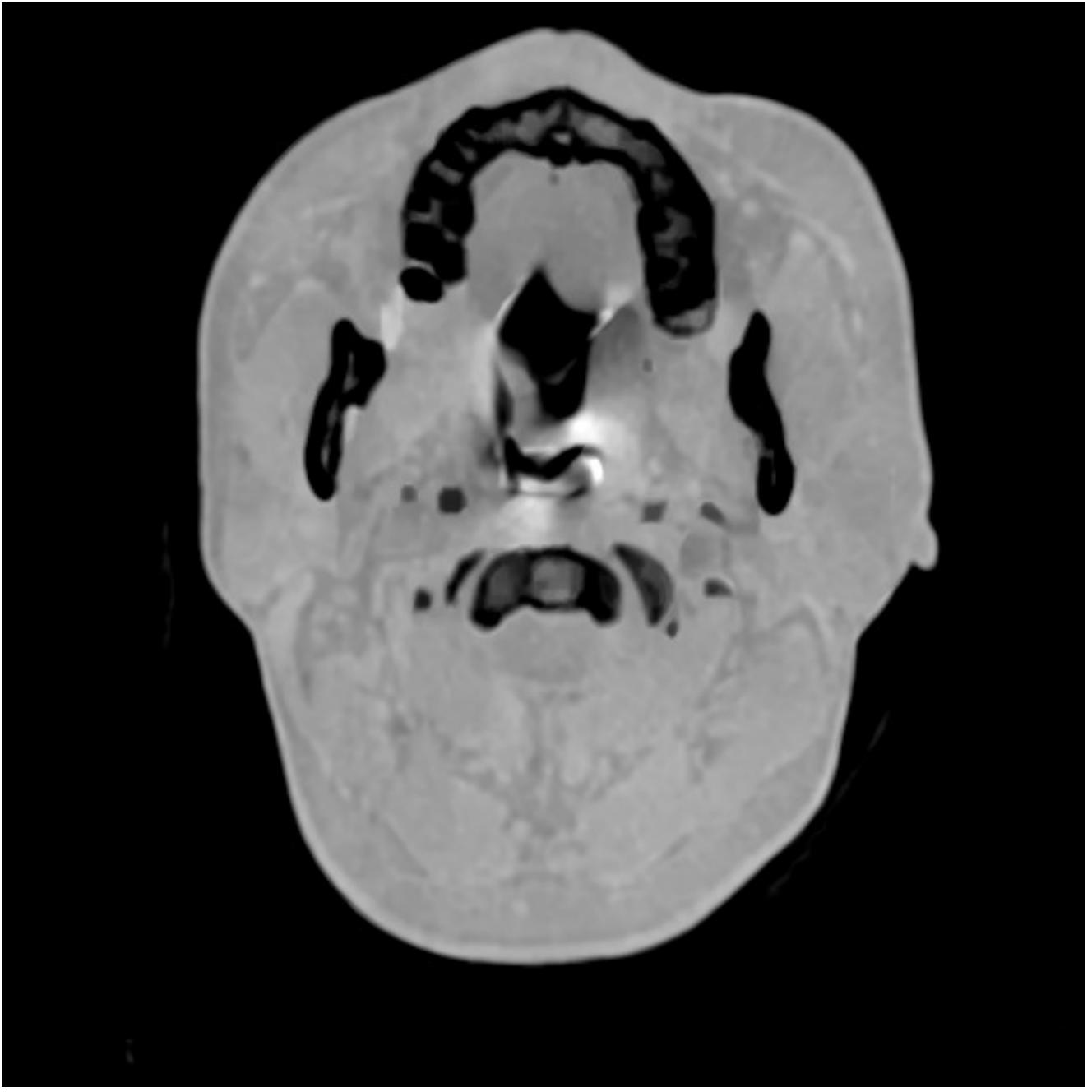}	};				
	\spy on (0.45,0.95) in node [right] at (1.1,-1.3);
	\spy [width=18mm, height =8mm] on (0.08,-0.15) in node [right] at (-2.0,-1.7);
	\draw[line width=1pt, thin, -latex, red] (0.69,0.69) -- node[xshift=0.05cm,yshift=0.09cm] {\tiny{1}} (0.4,0.67);
	\draw[line width=1pt, thin, -latex, red] (0.25,1.1) -- node[xshift=-0.05cm,yshift=0.13cm] {\tiny{2}} (0.41,0.98);	
	\end{tikzpicture} \\
	
	\begin{tikzpicture}
	[spy using outlines={rectangle,red,magnification=2,width=9mm, height =14mm, connect spies}]				
	\node {\includegraphics[scale=0.3]{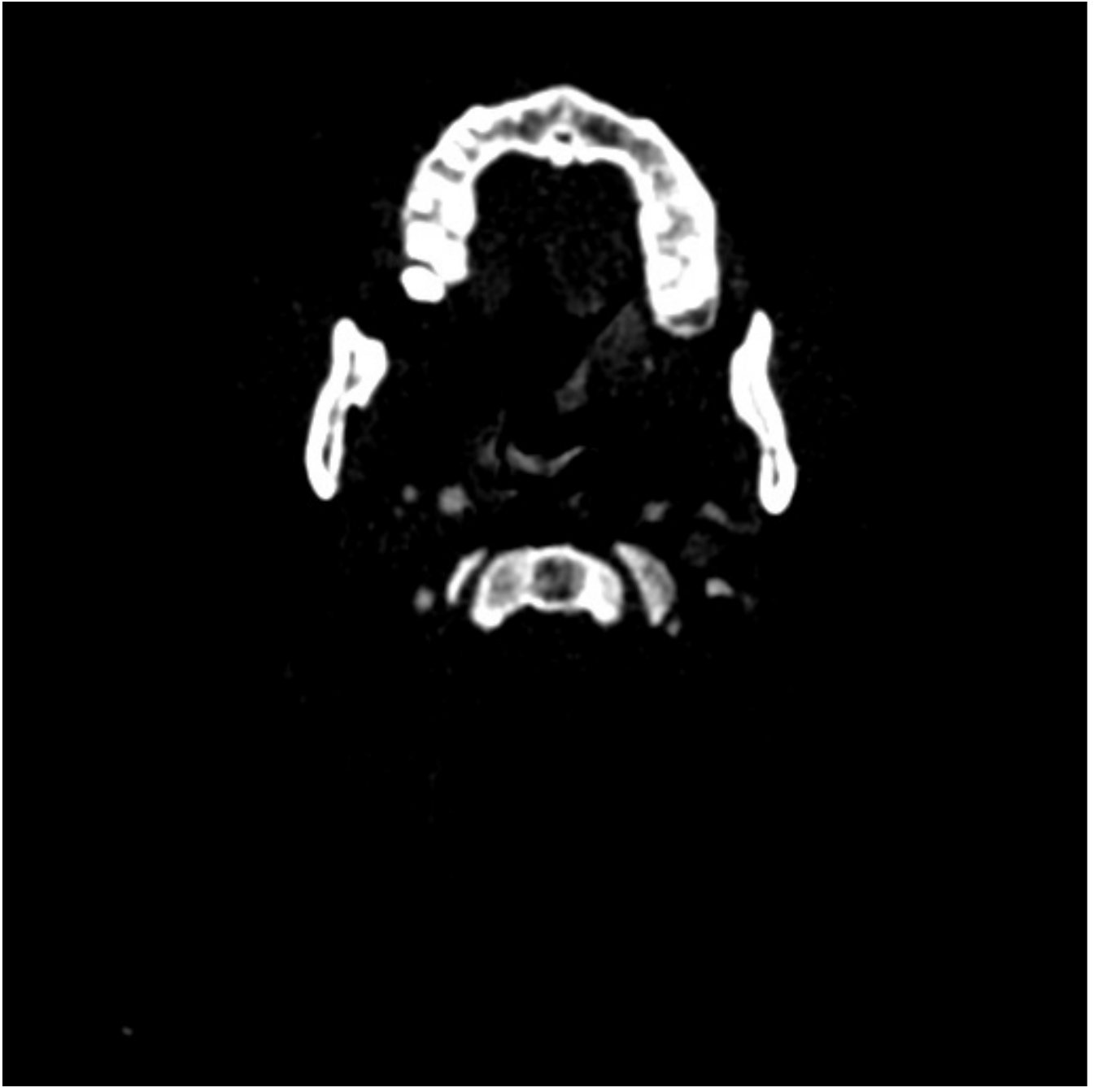}	};				
	\spy on (0.45,1.1) in node [right] at (1.1,-1.3);
	\spy [width=18mm, height =8mm] on (0.08,-0.15) in node [right] at (-2.0,-1.7);
	\end{tikzpicture}      	
	\begin{tikzpicture}
	[spy using outlines={rectangle,red,magnification=2,width=9mm, height =14mm, connect spies}]				
	\node {\includegraphics[scale=0.3]{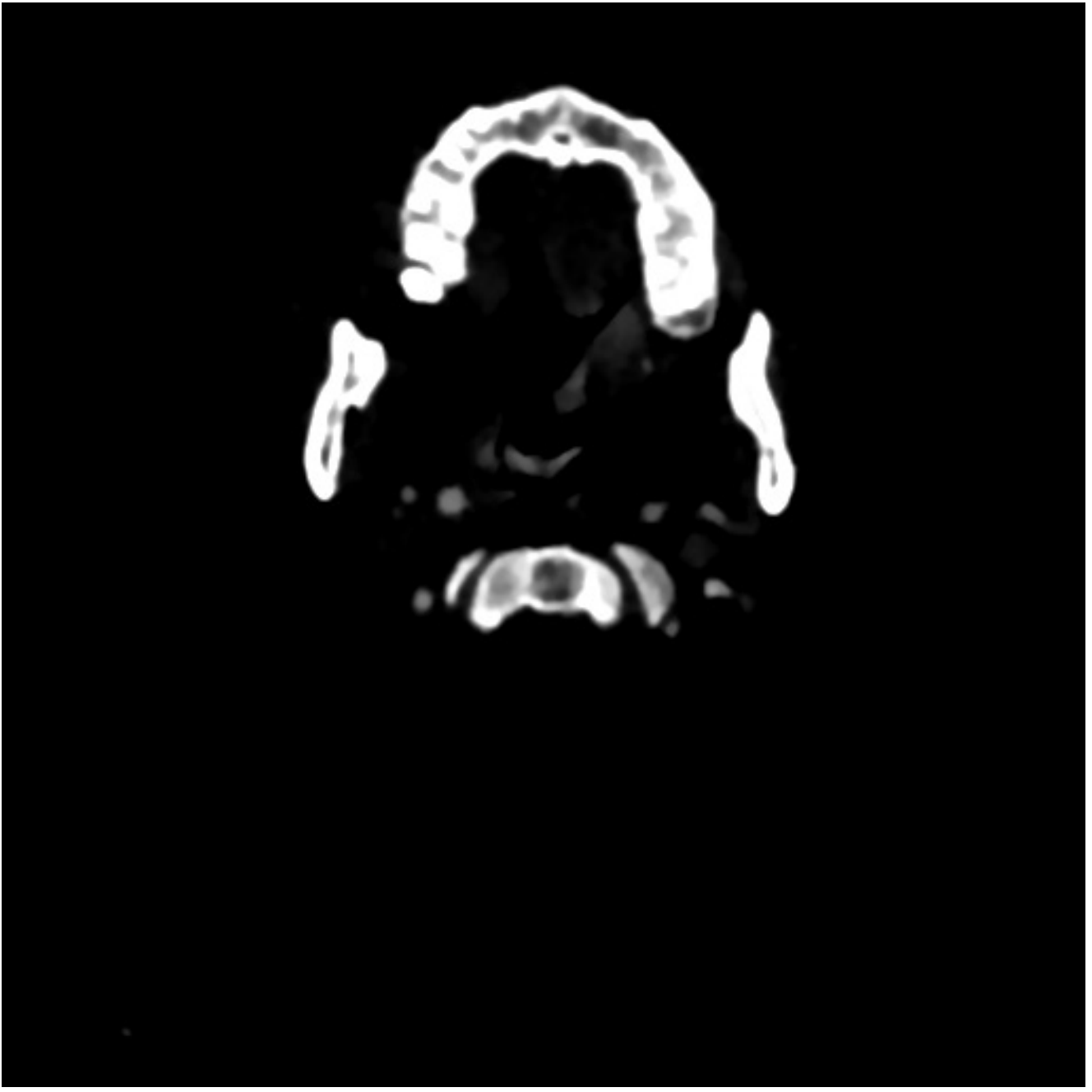}	};				
	\spy on (0.45,1.1) in node [right] at (1.1,-1.3);
	\spy [width=18mm, height =8mm] on (0.08,-0.15) in node [right] at (-2.0,-1.7);
	\end{tikzpicture}      	
	\begin{tikzpicture}
	[spy using outlines={rectangle,red,magnification=2,width=9mm, height =14mm, connect spies}]				
	\node {\includegraphics[scale=0.3]{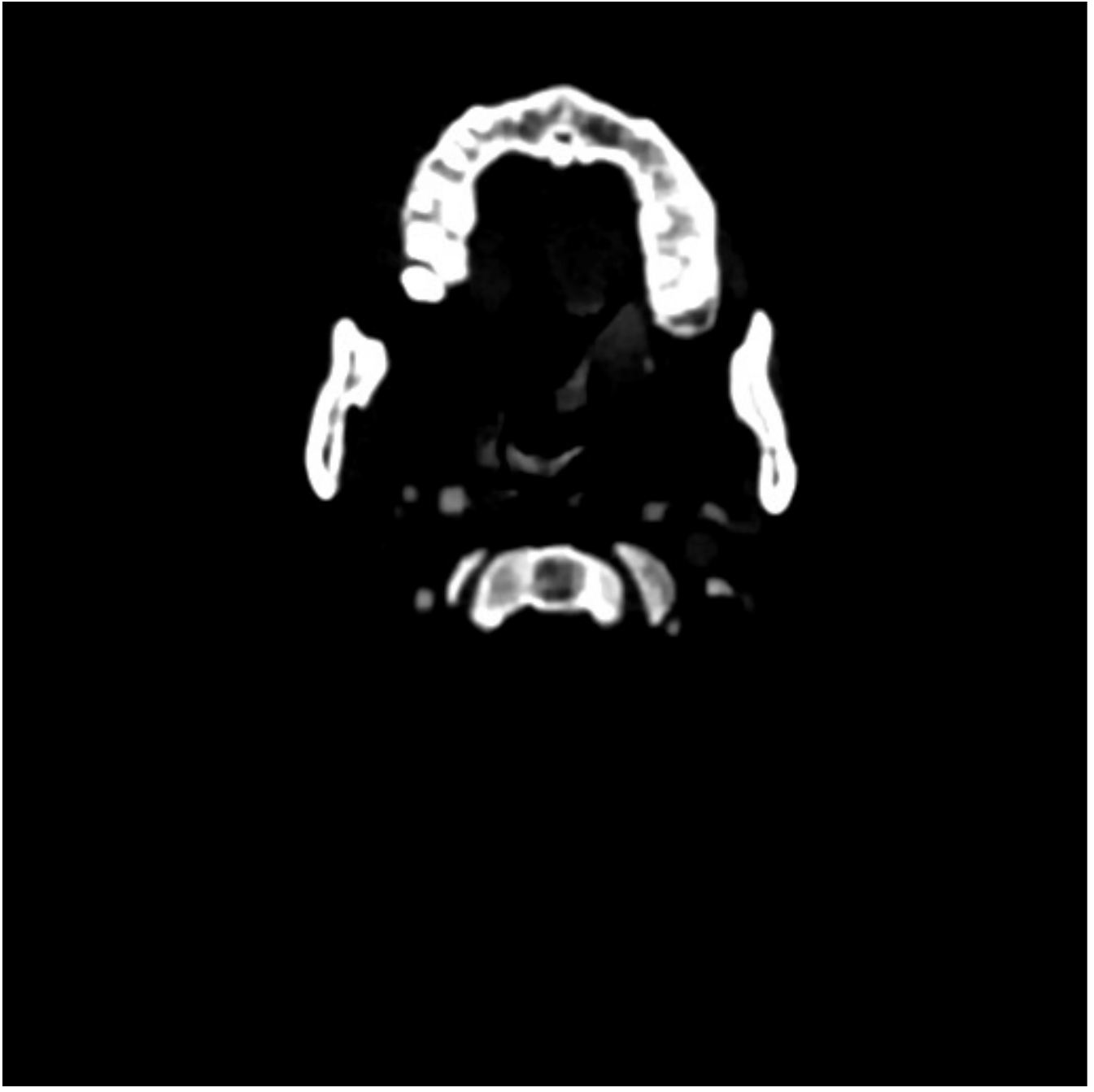}	};				
	\spy on (0.45,1.1) in node [right] at (1.1,-1.3);
	\spy [width=18mm, height =8mm] on (0.08,-0.15) in node [right] at (-2.0,-1.7);
	\end{tikzpicture}     	
	\begin{tikzpicture}
	[spy using outlines={rectangle,red,magnification=2,width=9mm, height =14mm, connect spies}]				
	\node {\includegraphics[scale=0.3]{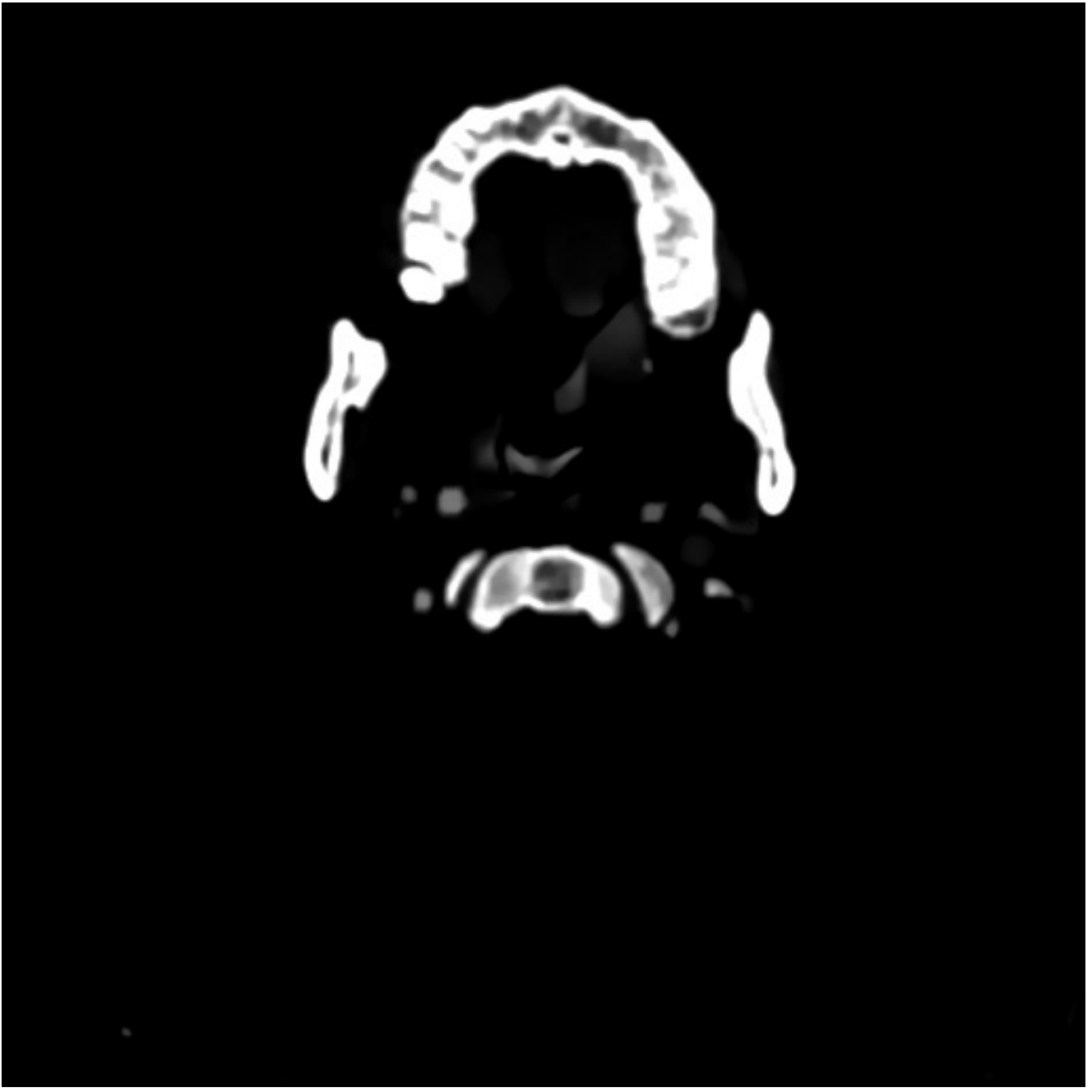}	};				
	\spy on (0.45,1.1) in node [right] at (1.1,-1.3);	
	\spy [width=18mm, height =8mm] on (0.08,-0.15) in node [right] at (-2.0,-1.7);
	\end{tikzpicture}      		
	\caption{ First to fourth column: material images obtained by Direct Matrix Inversion, DECT-EP, DECT-TDL, and DECT-MULTRA, respectively. The top and bottom rows show the water and bone images with display windows [0.5 1.3] g/cm$^3$ and [0.05 0.905] g/cm$^3$, respectively.}
	\label{Fig:head_comp}
	\vspace{-0.1in}
\end{figure*}

Fig.~\ref{Fig:head_comp} shows the water and bone material images decomposed by Direct Matrix Inversion,  DECT-EP, DECT-TDL, and DECT-MULTRA. 
DECT-MULTRA reduces artifacts at the boundaries of different materials and suppresses noise in the  material images much better than the other methods. 
One clearly noticeable improvement is seen in the rightmost zoom-ins in the water images, where  
Direct Matrix Inversion and DECT-EP both missed a dark spot (pointed by the red arrow numbered~1), while DECT-TDL and DECT-MULTRA preserved this feature that exists in the high and low energy attenuation maps in Fig.~\ref{Fig:head_atten}. 
The structure of the ``dark spot" is an artery (see the high and low attenuation images in Fig.~\ref{Fig:head_atten}) that contains diluted iodine solution caused by the angiogram.  Iodine is grouped into the bone image, while in the water image there are only some pixels with tiny values or values close to zero, thus it is a ``dark spot".
Moreover, DECT-MULTRA substantially improves the sharpness of edges in the soft tissues compared to DECT-TDL.
The rightmost zoom-ins in Fig.~\ref{Fig:head_comp} show that the marrow structures have sharper edges in the DECT-MULTRA water image than for DECT-TDL (pointed by the red arrow numbered~2). 
The clinical patient data is much more complex than the XCAT phantom, and has more structures (e.g., gum, teeth, artery, and so on). The results obtained by DECT-MULTRA demonstrate its ability to decompose pixels with mixed materials, and also the MULTRA model learned from the XCAT phantom generalized well to clinical head DECT data and outperformed the previous techniques.
The supplement additionally illustrates the superior performance of DECT-MULTRA over DECT-ST and DECT-CULTRA.

\section{Conclusions}
\label{Sec:conclusion}
This paper presented a new image-domain method dubbed DECT-MULTRA for DECT decomposition, and evaluated it relative to several competing methods. The proposed DECT-MULTRA framework combines conventional \mbox{PWLS} estimation with regularization based on a mixed union of learned sparsifying \mbox{transforms} model that exploits both the common properties among basis material images and their cross-dependencies. The various investigated sparsifying transform-based methods (DECT-MULTRA, DECT-CULTRA, and DECT-ST) reduce the high noise and artifacts observed in the decompositions obtained by nonadaptive methods such as Direct Matrix Inversion decomposition and DECT-EP. DECT-MULTRA successfully combines the advantages of both DECT-CULTRA and DECT-ST and reduces the artifacts at the boundaries of different materials and provides improved sharpness of edges in the soft tissue. 
In future work, we plan to apply DECT-MULTRA to more general multi-material (with several materials) decompositions. In a very recent work, we showed promise for DECT-CULTRA for this problem \cite{li:19:immmd}. 
We also plan to study the joint adaptation of the MULTRA model during the decomposition process, and explore extensions of the proposed approach to low-rank \footnote{While low-rank methods constrain a signal to lie in a low-dimensional subspace, sparsity-based methods often assume the signal lies approximately in a union of subspaces (a richer model).} + learned sparse models \cite{niu2018nonlocal} in future work.
Finally, improving image-domain decomposition methods to match the decomposition quality of the more accurate direct decomposition methods \cite{long:14:mmd}, while retaining the low runtimes of image-domain methods is an important area for future research.

\section{Acknowledgement}
The authors thank Dr. Tianye Niu, Zhejiang University, for providing clinical DECT images for our experiments.

\bibliographystyle{IEEEbib}
\bibliography{refs}

\begin{thebibliography}{10}

\bibitem{Mendonca2014A}
P.~R. Mendonca, P.~Lamb, and D.~V. Sahani,
\newblock ``A flexible method for multi-material decomposition of dual-energy
  {CT} images,''
\newblock {\em {IEEE Trans. Med. Imag.}}, vol. 33, no. 1, pp. {99--116}, 2014.

\bibitem{Mccollough2015Dual}
C.~H. McCollough, S.~Leng, L.~Yu, and J.~G. Fletcher,
\newblock ``Dual- and multi-energy {CT}: {P}rinciples, technical approaches,
  and clinical applications,''
\newblock {\em Radiology}, vol. 276, no. 3, pp. 637--653, 2015.

\bibitem{Li2013Iodine}
Y.~Li, G.~Shi, S.~Wang, S.~Wang, and R.~Wu,
\newblock ``Iodine quantification with dual-energy {CT}: phantom study and
  preliminary experience with {VX}2 residual tumour in rabbits after
  radiofrequency ablation,''
\newblock {\em British Journal of Radiology}, vol. 86, no. 1029, pp. 143 --
  151, 2013.

\bibitem{Chandarana2011Iodine}
H.~Chandarana, A.~J. Megibow, B.~A. Cohen, R.~Srinivasan, D.~Kim, C.~Leidecker,
  and M.~Macari,
\newblock ``Iodine quantification with dual-energy {CT}: phantom study and
  preliminary experience with renal masses.,''
\newblock {\em Am. J. Roentgenol}, vol. 196, no. 6, pp. W693 -- W700, 2011.

\bibitem{Primak2007Noninvasive}
A.~N. Primak, J.~G. Fletcher, T.~J. Vrtiska, O.~P. Dzyubak, J.~C. Lieske, M.~E.
  Jackson, J.~C. Williams, Jr, and C.~H. Mccollough,
\newblock ``Noninvasive differentiation of uric acid versus non-uric acid
  kidney stones using dual-energy {CT},''
\newblock {\em Academic Radiology}, vol. 14, no. 12, pp. 1441--1447, 2007.

\bibitem{Lifeng2011Virtual}
L.~Yu, J.~A. Christner, S.~Leng, J.~Wang, J.~G. Fletcher, and C.~H. McCollough,
\newblock ``Virtual monochromatic imaging in dual-source dual-energy {CT}:
  radiation dose and image quality,''
\newblock {\em {Med. Phys.}}, vol. 38, no. 12, pp. 6371--6379, 2011.

\bibitem{Liu2010Feasibility}
Y.~Liu, J.~Cheng, Z.~Chen, and Y.~Xing,
\newblock ``Feasibility study: Low-cost dual energy {CT} for security
  inspection,''
\newblock in {\em Nuclear Science Symposium Conference Record}, 2010, pp.
  879--882.

\bibitem{Lalonde2016}
A.~Lalonde and H.~Bouchard,
\newblock ``A general method to derive tissue parameters for {M}onte {C}arlo
  dose calculation with multi-energy {CT},''
\newblock {\em {Phys. Med. Biol.}}, vol. 61, no. 22, pp. 8044--69, 2016.

\bibitem{shen182}
C.~Shen, B.~Li, L.~Chen, M.~Yang, Y.~Lou, and X.~Jia,
\newblock ``Material elemental decomposition in dual and multi-energy {CT} via
  a sparsity-dictionary approach for proton stopping power ratio calculation,''
\newblock {\em {Med. Phys.}}, vol. 45, pp. 1491--503, 2018.

\bibitem{shen189}
C.~Shen, B.~Li, Y.~Lou, M.~Yang, L.~Zhou, and X.~Jia,
\newblock ``Multienergy element-resolved cone beam {CT (MEER-CBCT)} realized on
  a conventional {CBCT} platform,''
\newblock {\em {Med. Phys.}}, vol. 45, pp. 4461--70, 2018.

\bibitem{long:14:mmd}
Y.~Long and J.~A. Fessler,
\newblock ``Multi-material decomposition using statistical image reconstruction
  for spectral {CT},''
\newblock {\em {IEEE Trans. Med. Imag.}}, vol. 33, no. 8, pp. {1614--1626},
  Aug. 2014.

\bibitem{noh:09:ssr}
J.~Noh, J.~A. Fessler, and P.~E. Kinahan,
\newblock ``Statistical sinogram restoration in dual-energy {CT} for {PET}
  attenuation correction,''
\newblock {\em {IEEE Trans. Med. Imag.}}, vol. 28, no. 11, pp. {1688--1702},
  Nov. 2009.

\bibitem{xue:2017:statistical}
Y.~Xue, R.~Ruan, X.~Hu, Y.~Kuang, J.~Wang, Y.~Long, and T.~Niu,
\newblock ``Statistical image-domain multi-material decomposition for
  dual-energy {CT},''
\newblock {\em {Med. Phys.}}, vol. 44, no. 3, pp. {886--901}, 2017.

\bibitem{MM:11:accu}
M.~M. Goodsitt, E.~G. Christodoulou, and S.~C. Larson,
\newblock ``Accuracies of the synthesized monochromatic {CT} numbers and
  effective atomic numbers obtained with a rapid k{V}p switching dual energy
  {CT} scanner,''
\newblock {\em {Med. Phys.}}, vol. 38, no. 4, pp. {2222--2232}, Apr. 2011.

\bibitem{yu:16:spi}
Z.~Yu, S.~Leng, Z.~Li, and C.~H. McCollough,
\newblock ``Spectral prior image constrained compressed sensing (spectral
  {PICCS)} for photon-counting computed tomography,''
\newblock {\em {Phys. Med. Biol.}}, vol. 61, no. 18, pp. {6707--6732}, Sept.
  2016.

\bibitem{xu:12:ldx}
Q.~Xu, H.~Yu, X.~Mou, L.~Zhang, J.~Hsieh, and G.~Wang,
\newblock ``Low-dose {X-ray} {CT} reconstruction via dictionary learning,''
\newblock {\em {IEEE Trans. Med. Imag.}}, vol. 31, no. 9, pp. {1682--1697},
  Sept. 2012.

\bibitem{mechlem:16:dbi}
K.~Mechlem, S.~Allner, K.~Mei, F.~Pfeiffer, and P.~B. {No{\"e}l},
\newblock ``Dictionary-based image denoising for dual energy computed
  tomography,''
\newblock in {\em {Proc. SPIE 9783 Medical Imaging 2016: Phys. Med. Im.}},
  2016, pp. {97830E--1--97830E--7}.

\bibitem{Li2012Dual}
L.~Li, Z.~Chen, and P.~Jiao,
\newblock ``Dual-energy {CT} reconstruction based on dictionary learning and
  total variation constraint,''
\newblock in {\em {Proc. IEEE Intl. Symp. Biomed. Imag.}}, 2012, pp.
  {2358--2361}.

\bibitem{zhao:12:ddl}
B.~Zhao, H.~Ding, Y.~Lu, G.~Wang, J.~Zhao, and S.~Molloi,
\newblock ``Dual-dictionary learning-based iterative image reconstruction for
  spectral computed tomography application,''
\newblock {\em {Phys. Med. Biol.}}, vol. 57, no. 24, pp. {8217--8230}, Dec.
  2012.

\bibitem{zhang:15:tbd}
Y.~Zhang, X.~Mou, H.~Yu, G.~Wang, and Q.~Xu,
\newblock ``Tensor based dictionary learning for spectral {CT}
  reconstruction,''
\newblock {\em {IEEE Trans. Med. Imag.}}, vol. 36, no. 1, pp. 142--154, 2017.

\bibitem{WU2018538}
W.~Wu, Y.~Zhang, Q.~Wang, F.~Liu, P.~Chen, and H.~Yu,
\newblock ``Low-dose spectral {CT} reconstruction using image gradient
  $\ell_0$-norm and tensor dictionary,''
\newblock {\em {Applied Mathematical Modelling}}, vol. 63, pp. {538 -- 557},
  2018.

\bibitem{Ravishankar2015}
S.~Ravishankar and Y.~Bresler,
\newblock ``{$l_0$} sparsifying \mbox{transform} learning with efficient
  optimal updates and convergence guarantees,''
\newblock {\em {IEEE Trans. Sig. Proc.}}, vol. 63, no. 9, pp. {2389--2404}, May
  2015.

\bibitem{ravishankar:13:lst}
S.~Ravishankar and Y.~Bresler,
\newblock ``Learning sparsifying transforms,''
\newblock {\em {IEEE Trans. Sig. Proc.}}, vol. 61, no. 5, pp. {1072--1086},
  Mar. 2013.

\bibitem{Zheng2016Low}
X.~Zheng, Z.~Lu, S.~Ravishankar, Y.~Long, and J.~A. Fessler,
\newblock ``Low dose {CT} image reconstruction with learned sparsifying
  transform,''
\newblock in {\em Proc. IEEE Wkshp. on Image, Video, Multidim. Signal Proc.},
  July 2016, pp. 1--5.

\bibitem{Ye:17:adaptive}
S.~Ye, S.~Ravishankar, Y.~Long, and J.~A. Fessler,
\newblock ``Adaptive sparse modeling and shifted-poisson likelihood based
  approach for low-dose {CT} image reconstruction,''
\newblock in {\em Proc. IEEE Wkshp. Machine Learning for Signal Proc.}, 2017,
  pp. 1--6.

\bibitem{zheng2018}
X.~Zheng, S.~Ravishankar, Y.~Long, and J.~A. Fessler,
\newblock ``{PWLS-ULTRA}: An efficient clustering and learning-based approach
  for low-dose 3{D} {CT} image reconstruction,''
\newblock {\em {IEEE Trans. Med. Imag.}}, vol. 37, no. 6, pp. 1498--1510, June
  2018.

\bibitem{YuCNN}
Y.~Liao, Y.~Wang, S.~Li, J.~He, D.~Zeng, Z.~Bian, and J.~Ma,
\newblock ``Pseudo dual energy {CT} imaging using deep learning-based
  framework: basic material estimation,''
\newblock in {\em {Proc. SPIE}}, 2018, vol. 10573, pp. 105734N--1--105734N--5.

\bibitem{Niu:18:butterfly}
W.~Zhang, H.~Zhang, L.~Wang, X.~Wang, A.~Cai, L.~Li, T.~Niu, and B.~Yan,
\newblock ``Image domain dual material decomposition for dual-energy {CT} using
  butterfly network,''
\newblock {\em {Med. Phys.}}, vol. 46, no. 5, pp. 2037 -- 51, 2019.

\bibitem{li2018image}
Z.~Li, S.~Ravishankar, Y.~Long, and J.~A. Fessler,
\newblock ``Image-domain material decomposition using data-driven sparsity
  models for dual-energy {CT},''
\newblock in {\em {Proc. IEEE Intl. Symp. Biomed. Imag.}}, Apr. 2018, pp.
  52--56.

\bibitem{niu2014iterative}
T.~Niu, X.~Dong, M.~Petrongolo, and L.~Zhu,
\newblock ``Iterative image-domain decomposition for dual-energy {CT},''
\newblock {\em {Med. Phys.}}, vol. 41, no. 4, pp. 041901, Apr. 2014.

\bibitem{Wen2015Structured}
B.~Wen, S.~Ravishankar, and Y.~Bresler,
\newblock ``Structured overcomplete sparsifying transform learning with
  convergence guarantees and applications,''
\newblock {\em {Intl. J. Comp. Vision}}, vol. 114, no. 2-3, pp. {137--166},
  Sept. 2015.

\bibitem{Zhang2014Model}
R.~Zhang, J.~B. Thibault, C.~A. Bouman, K.~D. Sauer, and J.~Hsieh,
\newblock ``Model-based iterative reconstruction for dual-energy {X-ray} {CT}
  using a joint quadratic likelihood model,''
\newblock {\em {IEEE Trans. Med. Imag.}}, vol. 33, no. 1, pp. {117--134}, Jan.
  2014.

\bibitem{Segars2008Realistic}
W.~P. Segars, M.~Mahesh, T.~J. Beck, E.~C. Frey, and B.~M.~W. Tsui,
\newblock ``Realistic {CT} simulation using the {4D} {XCAT} phantom,''
\newblock {\em {Med. Phys.}}, vol. 35, no. 8, pp. {3800--3808}, Aug. 2008.

\bibitem{ravishankar:16:ddl}
S.~Ravishankar and Y.~Bresler,
\newblock ``Data-driven learning of a union of sparsifying transforms model for
  blind compressed sensing,''
\newblock {\em {IEEE Trans. Computational Imaging}}, vol. 2, no. 3, pp.
  {294--309}, Sept. 2016.

\bibitem{aharon:06:ksa}
M.~Aharon, M.~Elad, and A.~Bruckstein,
\newblock ``{K-SVD:} an algorithm for designing overcomplete dictionaries for
  sparse representation,''
\newblock {\em {IEEE Trans. Sig. Proc.}}, vol. 54, no. 11, pp. {4311--22}, Nov.
  2006.

\bibitem{K-CPD}
G.~Duan, H.~Wang, Z.~Liu, J.~Deng, and Y.-W. Chen,
\newblock ``{K-CPD}: Learning of overcomplete dictionaries for tensor sparse
  coding,''
\newblock in {\em Proc. Int. Conf. Pattern Recognit. (ICPR)}, Nov. 2012, pp.
  493--496.

\bibitem{li:19:immmd}
Z.~Li, S.~Ravishankar, and Y.~Long,
\newblock ``Image-domain multi-material decomposition using a union of
  cross-material models,''
\newblock in {\em {Proc. Intl. Mtg. on Fully 3D Image Recon. in Rad. and Nuc.
  Med}}, 2019, vol. 11072, pp. {1107210--1 -- 1107210--5}.

\bibitem{niu2018nonlocal}
S.~Niu, G.~Yu, J.~Ma, and J.~Wang,
\newblock ``Nonlocal low-rank and sparse matrix decomposition for spectral {CT}
  reconstruction,''
\newblock {\em Inverse Problems}, vol. 34, no. 2, pp. 024003, 2018.

\end{thebibliography}


\begin{thebibliography}{1}

\bibitem{li:18:tmi}
Z.~Li, S.~Ravishankar, Y.~Long, and J.~A. Fessler,
\newblock ``{DECT-MULTRA}: Dual-energy {CT} image decomposition with learned
  mixed material models and efficient clustering,''
\newblock {\em {IEEE Trans. Med. Imag.}}, 2018,
\newblock submitted.

\end{thebibliography}

\newpage

%
%
\end{document}


%
\title{DECT-MULTRA: Dual-Energy CT Image Decomposition With Learned Mixed Material Models and  Efficient Clustering -- Supplementary Material}
%
%

\author{Zhipeng~Li,~\IEEEmembership{Student Member,~IEEE,} Saiprasad Ravishankar,~\IEEEmembership{Member,~IEEE,} Yong~Long$^*$,~\IEEEmembership{Member,~IEEE,} and~Jeffrey~A.~Fessler,~\IEEEmembership{Fellow,~IEEE}

%
%
}

\maketitle
This supplement provides additional results to accompany our manuscript \cite{li:18:tmi}.

\setcounter{section}{6}
\section{Additional Results}
\subsection{Decomposition Error Images for the XCAT Phantom}
Section \uppercase\expandafter{\romannumeral4}.C of \cite{li:18:tmi} compared the performance of various methods for decomposing several slices of the XCAT phantom.
Fig.~\ref{Fig:error} compares the decomposition error images (shown for Slice 77) for DECT-TDL and DECT-MULTRA. DECT-MULTRA produces smaller decomposition errors than DECT-TDL that are clearly noticeable in the regions pointed by the red arrows in the water and bone error images. 
\setcounter{figure}{10}
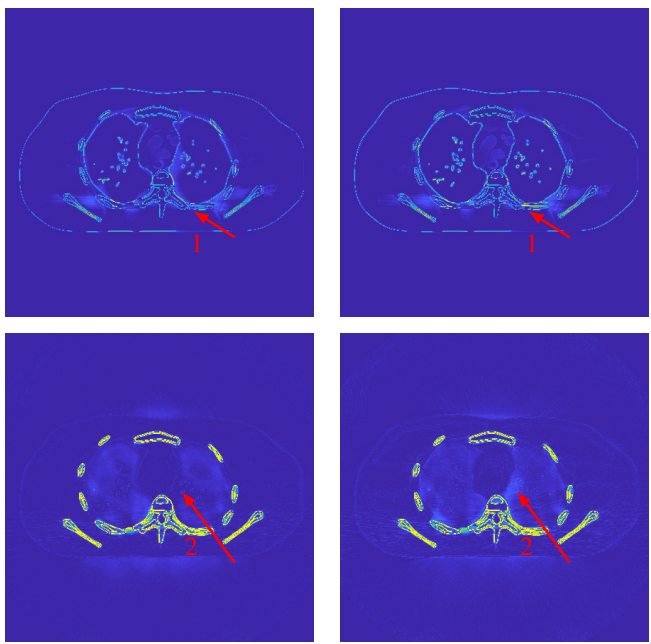
\begin{figure}[hbt]
	\centering
	\begin{tikzpicture}
	[spy using outlines={rectangle,red,magnification=2,width=14mm, height =10mm, connect spies}]				
	\node {\includegraphics[scale=0.36,trim=30 30 30 30,clip]{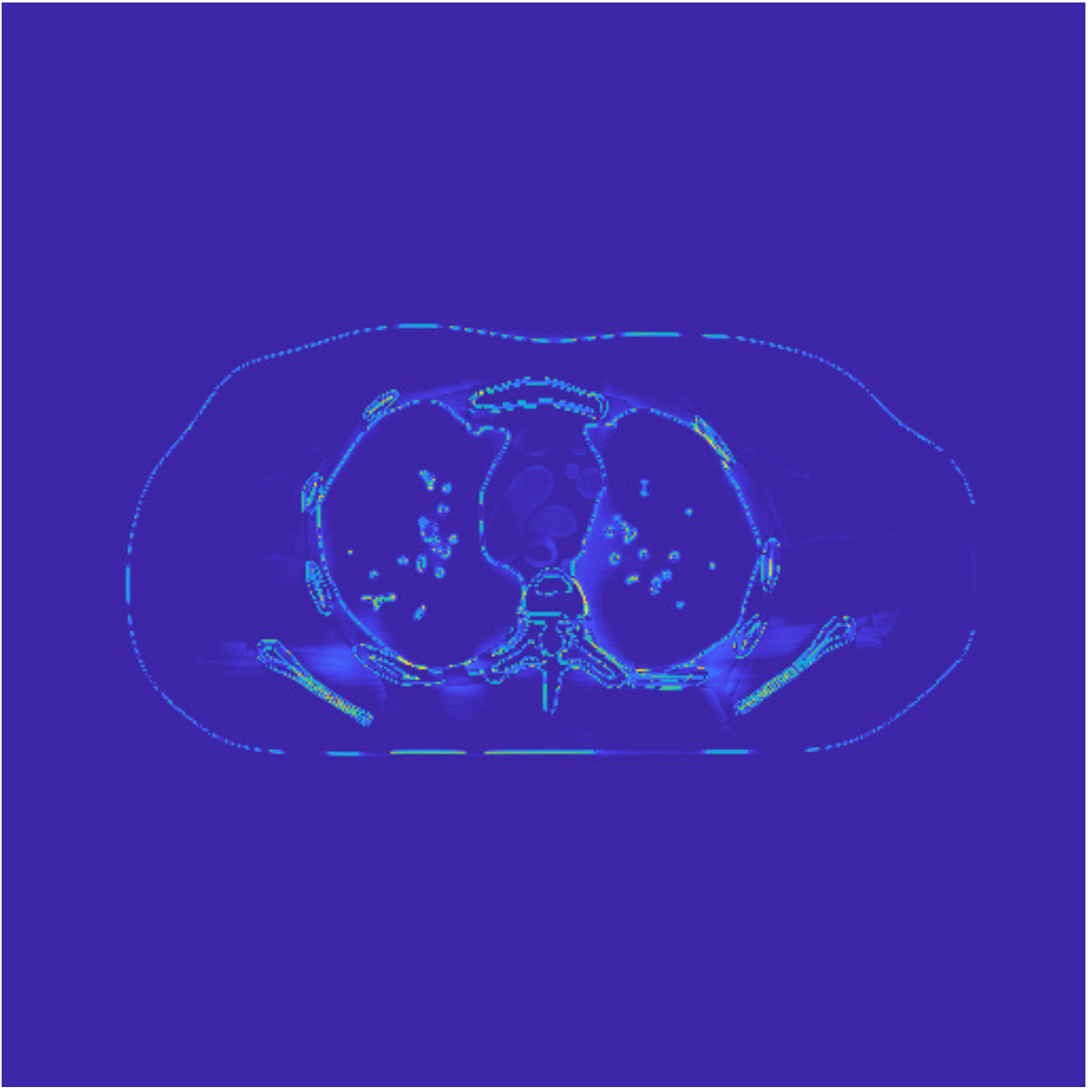}	};	
	\draw[line width=1pt,  -latex, red] (1,-1) -- node[auto] {1} (0.45,-0.65);
	\end{tikzpicture} 
	\begin{tikzpicture}
	[spy using outlines={rectangle,red,magnification=2,width=14mm, height =10mm, connect spies}]				
	\node {\includegraphics[scale=0.36,trim=30 30 30 30,clip]{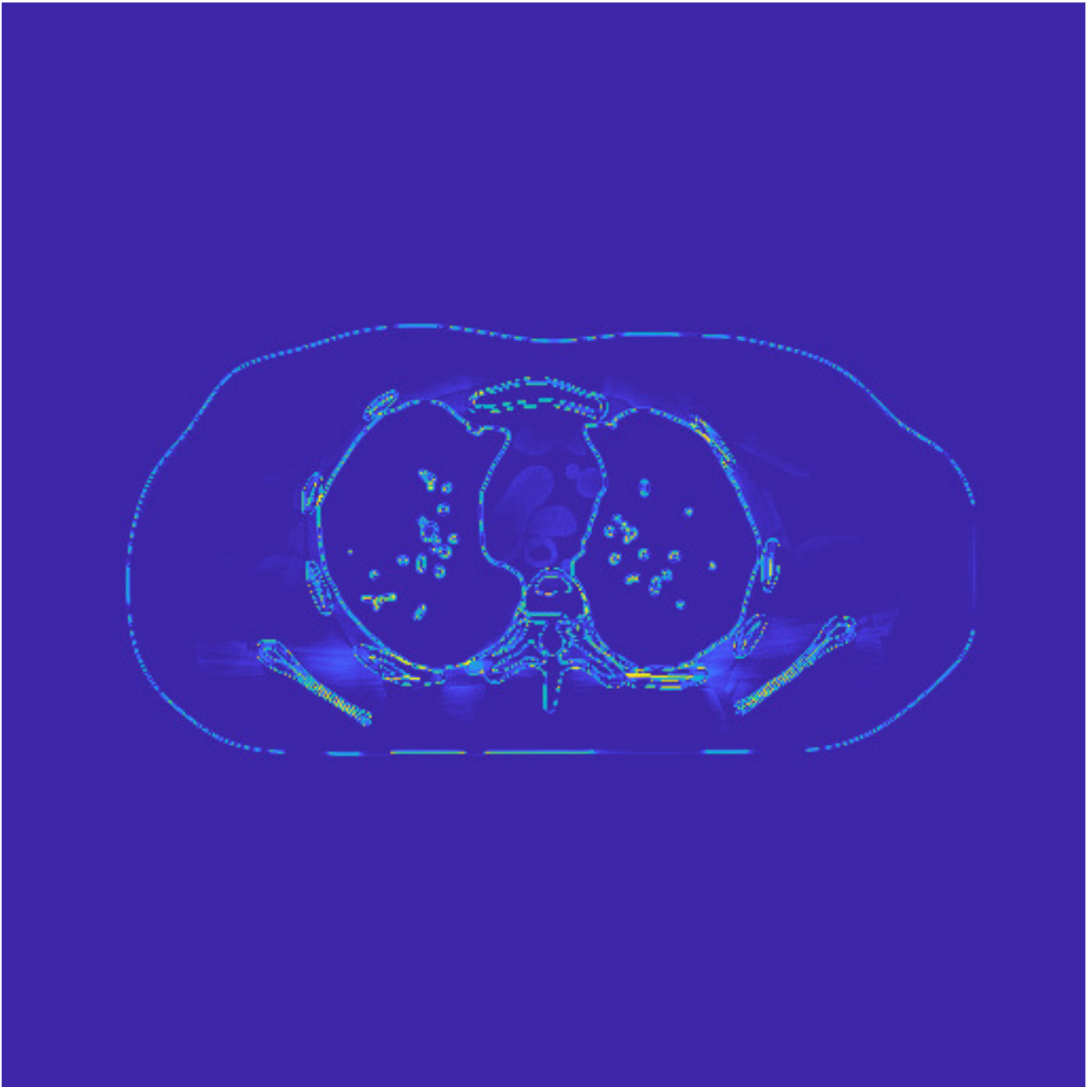}	};
	\draw[line width=1pt,  -latex, red] (1,-1) -- node[auto] {1} (0.45,-0.65);						
	\end{tikzpicture} \\ 
	\begin{tikzpicture}
	[spy using outlines={rectangle,red,magnification=2,width=14mm, height =10mm, connect spies}]				
	\node {\includegraphics[scale=0.355,trim=30 30 30 30,clip]{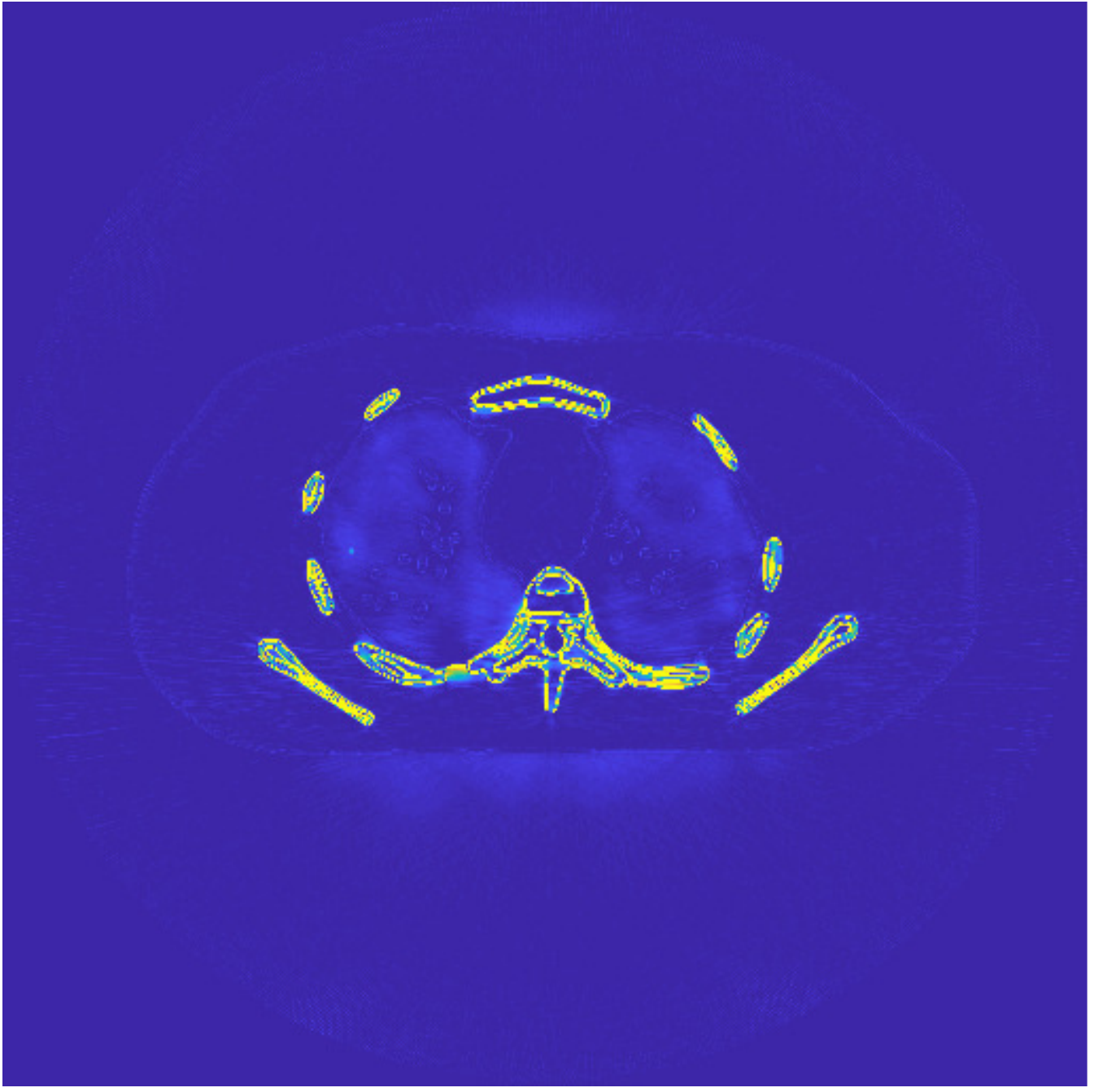}	};
	\draw[line width=1pt,  -latex, red] (1,-1) -- node[auto] {2} (0.3,-0.05);				
	\end{tikzpicture} 
	\begin{tikzpicture}
	[spy using outlines={rectangle,red,magnification=2,width=14mm, height =10mm, connect spies}]				
	\node {\includegraphics[scale=0.355,trim=30 30 30 30,clip]{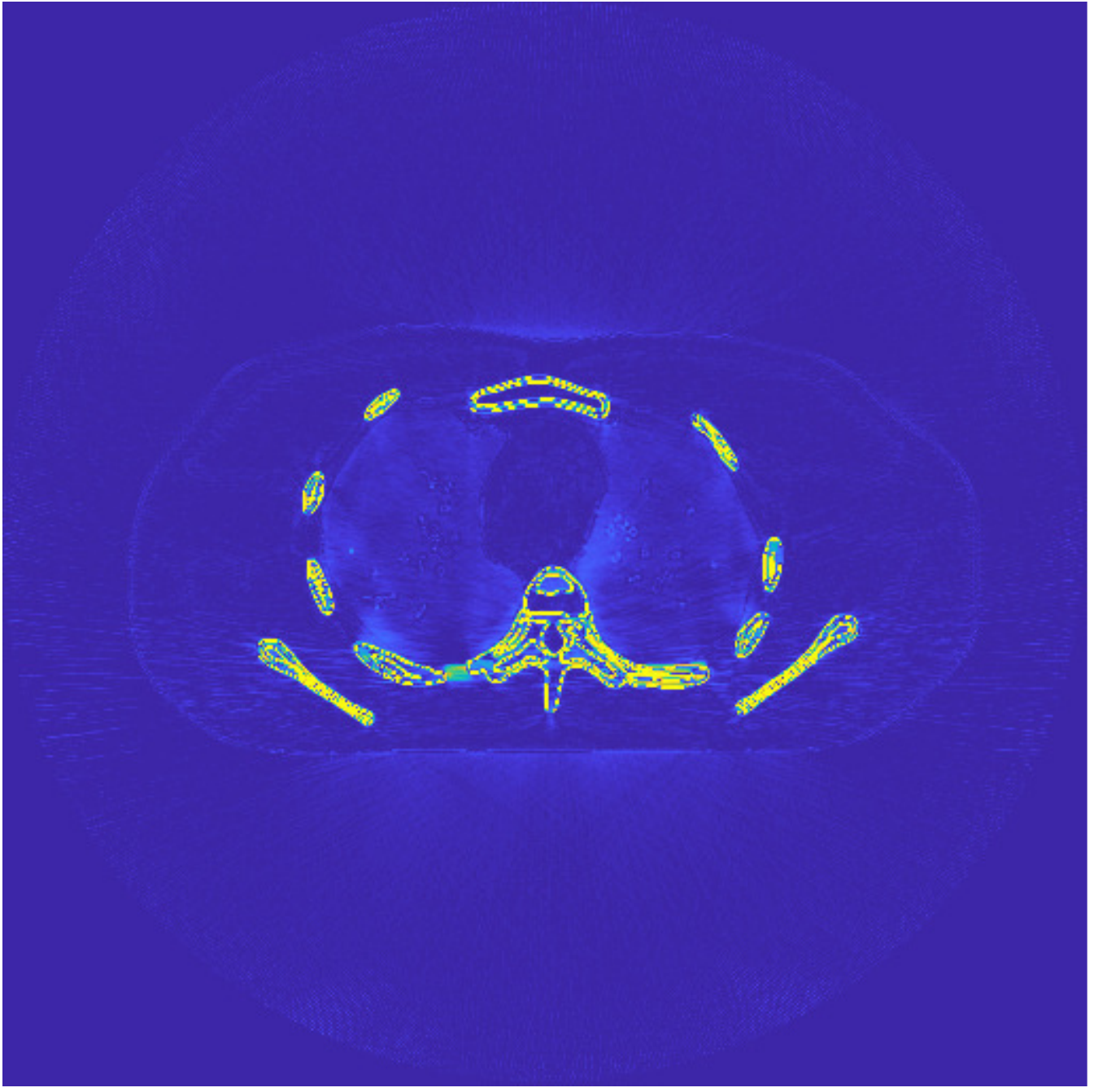}	};
	\draw[line width=1pt,  -latex, red] (1,-1) -- node[auto] {2} (0.3,-0.05);							
	\end{tikzpicture}
	\caption{Material decomposition error images (cropped) for DECT-MULTRA (left column) and DECT-TDL (right \mbox{column}). The top and bottom rows show the error images for water and bone with display windows [0.03  0.5] and [0  0.3]\,g/cm$^3$, respectively. }
	\label{Fig:error}
\end{figure}

\begin{figure}[htb]
	\centering
	\begin{tikzpicture}
	[spy using outlines={rectangle,red,magnification=2,width=9mm, height =14mm, connect spies}]
	\spy on (0.45,0.95) in node [right] at (1.1,-1.3);
	\spy [width=18mm, height =8mm] on (0.08,-0.15) in node [right] at (-2.0,-1.7);			
	\node {\includegraphics[scale=0.3]{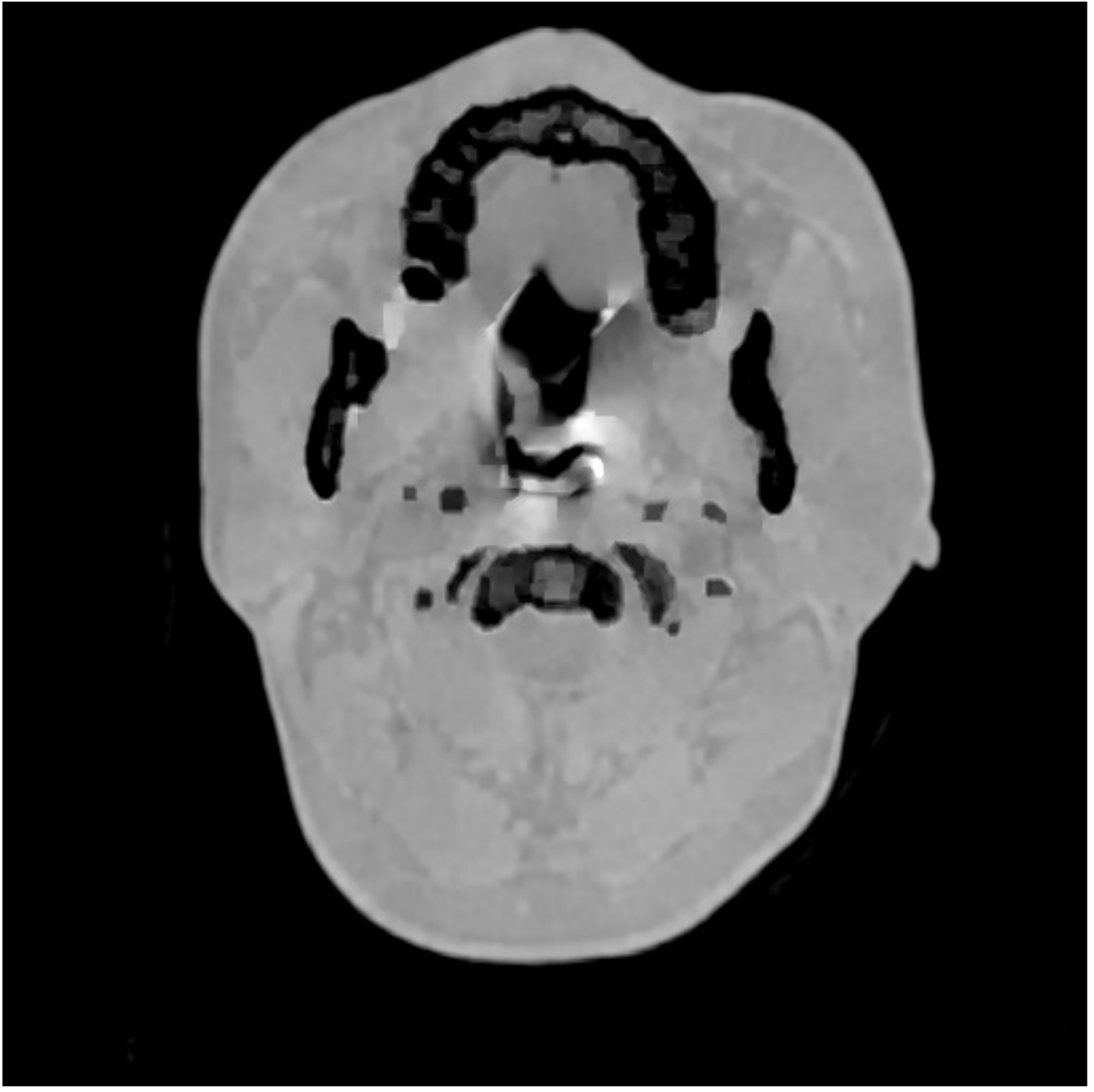}	};
	\draw[line width=1pt, thin, -latex, red] (0.69,0.69) -- node[auto] {} (0.4,0.67);
	\draw[line width=1pt, thin, -latex, red] (0.25,1.1) -- node[xshift=0.2cm,yshift=0.17cm] {} (0.41,0.98);				
	\end{tikzpicture} 
	\begin{tikzpicture}
	[spy using outlines={rectangle,red,magnification=2,width=9mm, height =14mm, connect spies}]				
	\node {\includegraphics[scale=0.3]{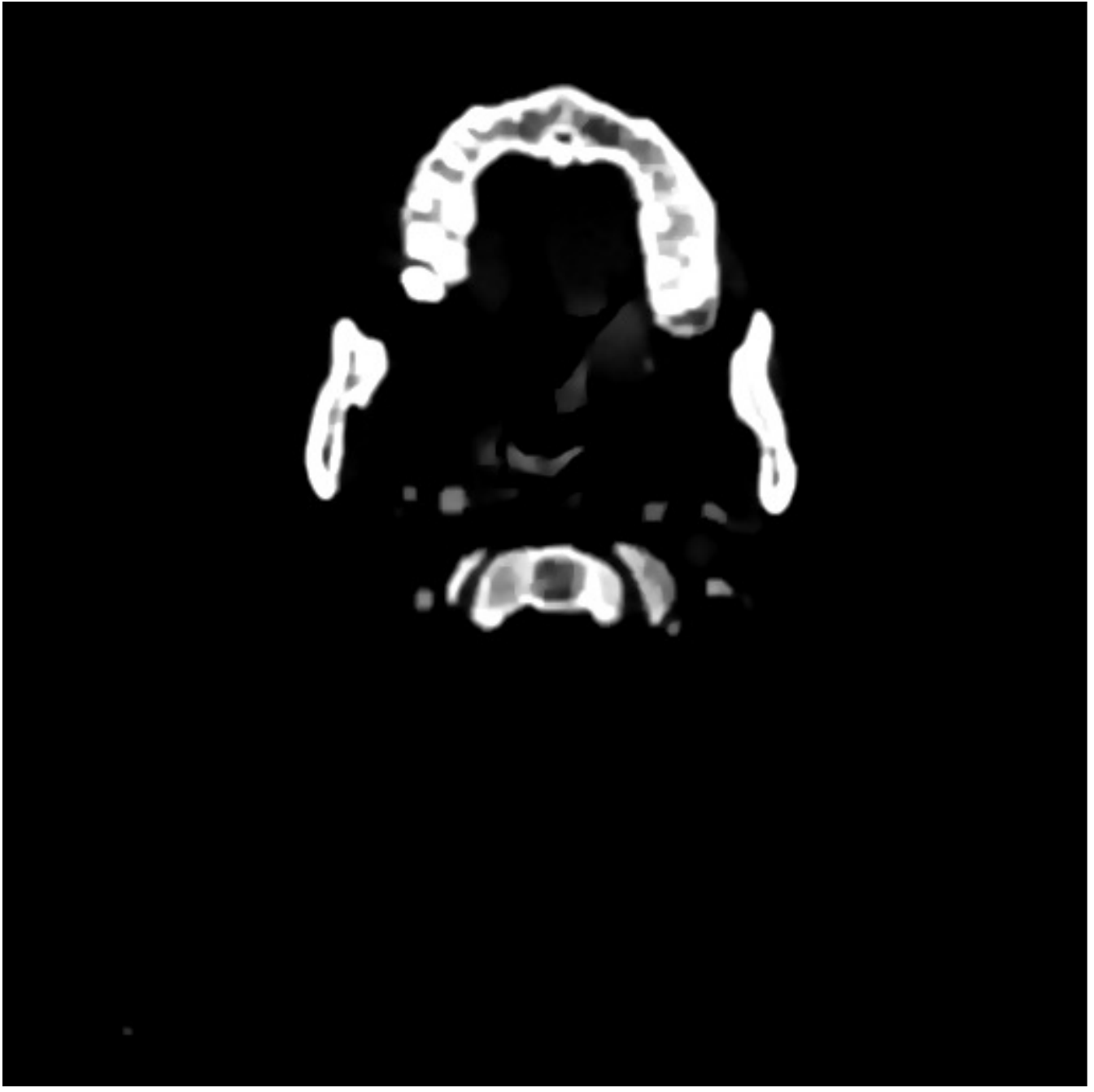}	};				
	\spy on (0.45,1.) in node [right] at (1.1,-1.3);
	\spy [width=18mm, height =8mm] on (0.08,-0.15) in node [right] at (-2.0,-1.7);	
	\end{tikzpicture}  \\	\vspace{-0.05in}
	\begin{tikzpicture}
	[spy using outlines={rectangle,red,magnification=2,width=9mm, height =14mm, connect spies}]				
	\node {\includegraphics[scale=0.3]{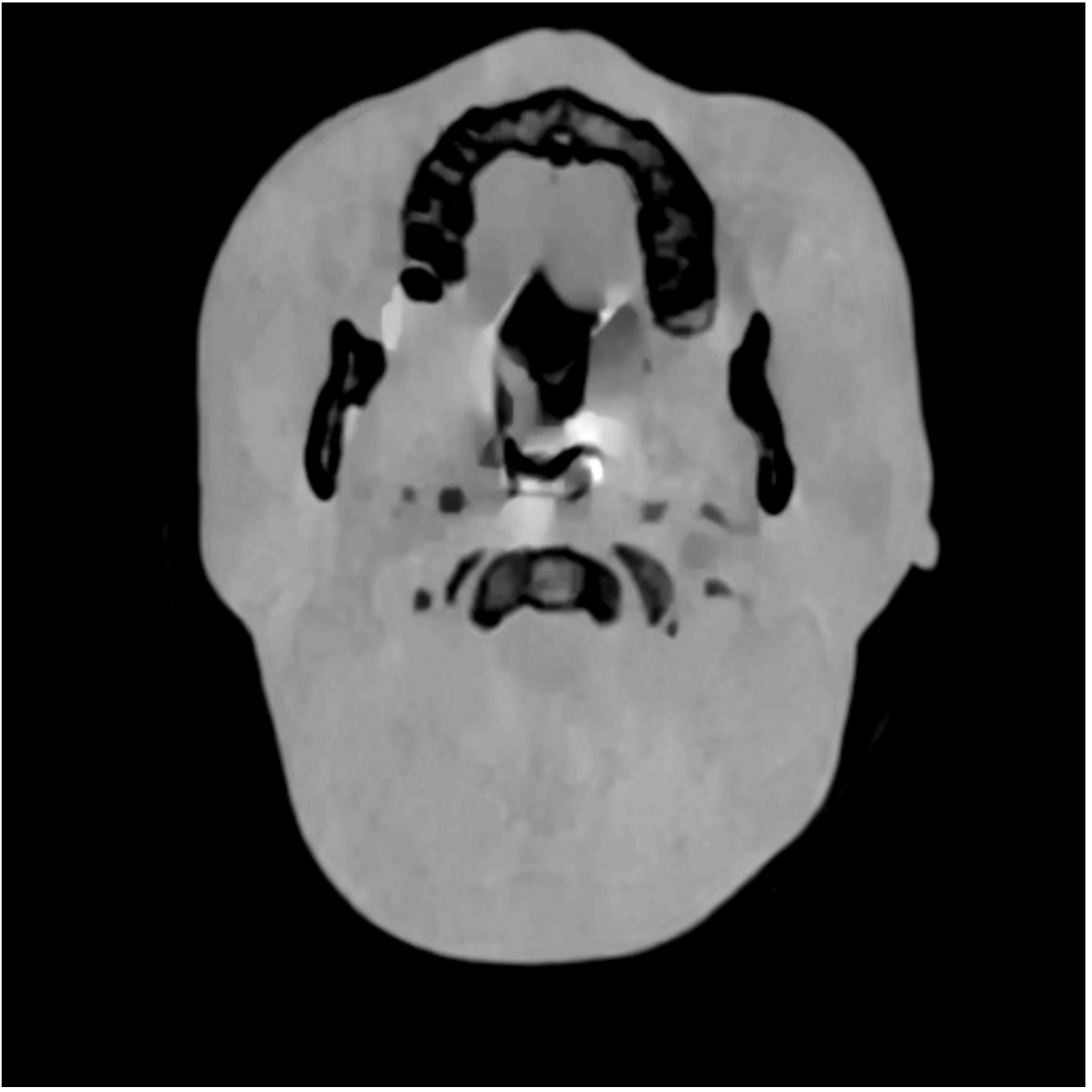}	};				
	\spy on (0.45,0.95) in node [right] at (1.1,-1.3);
	\spy [width=18mm, height =8mm] on (0.08,-0.15) in node [right] at (-2.0,-1.7);
	\draw[line width=1pt, thin, -latex, red] (0.69,0.69) -- node[auto] {} (0.4,0.67);
	\draw[line width=1pt, thin, -latex, red] (0.25,1.1) -- node[xshift=0.2cm,yshift=0.17cm] {} (0.41,0.98);
	\end{tikzpicture}   		  
	\begin{tikzpicture}
	[spy using outlines={rectangle,red,magnification=2,width=9mm, height =14mm, connect spies}]				
	\node {\includegraphics[scale=0.3]{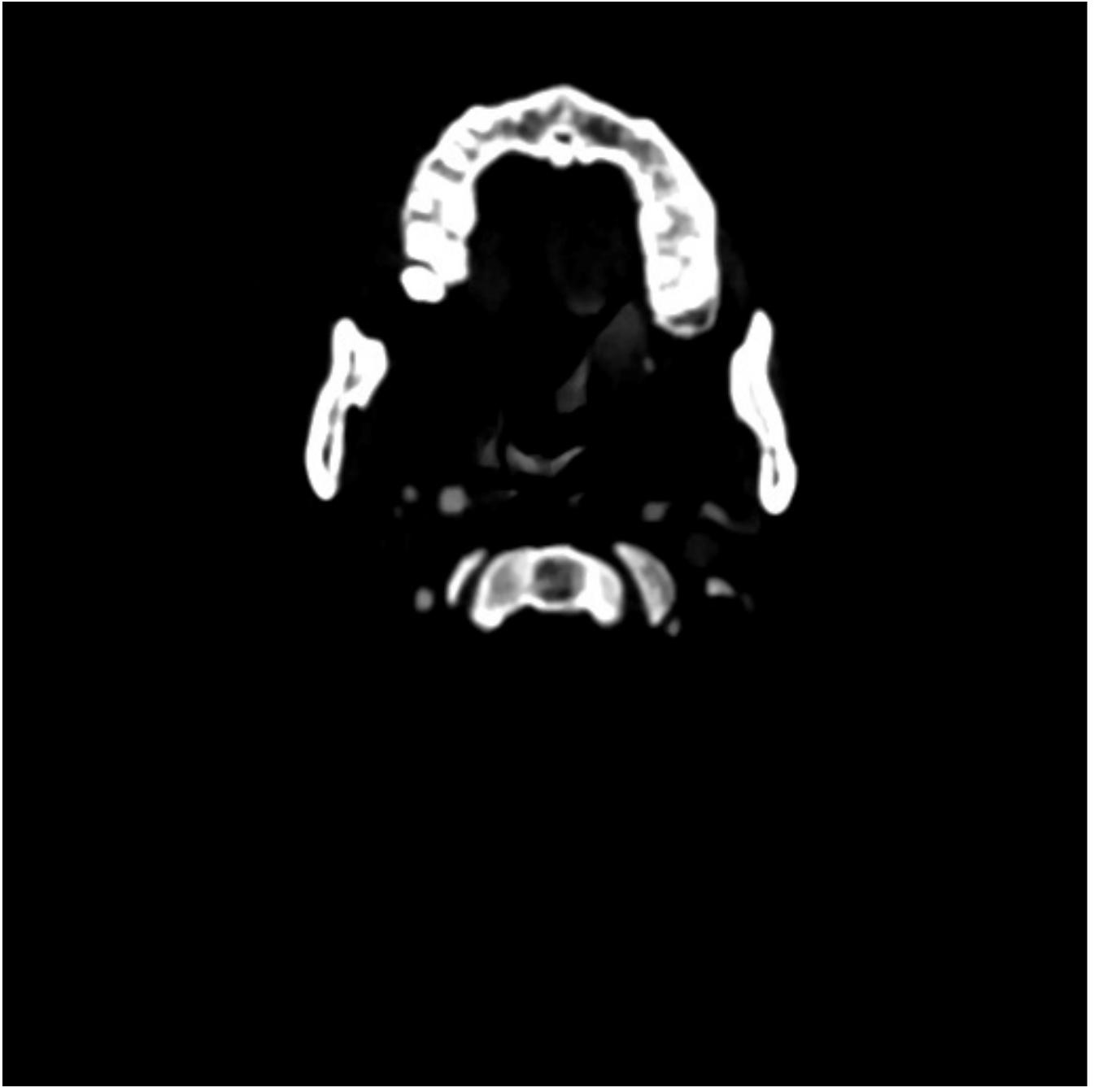}	};				
	\spy on (0.45,1.) in node [right] at (1.1,-1.3);
	\spy [width=18mm, height =8mm] on (0.08,-0.15) in node [right] at (-2.0,-1.7); 
	\end{tikzpicture}\\ \vspace{-0.05in}
	\begin{tikzpicture}
	[spy using outlines={rectangle,red,magnification=2,width=9mm, height =14mm, connect spies}]				
	\node {\includegraphics[scale=0.3]{./Fig4/multra_head_water.pdf}	};				
	\spy on (0.45,0.95) in node [right] at (1.1,-1.3);
	\spy [width=18mm, height =8mm] on (0.08,-0.15) in node [right] at (-2.0,-1.7);
	\draw[line width=1pt, thin, -latex, red] (0.69,0.69) -- node[auto] {} (0.4,0.67);
	\draw[line width=1pt, thin, -latex, red] (0.25,1.1) -- node[xshift=0.2cm,yshift=0.17cm] {} (0.41,0.98);
	\end{tikzpicture}   		  
	\begin{tikzpicture}
	[spy using outlines={rectangle,red,magnification=2,width=9mm, height =14mm, connect spies}]				
	\node {\includegraphics[scale=0.3]{./Fig4/multra_head_bone.pdf}	};				
	\spy on (0.45,1.) in node [right] at (1.1,-1.3);
	\spy [width=18mm, height =8mm] on (0.08,-0.15) in node [right] at (-2.0,-1.7); 		
	\end{tikzpicture} 	      		
	\caption{ Material images decomposed by DECT-ST (top row), DECT-CULTRA (middle row), and DECT-MULTRA (bottom row), respectively. The left and right columns show the water and bone images with display windows [0.5 1.3] g/cm$^3$ and [0.05 0.905] g/cm$^3$, respectively.}
	\label{Fig:head_comp_directEP}
\end{figure}

	\begin{figure*}[htb]
		\centering
		\begin{minipage}{0.1\linewidth}
			\centerline{\includegraphics[scale=0.45,trim=40 40 40 40,clip]{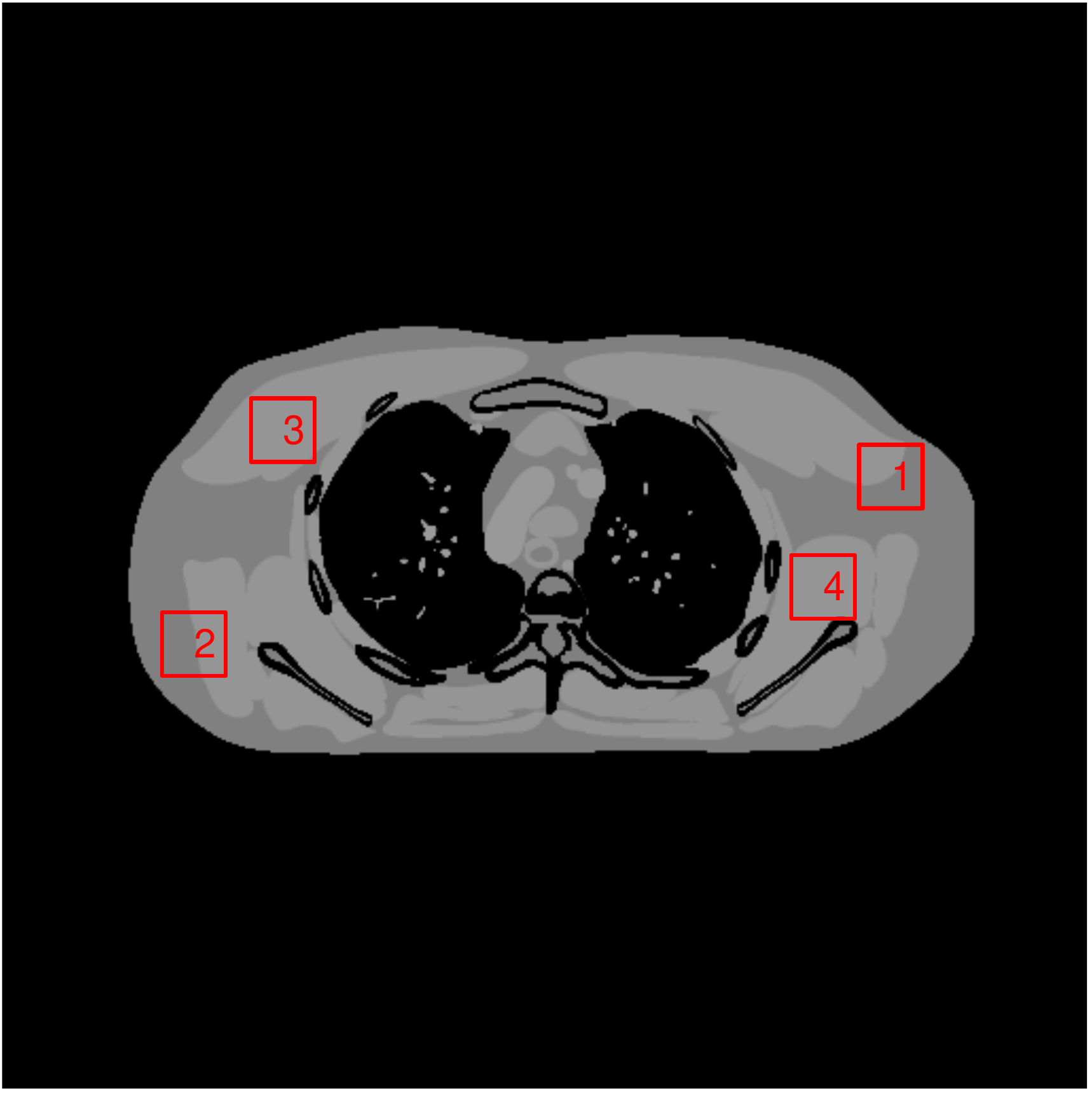}}
			\centerline{(a)}
		\end{minipage} \hspace{3in} 
		\begin{minipage}{0.1\linewidth}
			\centerline{\includegraphics[scale=0.225]{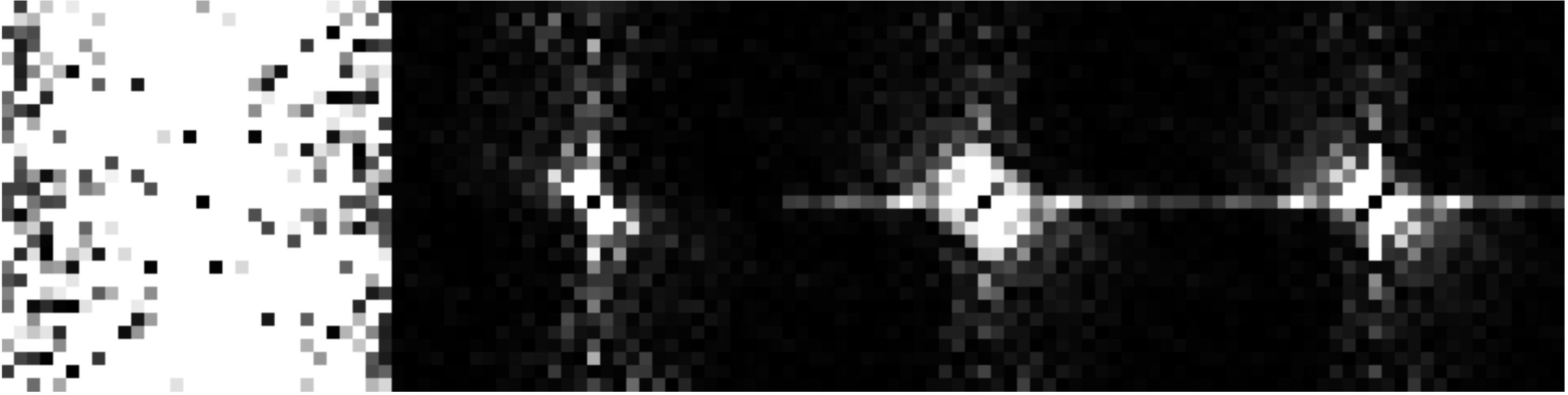}}
			\centerline{\includegraphics[scale=0.225]{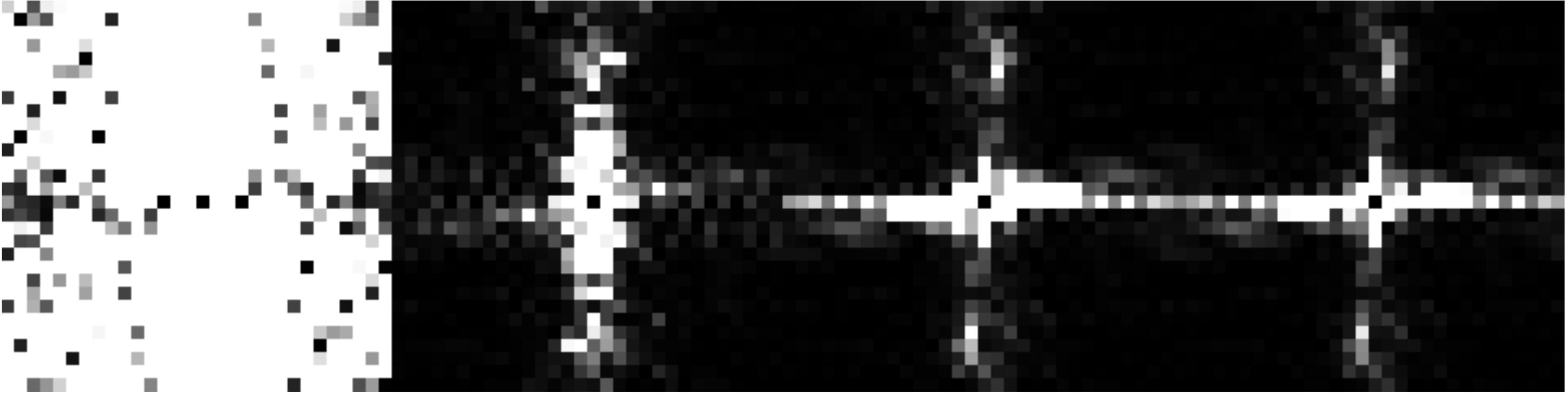}}
			\centerline{\includegraphics[scale=0.225]{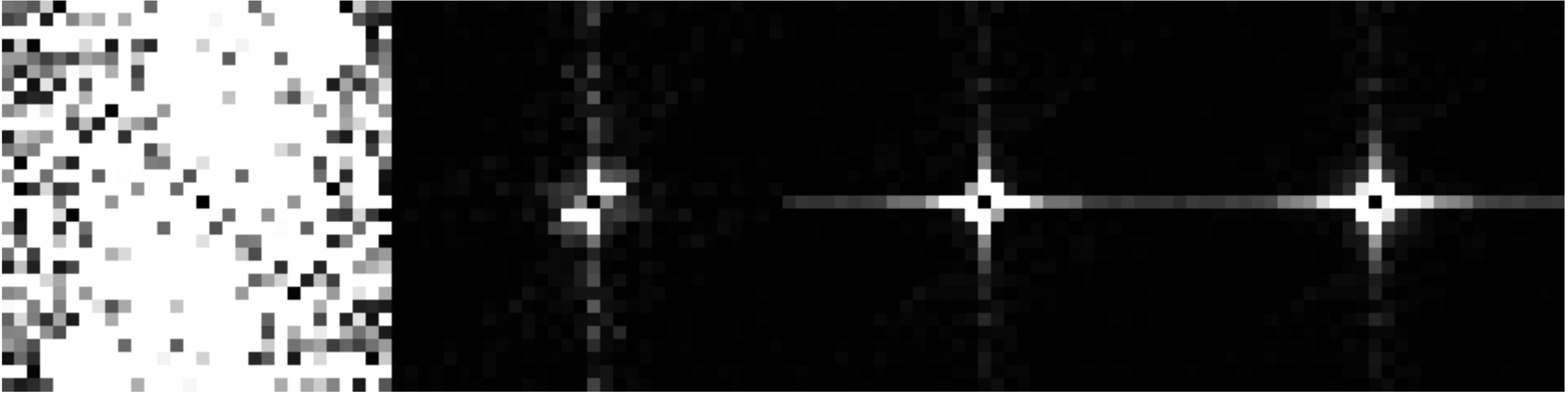}}
			\centerline{\includegraphics[scale=0.225]{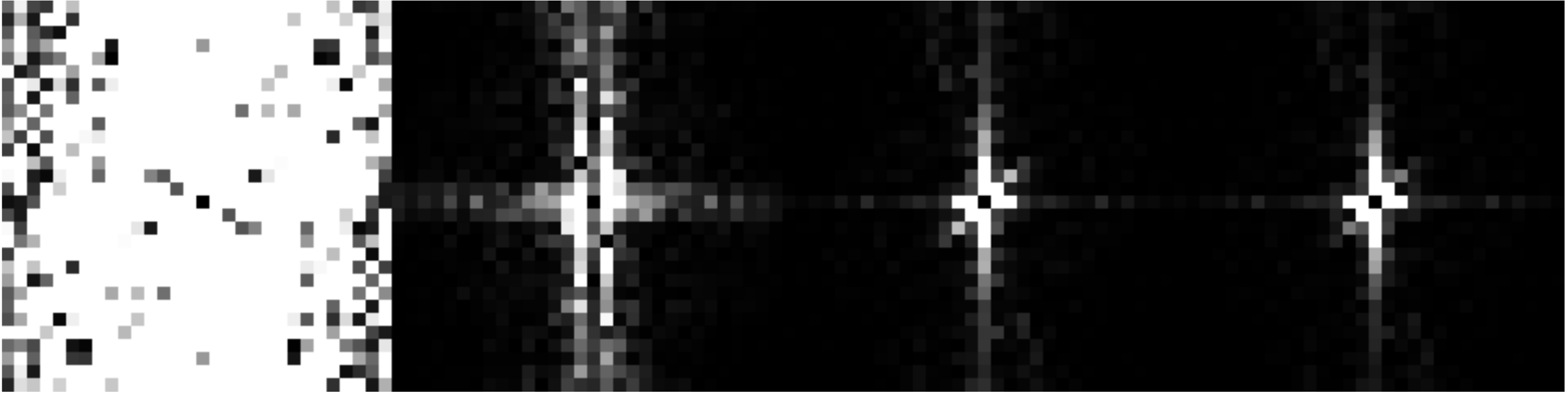}}
			\centerline{(b)}
		\end{minipage}
		\caption{(a) Selected four ROIs indicated by red squares on the true water image. (b) Left to right: NPS measured within ROIs of water error images obtained by direct matrix inversion, DECT-EP, DECT-TDL, and DECT-MULTRA. The first to the fourth rows in (b) show the NPS of the first to fourth ROIs respectively, with display windows [0 0.5]~g$^2$/cm$^6$.}
		\label{fig:nps}
	\end{figure*}

\begin{figure*}[htb]
	\centering
	\includegraphics[width=0.22\textwidth]{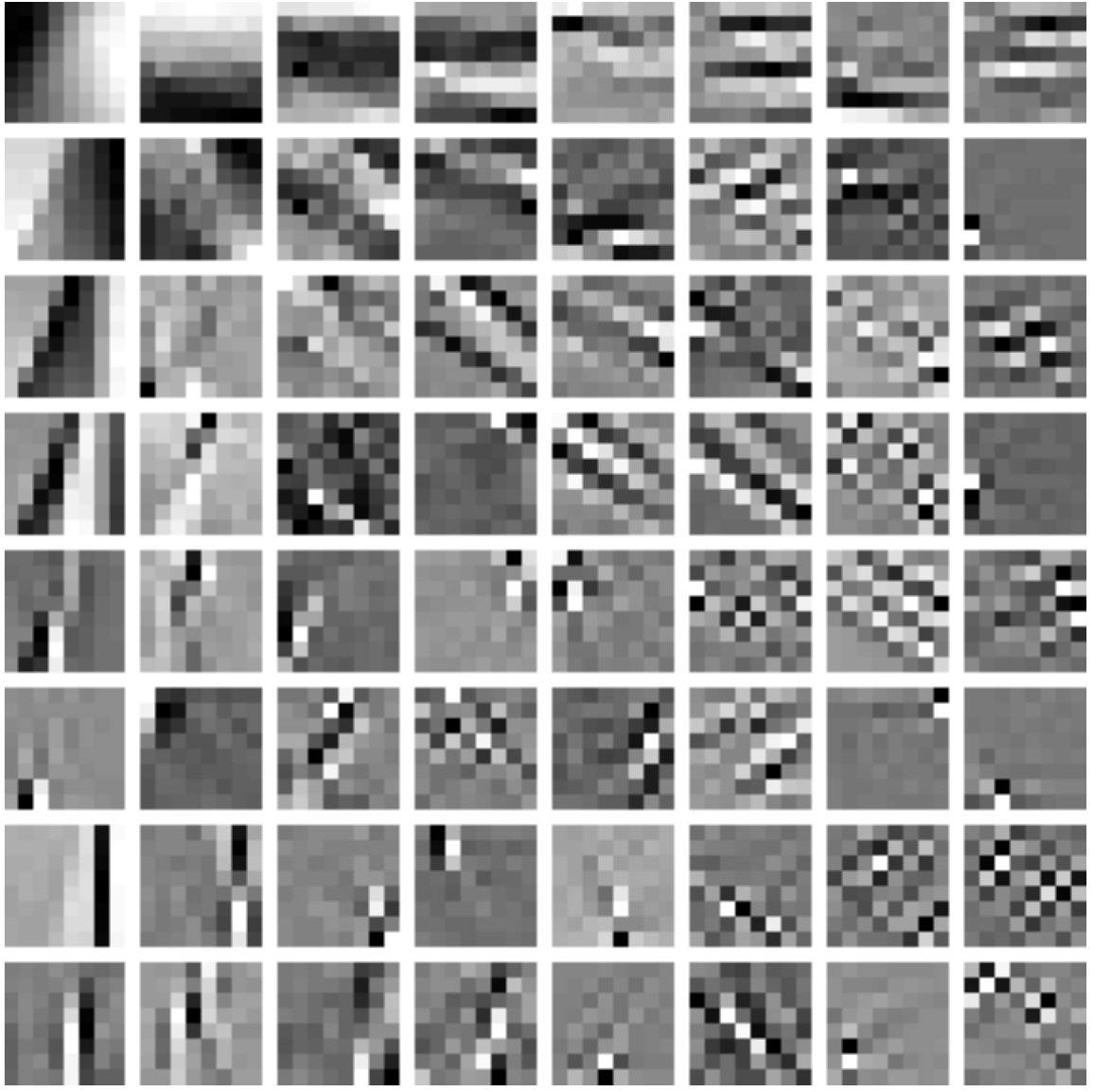}
	\includegraphics[width=0.22\textwidth]{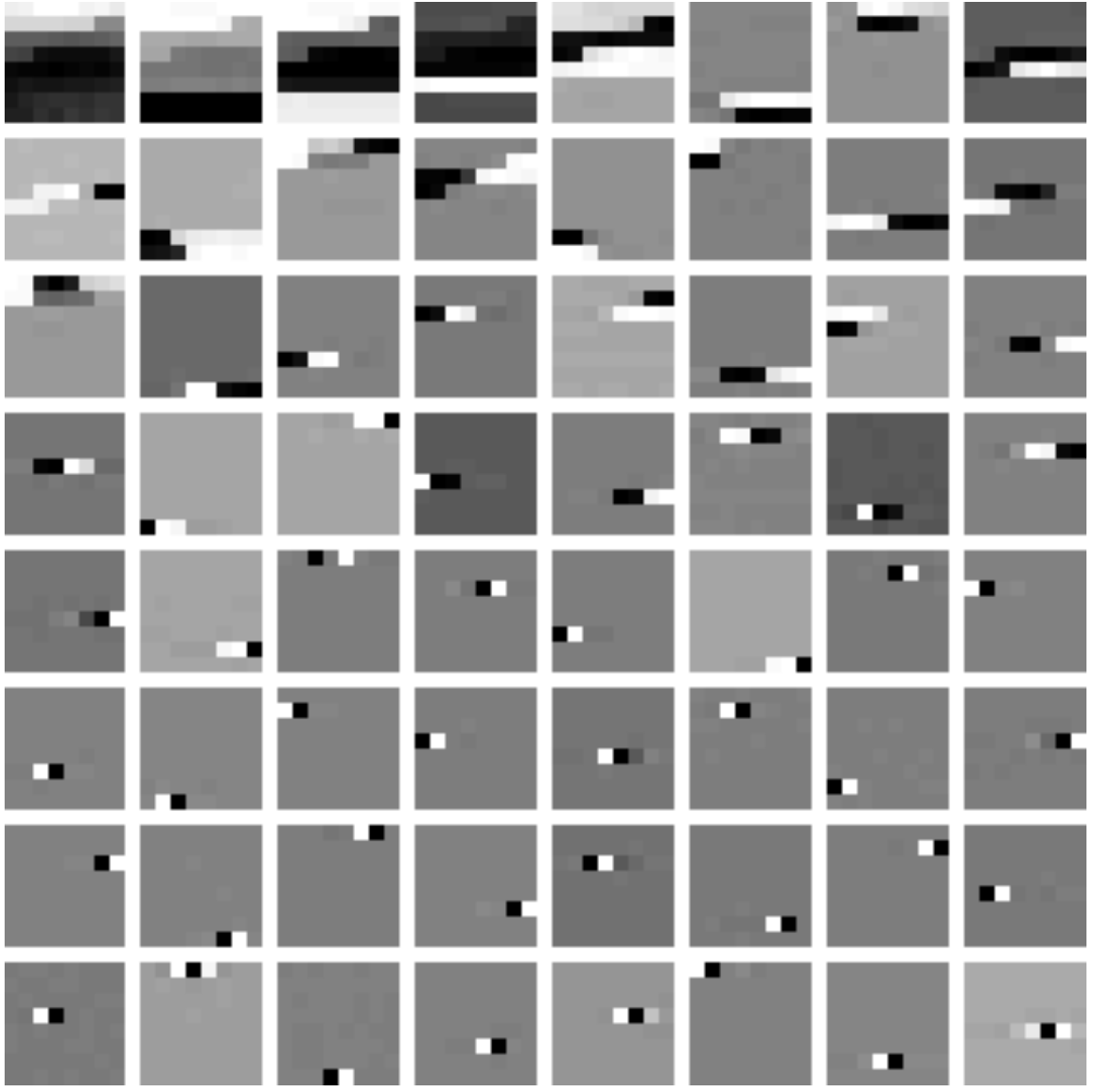}
	\includegraphics[width=0.22\textwidth]{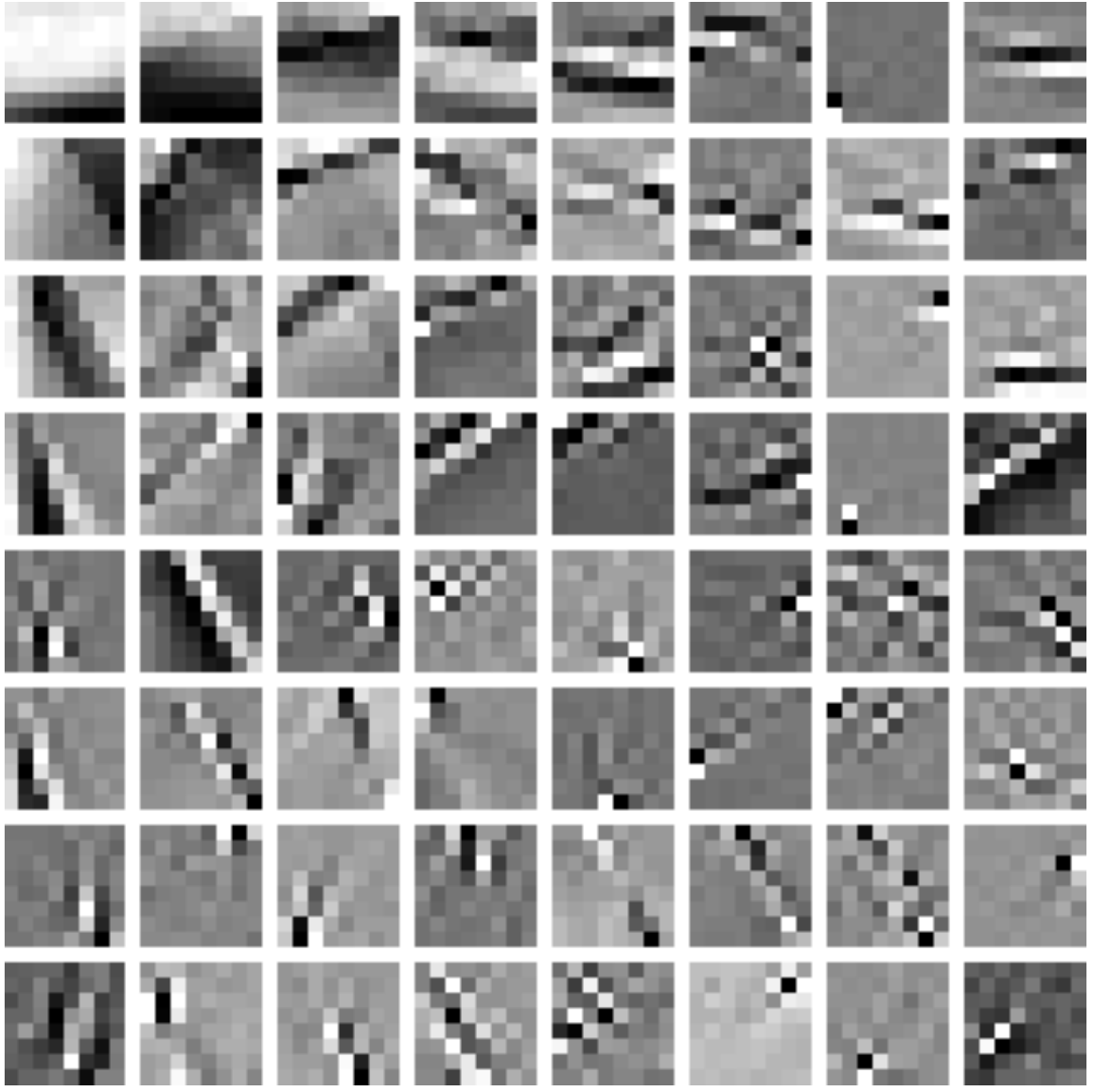}
	\includegraphics[width=0.22\textwidth]{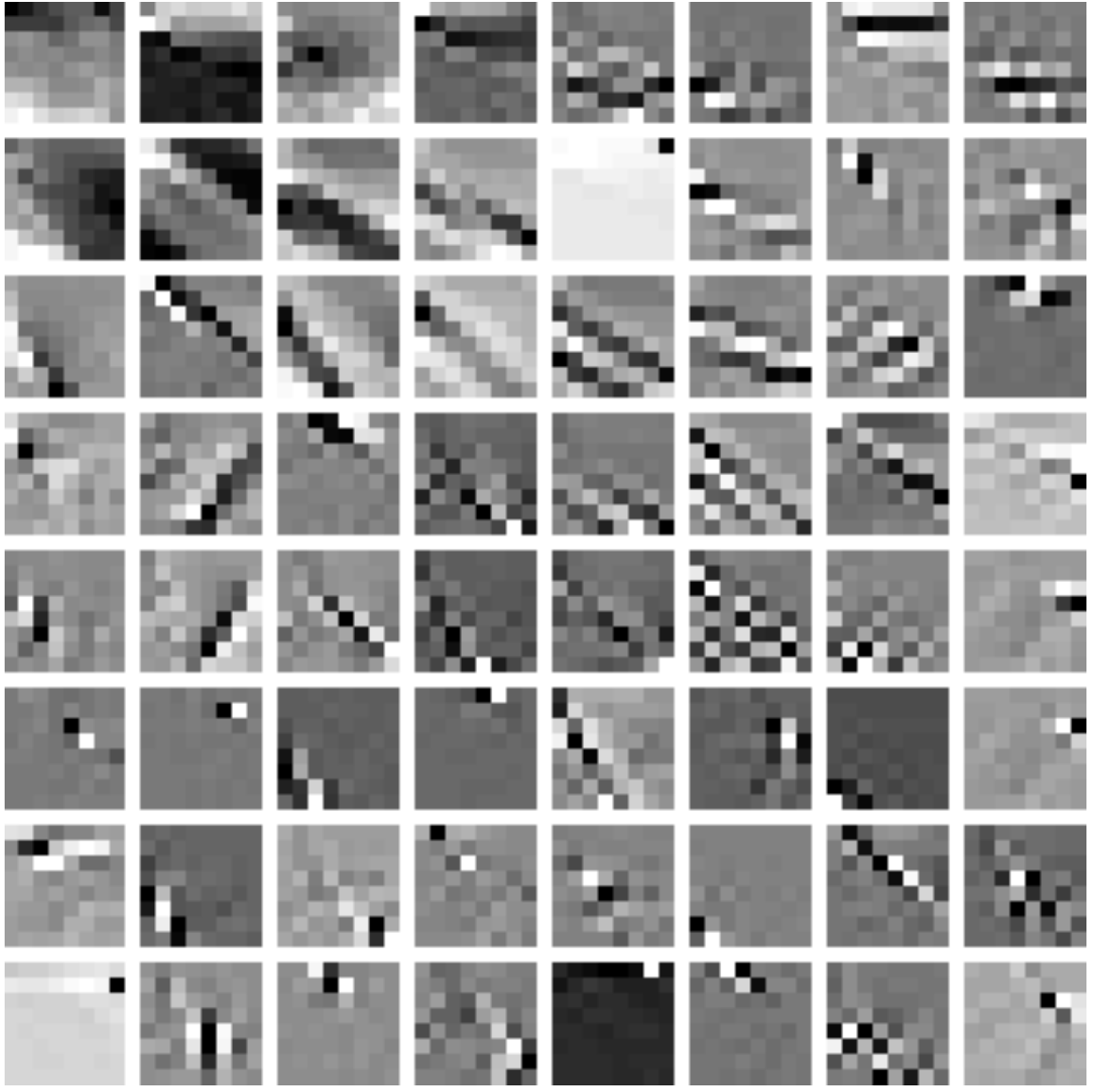}  \\
	\vspace{0.02in}
	\begin{overpic}[width=0.22\textwidth]{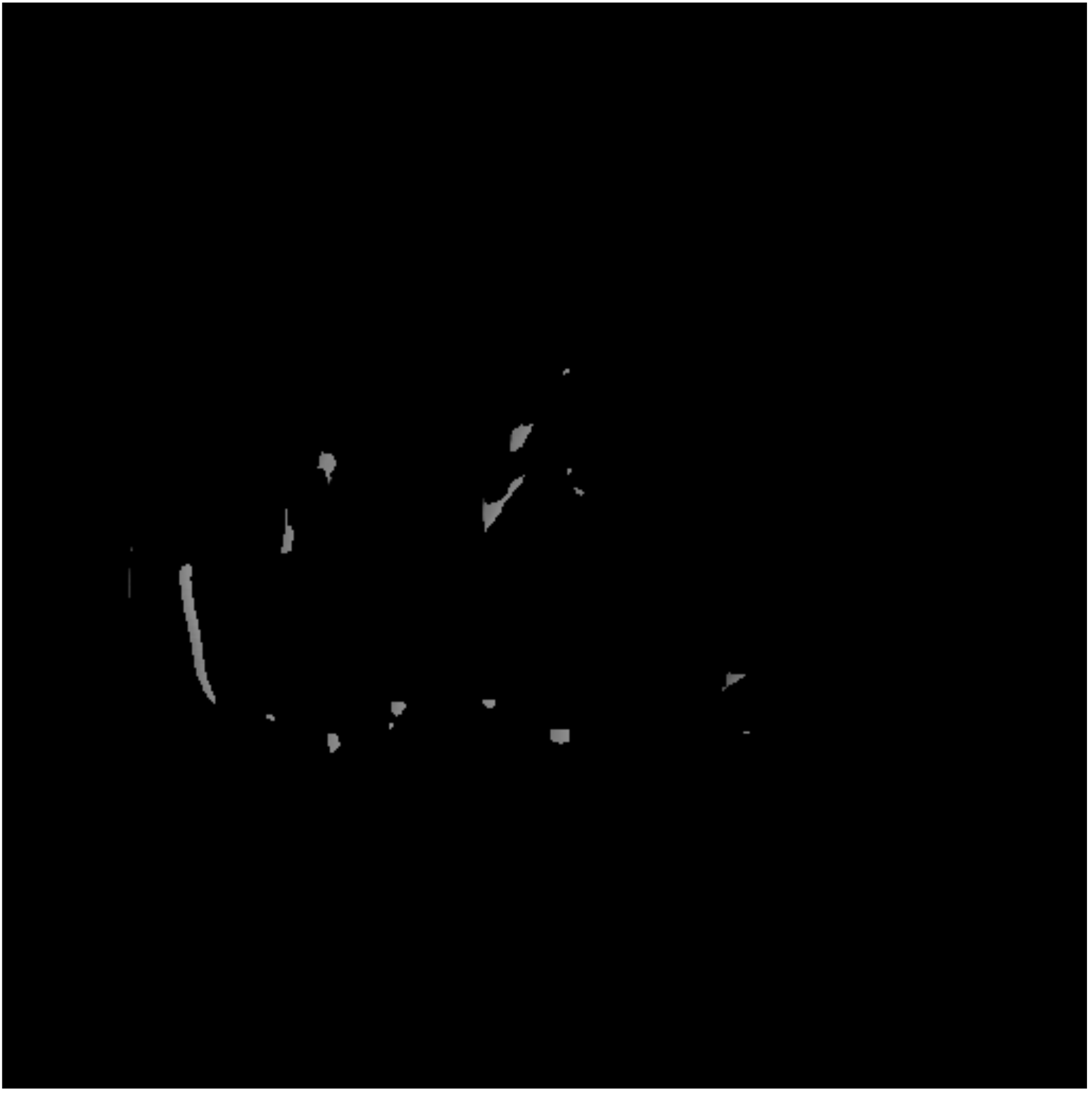}
		\put(30,85){ \color{white}{\bf \large{Class 1}}}     
	\end{overpic}
	\begin{overpic}[width=0.22\textwidth]{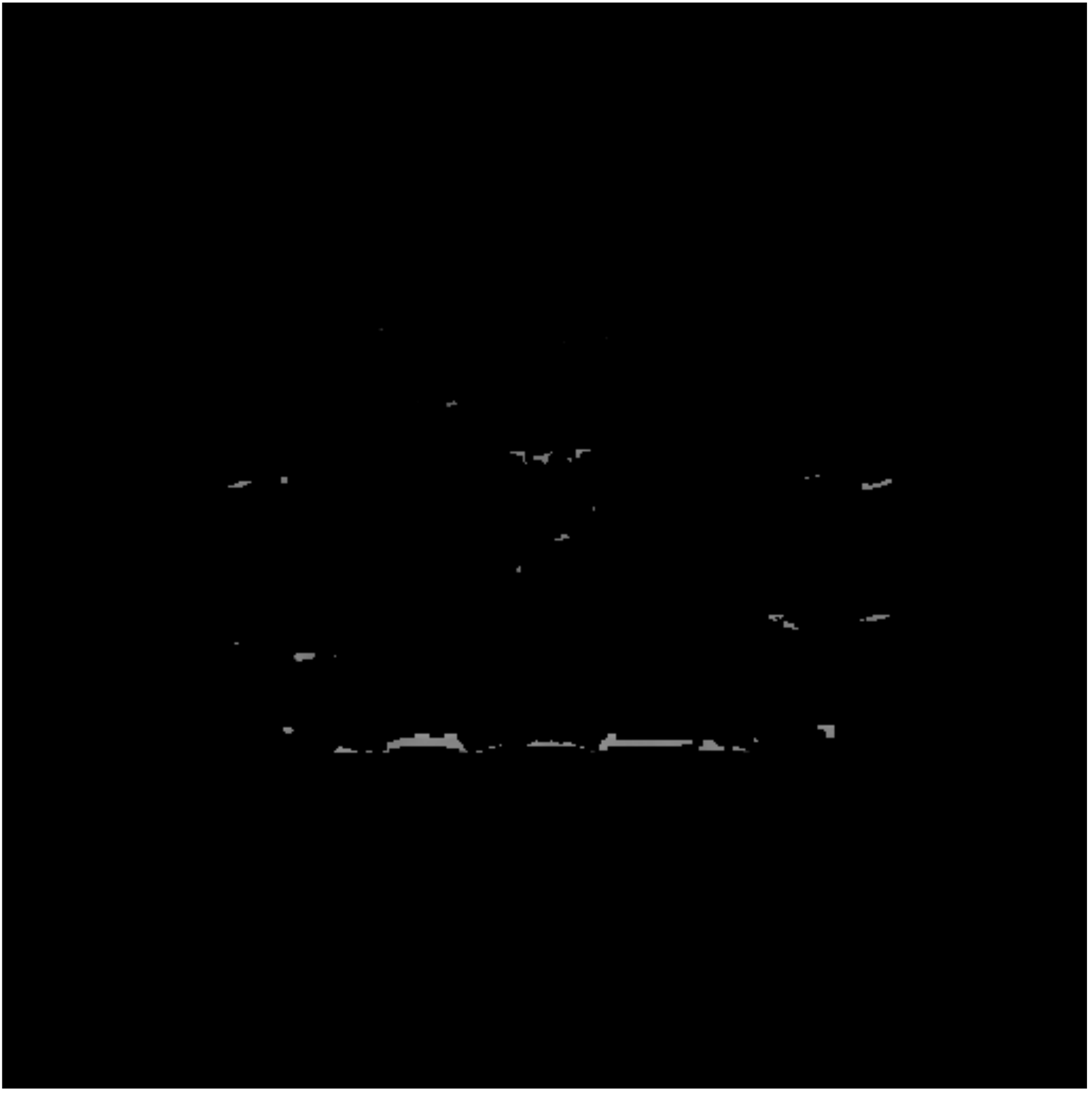}
		\put(30,85){ \color{white}{\bf \large{Class 3}}}     
	\end{overpic}
	\begin{overpic}[width=0.22\textwidth]{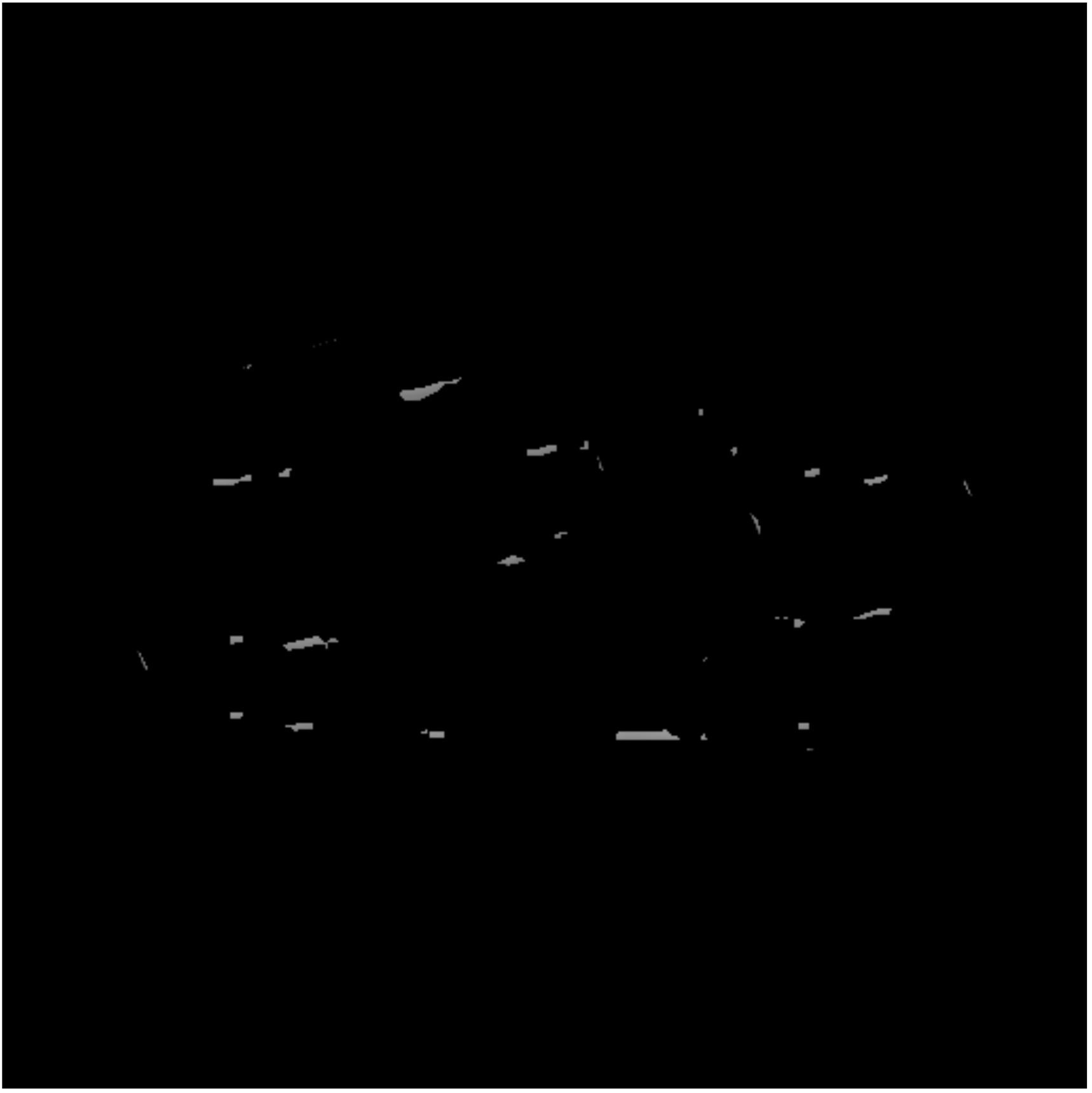}
		\put(30,85){ \color{white}{\bf \large{Class 7}}}     
	\end{overpic}
	\begin{overpic}[width=0.22\textwidth]{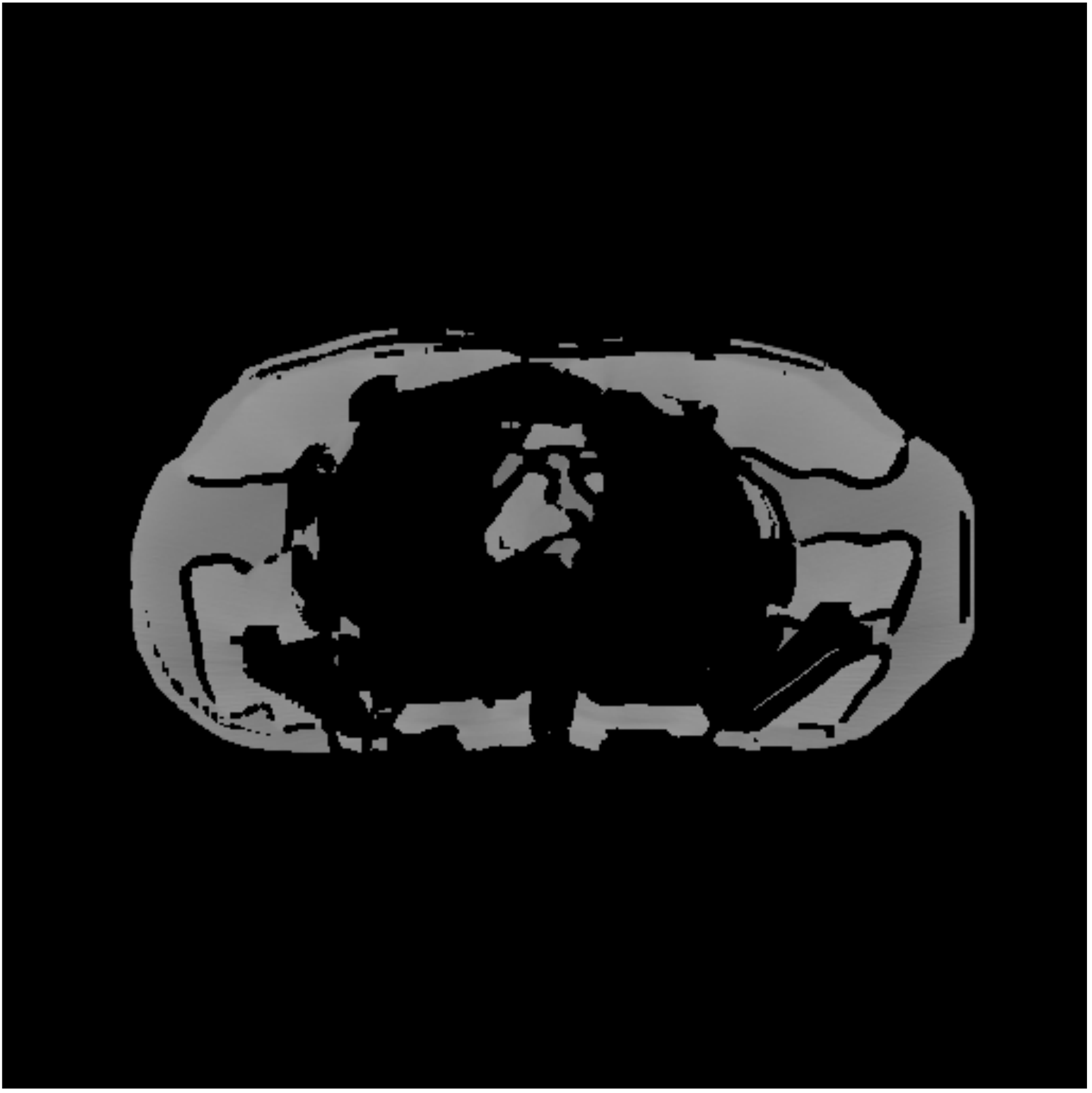}
		\put(30,85){ \color{white}{\bf \large{Class 11}}}     
	\end{overpic}				
	\caption{Pixel-level clustering in slice 77 with the DECT-MULTRA common-material model: The top row shows the individual common-material transforms for classes $1$, $3$, $7$, and $11$, with the transform rows shown as $8 \times 8$ patches. The bottom row shows the corresponding water pixels (using estimated densities in the decomposition) grouped into each class, with display windows [0.7  1.3]\,g/cm$^3$. }
	\label{Fig:Common_Clustering}
\end{figure*}

\subsection{Decompositions of Head Data Using ST-based Methods}
Section \uppercase\expandafter{\romannumeral4}.D of \cite{li:18:tmi} evaluated multiple image-domain material decomposition methods using clinical DECT head data. 
Here, we show further comparisons of decompositions obtained by DECT-ST, DECT-CULTRA, and DECT-MULTRA for the clinical head data. 
For DECT-ST, the parameters $\{\beta,\,\gamma\}$ were set as $\{150,\,0.012\}$ and $\{200,\,0.024\}$ for water and bone, respectively.
For DECT-CULTRA, the parameters $\{\beta,\,\gamma\}$ were set as $\{200,\,0.024\}$. These parameters provided good visual quality of the decompositions.
Fig.~\ref{Fig:head_comp_directEP} shows the material density images decomposed by DECT-ST, DECT-CULTRA, and DECT-MULTRA. 
DECT-MULTRA reduces artifacts (e.g., blocky artifacts) at the boundaries of different materials compared to DECT-ST. 
It also improves the sharpness and contrast of edges in the soft-tissue compared to DECT-CULTRA.

\subsection{NPS of Water Images Obtained by Different Methods}
To evaluate the noise texture with DECT-MULTRA, we selected several areas (whose positions are indicated by red squares in  Fig.~\ref{fig:nps}a) in the water error image as regions of interest (ROIs). The noise power spectrum (NPS) is then measured within each ROI of $30$ by $30$ pixels. The 2D NPS is defined as $NPS = |DFT_2 \{f\}|^2$, where $f$ denotes the ROI of the error image in which gray values are offset to achieve zero mean, and $DFT_2\{f\}$ is the 2D Discrete Fourier Transform (DFT) of $f$.  The NPS comparison for different method is shown in Fig.~\ref{fig:nps}b.	
It is obvious that DECT-MULTRA achieves a better NPS than DECT-TDL, especially in ROI \#1 and ROI \#2. What's more, the overall noise in the ROIs of the DECT-MULTRA decomposition is much less than that for DECT-EP, and the direct matrix inversion method. This shows the superiority of the proposed MULTRA approach.

\subsection{Examples of Common-material Transforms and Corresponding Clustering Results }
Fig.~\ref{Fig:Common_Clustering} shows common-material transforms for four classes along with the pixels grouped with them in slice 77.
Common-material transforms clearly appear different from the cross-material transforms, and they capture most of the water areas.
Because patches overlapping the bone areas usually also overlap water areas, obviously these areas are mainly grouped with the cross-material transforms (e.g., clustering results shown in Fig.~5 of the main paper \cite{li:18:tmi}). So the clustering results for the common-material classes primarily show one material rather than mixed materials.  
 

\subsection{Decompositions of XCAT phantom obtained by DECT-MULTRA with $\beta_1=99$ and $\beta_2=1$}
\begin{figure}[htb]
	\centering
	\begin{tikzpicture}
	[spy using outlines={rectangle,red,magnification=2,width=15mm, height =10mm, connect spies}]				
	\node {\includegraphics[width=0.22\textwidth]{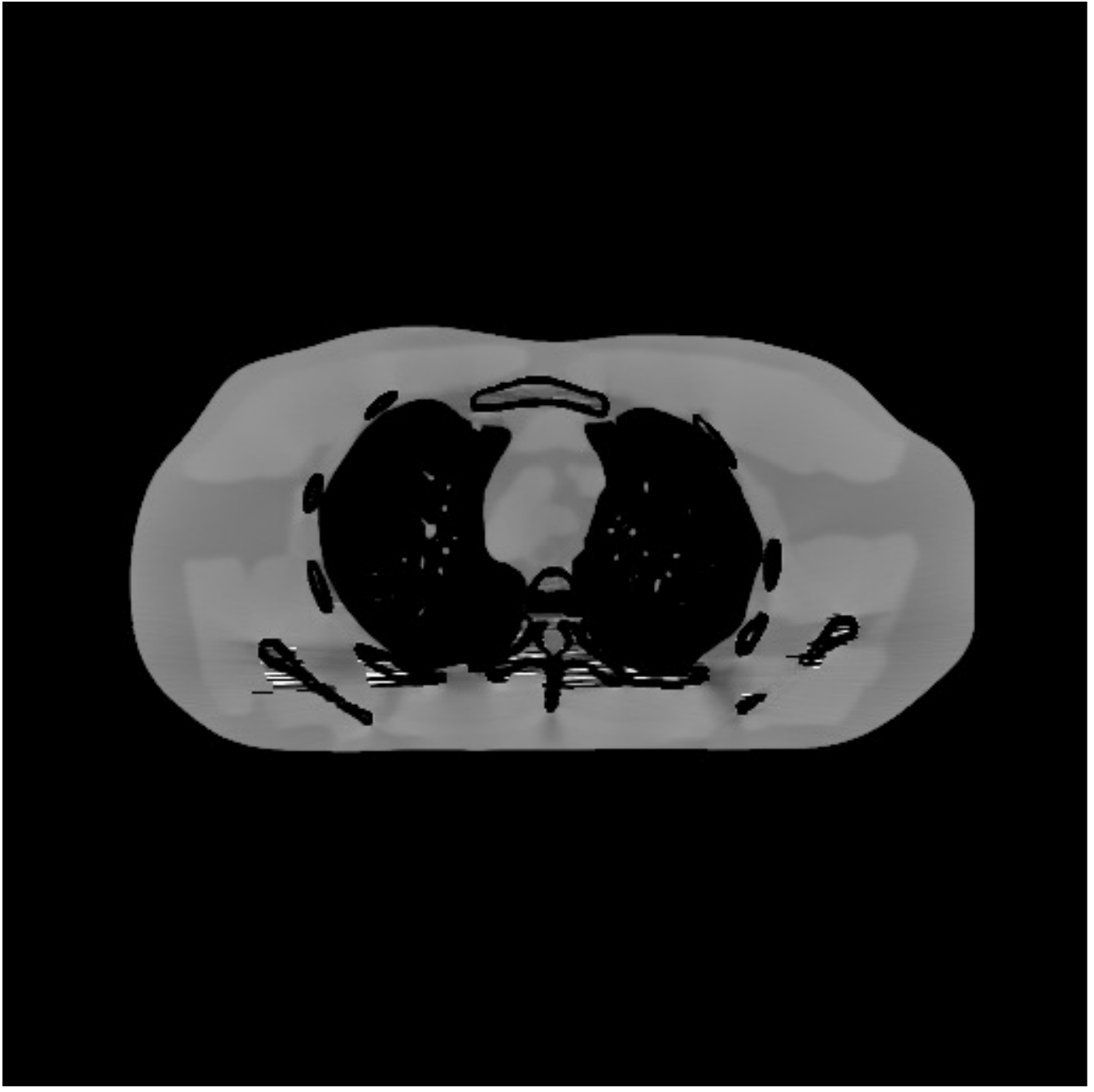}};
	\spy on (0.6,-0.5) in node [right] at (0.1,-1.35);	
	\spy on (-0.7,-0.5) in node [right] at (-1.8,-1.35);
	\draw[line width=1pt, thin, -latex, red] (0.55,-0.70) -- node[auto] {} (0.35,-0.53);
	\draw[line width=1pt, thin, -latex, red] (-0.45,-0.70) -- node[auto] {} (-0.53,-0.53);					
	\end{tikzpicture}
	\begin{tikzpicture}
	[spy using outlines={rectangle,red,magnification=2,width=15mm, height =10mm, connect spies}]				
	\node {\includegraphics[width=0.22\textwidth]{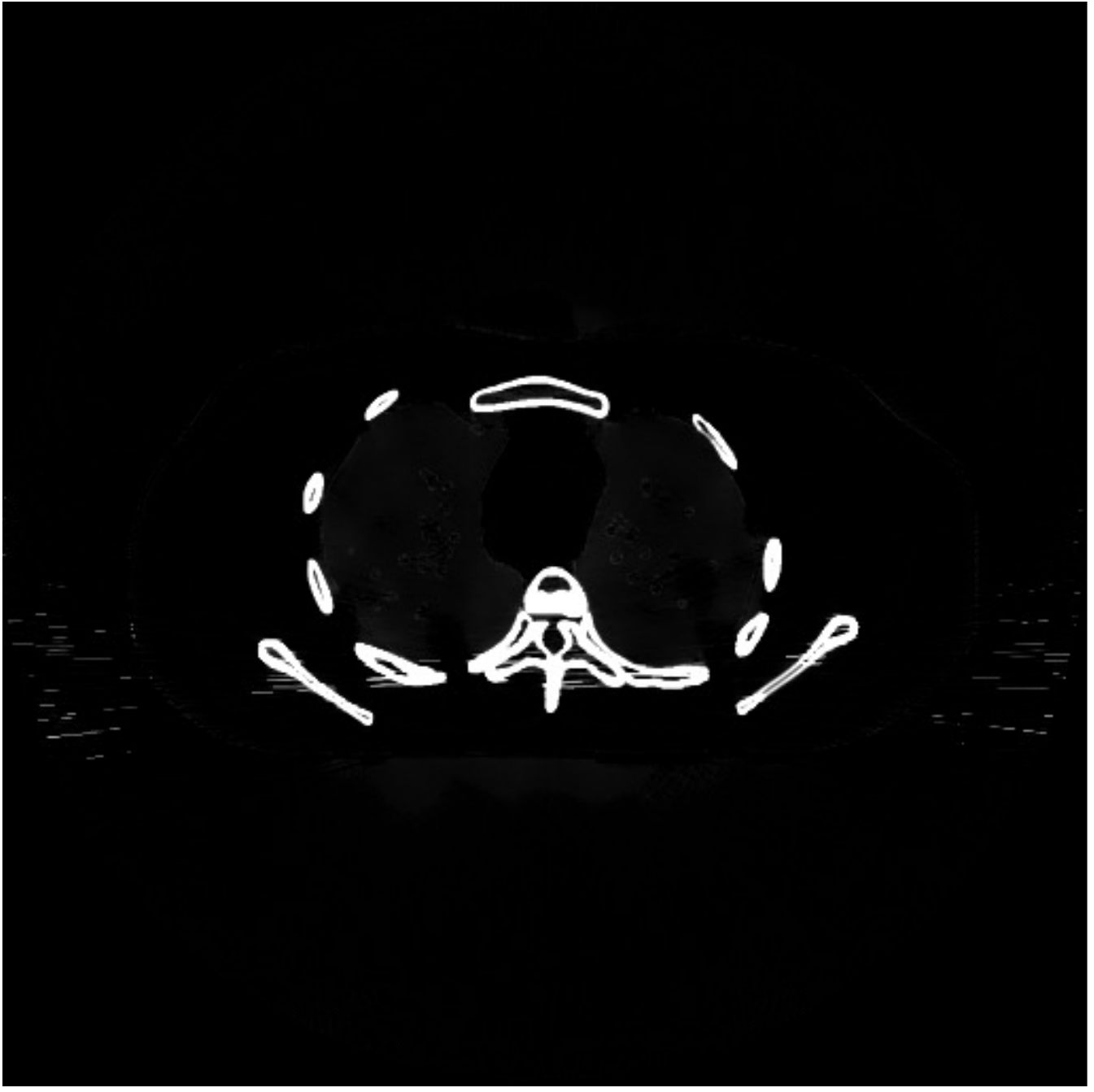}};
	\spy on (0.6,-0.5) in node [right] at (0.1,-1.35);	
	\spy on (-0.7,-0.5) in node [right] at (-1.8,-1.35);	
	\draw[line width=1pt, thin, -latex, red] (0.55,-0.7) -- node[auto] {} (0.35,-0.53);
	\draw[line width=1pt, thin, -latex, red] (-0.45,-0.7) -- node[auto] {} (-0.53,-0.53);					
	\end{tikzpicture}
	\caption{Material images decomposed by DECT-MULTRA with $\{\beta_1\,,\beta_2\}$ set as $\{99,\,1\}$. The water and bone images are shown with display windows [0.7  1.3]\,g/cm$^3$ and [0 0.8]\,g/cm$^3$, respectively. The RMSE of water and bone images are $55.2\times 10^{-3}$~g/cm$^3$ and $61.0\times10^{-3}$~g/cm$^3$, respectively.} 
	\label{fig:weight_991}
\end{figure}
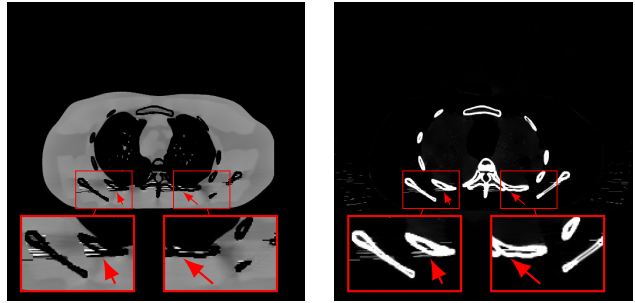

Fig.~\ref{fig:weight_991} shows water and bone material images decomposed by DECT-MULTRA with $\beta_1=99$ and $\beta_2=1$. The weight combination $\{\beta_1=99,\,\beta_2=1 \}$ improves the contrast but leads to artifacts near the boundaries of water and bone.

\bibliographystyle{IEEEbib}
\bibliography{refs}